\documentclass[sensors,article,accept,pdftex,moreauthors]{Definitions/mdpi} 
\usepackage{longtable}
\firstpage{1} 
\pubvolume{1}
\issuenum{1}
\articlenumber{0}
\pubyear{2024}
\copyrightyear{2024}
\externaleditor{Jorn Mehnen}
\datereceived{26 October 2024} 
\daterevised{19 December 2024} 
\dateaccepted{20 December 2024} 
\datepublished{ } 
\hreflink{https://doi.org/} 

\Title{{Securing Cloud-Based Internet of Things: Challenges and Mitigations}}

\TitleCitation{Securing Cloud-Based Internet of Things: Challenges and Mitigations}

\Author{Nivedita 
 Singh $^{1}$\orcidA{}, Rajkumar Buyya $^{2}$\orcidB{} and Hyoungshick Kim $^{1,*}$\orcidC{}}

\AuthorNames{Nivedita Singh, Rajkumar Buyya and Hyoungshick Kim}

\AuthorCitation{Singh, N.; Buyya, R.;\\ Kim, H.}

\address{%
$^{1}$ 
Department 
 of Computer Science \& Engineering, Sungkyunkwan University, Suwon, 16419, Republic of Korea; 
 singhnivvy@g.skku.edu\\
$^{2}$ \quad Cloud Computing and Distributed Systems (CLOUDS) Lab, School of Computing and Information Systems, The University of Melbourne, Victoria 3010 Australia; rbuyya@unimelb.edu.au\\
}

\corres{Correspondence: hyoung@skku.edu}

\abstract{The Internet of Things (IoT) has seen remarkable advancements in recent years, leading to a paradigm shift in the digital landscape. However, these technological strides have introduced new challenges, particularly in cybersecurity. IoT devices, inherently connected to the internet, are susceptible to various forms of attacks. Moreover, IoT services often handle sensitive user data, which could be exploited by malicious actors or unauthorized service providers. As IoT ecosystems expand, the convergence of traditional and cloud-based systems presents unique security threats in the absence of uniform regulations. Cloud-based IoT systems, enabled by Platform-as-a-Service (PaaS) and Infrastructure-as-a-Service (IaaS) models, offer flexibility and scalability but also pose additional security risks. The intricate interaction between these systems and traditional IoT devices demands comprehensive strategies to protect data integrity and user privacy. This paper highlights the pressing security concerns associated with the widespread adoption of IoT devices and services. We propose viable solutions to bridge the existing security gaps while anticipating and preparing for future challenges. This paper provides a detailed survey of the key security challenges that IoT services are currently facing. We also suggest proactive strategies to mitigate these risks, thereby strengthening the overall security of IoT devices and services.}


\begin{document}
\section{Introduction}
Imagine your morning coffee starts brewing as you wake up, your thermostats adjust automatically based on the weather, and a notification pops up reminding you to take your medication. The ubiquitous Internet of Things (IoT) is weaving itself into the fabric of our daily lives, promising a seamless and convenient future.
Applications range from smart homes to AI-powered medical equipment, intelligent agriculture, and advancements in the automotive industry~\citep{kumar2019internet}. 
Notably, IoT health monitoring systems have significantly improved the quality of life for the elderly and disabled, with the COVID-19 outbreak highlighting IoT's potential in health~\citep{kamal2020iot}.
Unlike traditional IoT, which relies on localized processing and storage, cloud-based IoT utilizes external cloud infrastructure for scalability, advanced analytics, and seamless device connectivity. This shift introduces unique challenges, including reliance on third-party providers, expanded attack surfaces, and privacy risks from centralized data storage, necessitating specialized security and privacy solutions.

With every connected IoT device comes a question: are we prioritizing comfort over security? 
This concern becomes even more serious as we delve into the world of cloud-based IoT.
This survey delves into the critical security and privacy risks emerging from the integration of IoT devices with cloud-based services, which facilitate data processing, storage, and communication. These challenges, central to this study, highlight the need for robust strategies to safeguard IoT systems as they increasingly depend on cloud infrastructure.
Traditional IoT systems come with inherent limitations, particularly in terms of memory, storage, processing, and communication capabilities. This has prompted many cloud-based third-party service companies to offer external solutions like Amazon Web Services (AWS) IoT and Azure IoT. These services provide a spectrum of capabilities, including data storage, data processing, and application hosting, which can help IoT devices collect, analyze, and act on data more effectively~\citep{tabrizi2017review}. However, entrusting sensitive data and device functionality to external providers introduces a new set of security challenges apart from already existing challenges in IoT that demand immediate attention. 

\begin{figure}[H]
\includegraphics[width=4in, height =2in]{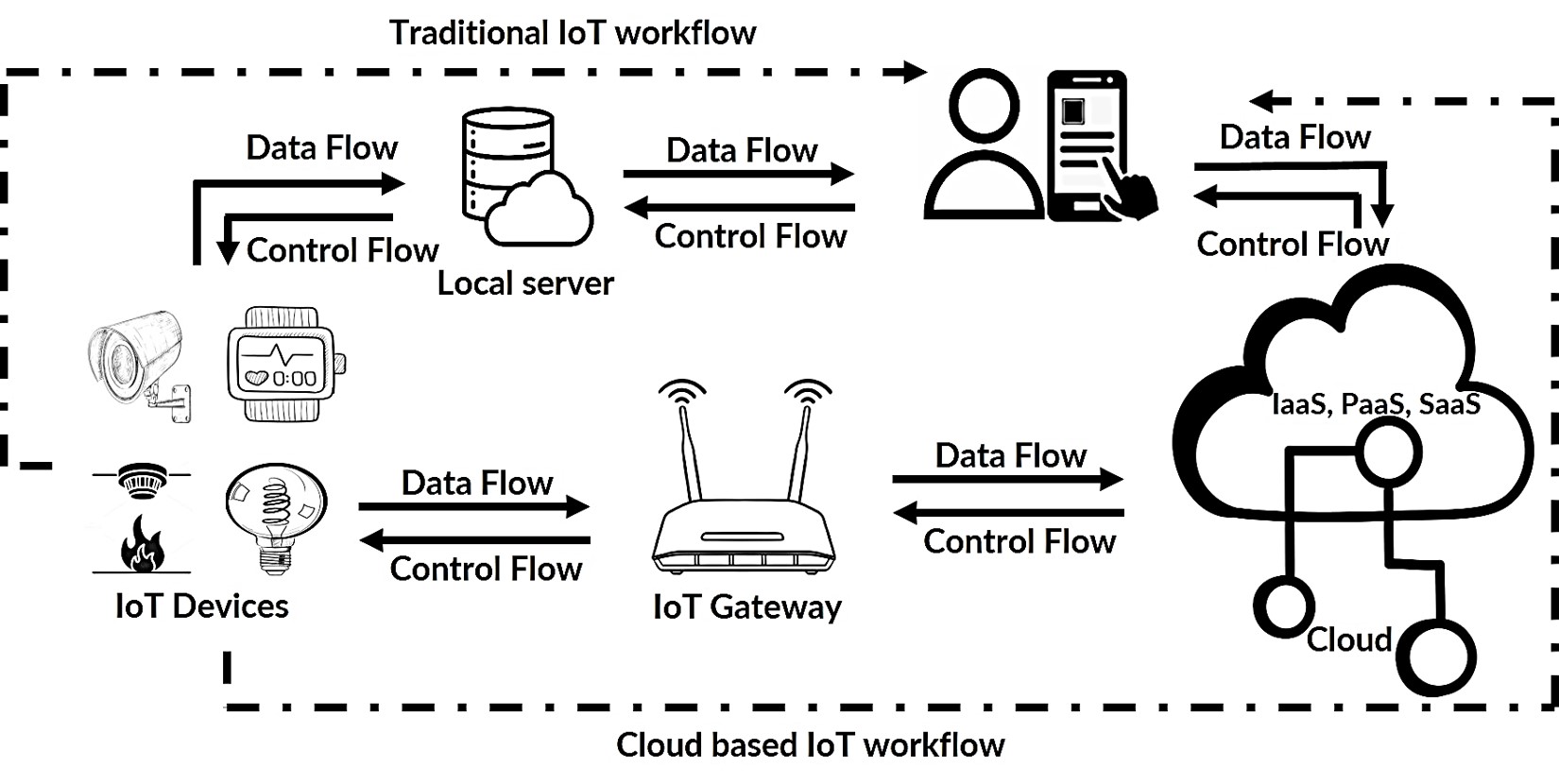}
\caption{IoT 
 cloud ecosystem model.}
\label{fig:IoTCloudModel}
\end{figure}

In Figure~\ref{fig:IoTCloudModel}, we illustrate the comparison between traditional and cloud-based IoT systems, denoted by a dashed line. Essential components include IoT devices like temperature sensors and smart air conditioners, a local server in traditional IoT setups, an IoT gateway, and the cloud in cloud-based systems, all interconnected to end users through mobile apps. The IoT system is a complex ecosystem of interconnected components that collect, process, store, and control data from IoT devices. 
This interaction is fundamental to the seamless functioning of the entire system. However, the involvement of third-party systems introduces a new dimension to this ecosystem, one that requires careful consideration, particularly in terms of security. For example, inadequately secure third-party systems may lead to data leaks, compromising the integrity of the IoT ecosystem. Moreover, the diversity of IoT components introduces additional security challenges, as each component may possess unique vulnerabilities. Consequently, prioritizing robust security measures becomes imperative for ensuring the resilience of IoT systems, especially since these devices frequently interact with human users and are sometimes connected to critical infrastructure.
\vspace{-16pt}

Cloud-based IoT has transformed our lives, making homes smarter and more convenient. However, as it expands into critical areas like consumer devices and transportation, addressing security vulnerabilities becomes crucial. Despite efforts by major cloud providers like AWS, Microsoft Azure, Google Cloud, and IBM Cloud, security remains a significant concern. This has drawn substantial attention, prompting researchers to actively identify vulnerabilities in the existing IoT infrastructure.~\citep{alrawi2019sok, apthorpe2017smart, bradley2018security, celik2018soteria, chu2018security, denning2013computer, ding2018safety}.

For example, a hacker could exploit a vulnerability in a wearable device to access users' personal health data. Jin et al.~\citep{jin2022p} discovered vulnerabilities in 36 cloud-based IoT devices, including the 'Hippokura' medical IoT system, where an attacker could gain unauthorized access and read conversations between patients and doctors. Additionally, incidents where an attacker hacks into a smart car's system to take control of the vehicle could become common in the future. To make matters worse, an attacker can lure a vehicle to a different location than intended using a `GPS deception attack'~\citep{samad2018internet}.

The integration of IoT with cloud systems involves continuous communication, making them susceptible to adversarial attacks, eavesdropping, MITM attacks, and packet tampering ~\citep{michelena2024novel, sivasankari2022detection}. These security concerns are critical, especially for IoT devices in essential systems like power grids, where a cyber attack could cause widespread outages. Multiple IoT systems relying on cloud infrastructure pose significant security risks. 

The rapid development of IoT systems often prioritizes cost over security, creating exploitable vulnerabilities. Manufacturers frequently neglect device security, viewing the cloud merely as a platform. IoT's distributed architecture and communication complexities pose unique security challenges~\citep{jin2022p}. 
Despite ongoing research, there are a lack of frameworks to systematically categorize IoT devices for addressing security issues, hindering the development of effective security measures.

We aim to review security issues in cloud-based IoT systems and address gaps in the literature, motivated by the need for a targeted security approach. Unlike existing \mbox{studies~\citep{chaudhary2023ddos, yang2022review}} that focus on isolated IoT scenarios, this paper categorizes cloud-based IoT into ten distinct types, each with unique challenges, and proposes tailored mitigation strategies. We highlight the importance of addressing these issues amid evolving technologies and regulatory demands. This leads to the following research questions:

\begin{itemize}
    \item RQ1: can we comprehensively categorize the various cloud-based IoT devices based on their specific purpose and address their existing or future security/privacy issues?
    \item RQ2: what mitigation approaches can be employed in each category to address the identified security/privacy issues?
    \item { RQ3: how can standardized verification protocols be customized to balance security and usability across all distinct IoT categories?}
  \end{itemize}

\begin{figure}[H]
\includegraphics[width=3.2in ]{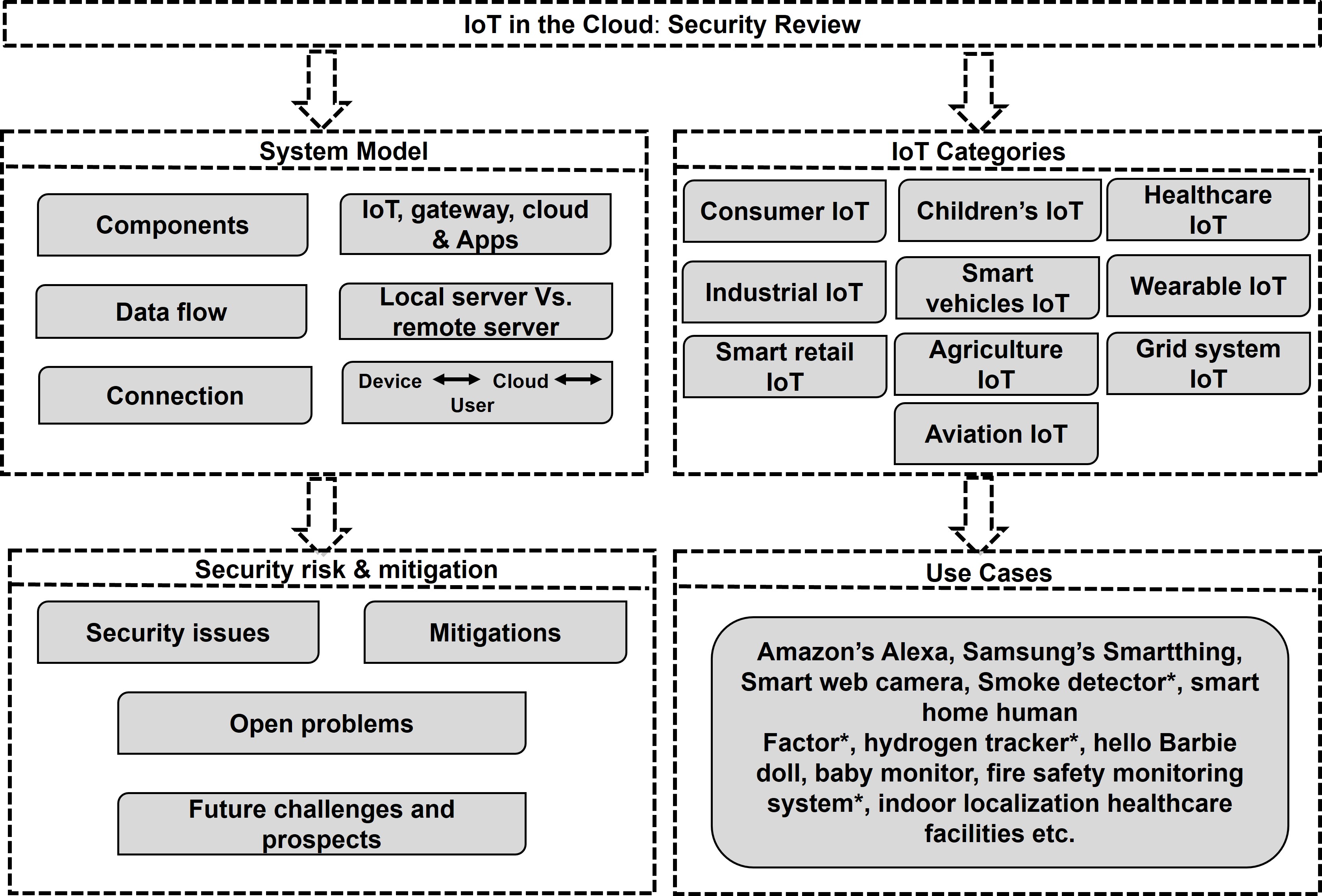}
\caption{Article 
 framework. {This 
 figure outlines the system model, IoT categories, security risks, mitigations, and related use cases. The 'Use Cases' represent real-world scenarios that illustrate the identified security issues or demonstrate the application of proposed mitigation strategies. While some use cases highlight cloud-based IoT systems, others showcase IoT systems transitioning toward cloud integration, emphasizing the relevance of the cloud in their future evolution.}}
\label{fig_2}
\end{figure}

\begin{figure}[H]
 \includegraphics[width=1\textwidth]{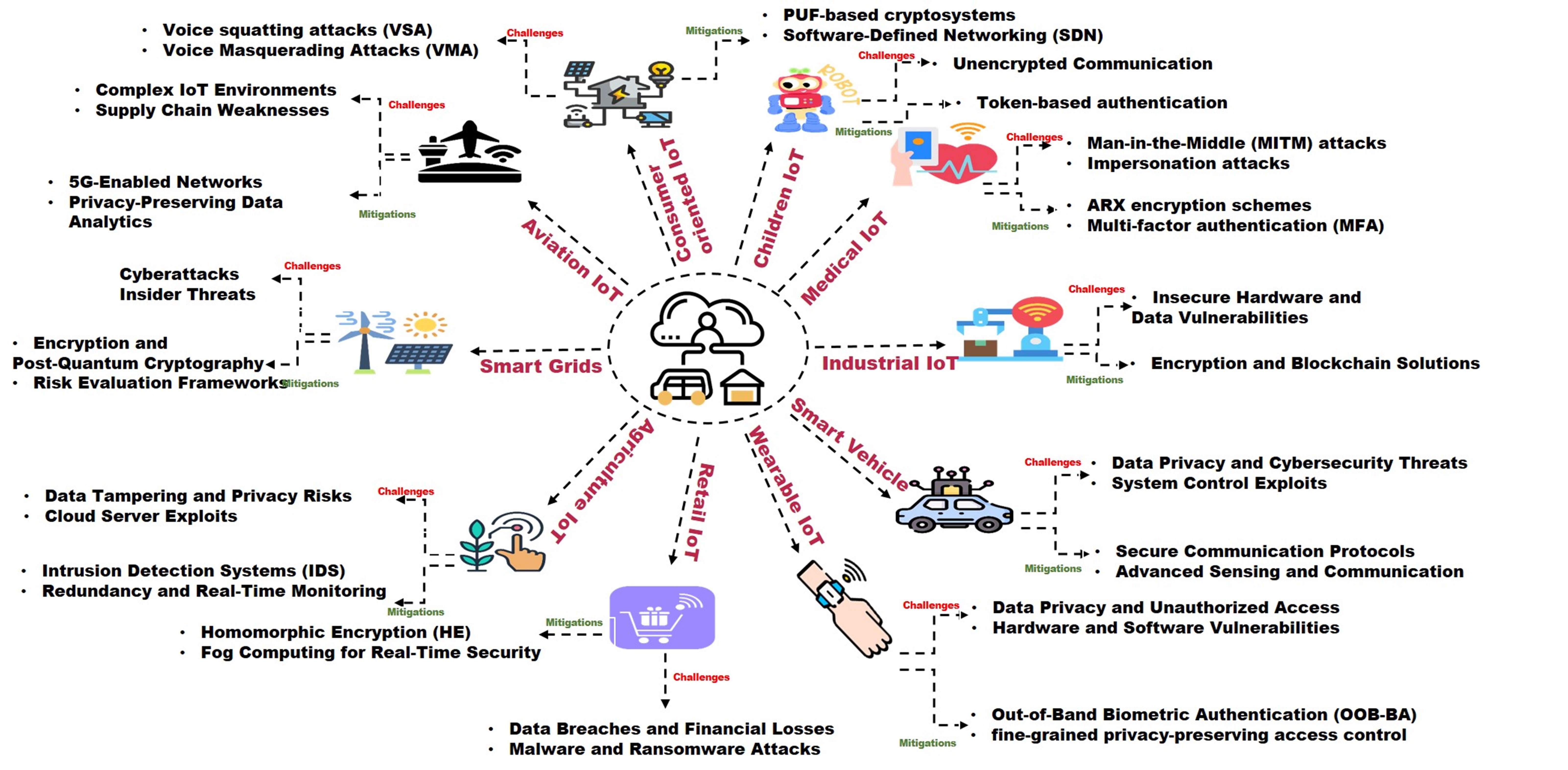}
\caption{Overview of the article, highlighting key security challenges and providing examples of mitigation strategies.}
\label{fig:Fig_New}
\end{figure}

In our article, Figure~\ref{fig_2} outlines the key sections, providing an overview of the structure, while Figure~\ref{fig:Fig_New} provides an overview of the main content, emphasizing key challenges and corresponding mitigation strategies. We begin with a system model to understand security challenges, followed by blocks that examine security/privacy risks and mitigation approaches. The final block showcases real use cases addressing these issues. 

The rest of this article is structured as follows: Section 
 \ref{sec2} covers the IoT cloud ecosystem model, literature review (Table~\ref{tab:mytable}), and our novel contributions. Section \ref{sec3} categorizes IoT challenges and solutions, addressing RQ1 and RQ2. In Section \ref{sec4}, we explore the way
forward (RQ3), culminating in our conclusions. We have included a summary table (Table~\ref{tab:SS}) at the end that consolidates the key points from each category for quick reference.
\vspace{-4pt}
\unskip
\section {Background and Related Work}\label{sec2}
This section delves into the fundamental concepts of IoT and cloud computing. We explore how these separate domains mutually benefit. However, it is noteworthy to acknowledge that this integration also brings forth concerns regarding security and \mbox{privacy~\citep{surya2016security}}.
\subsection{IoT Cloud Ecosystem Model}
Traditional IoT relies on localized data processing, offering better control but limited scalability. In contrast, cloud-based IoT enables real-time analytics and high scalability via cloud storage but introduces risks such as reliance on central storage, trust issues, and amplified privacy concerns. While data breaches affect both systems, the aggregation of sensitive user data in cloud-based IoT increases the attack surface, necessitating robust trust and privacy safeguards.

IoT devices form a connected system enabling electronic devices and sensors to communicate via the internet, enhancing daily life. Modern IoT cloud systems go beyond traditional frameworks, supporting large-scale implementations that serve millions of users and manage vast amounts of data from IoT devices~\citep{georgakopoulos2016internet}. In this context, the IoT gateway (refer to Figure~\ref{fig:IoTCloudModel}) connects IoT devices to the cloud, enabling local communication and edge data processing. 
A key feature of cloud-based IoT systems is interoperability through standard protocols like HTTP PUT and GET requests, allowing user access. The IoT hub is another essential component, enabling various devices, manufacturers, and users to join the IoT ecosystem~\citep{zhou2013cloudthings}.
\subsubsection{IoT Systems}
The concept of IoT revolves around the idea of ``smartness'', enabling devices and sensors to autonomously acquire and apply knowledge~\citep{shafique2020internet}. An IoT system is a network of interconnected devices, sensors, and software that collect, exchange, and analyze data, each presenting unique security challenges.
\subsubsection{Cloud Computing in IoT} 
Cloud computing's three service models---Infrastructure as a Service (IaaS, e.g., Amazon Web Services), Platform as a Service (PaaS, e.g., Microsoft Azure), and Software as a Service (SaaS, e.g., Google Apps)---benefit IoT devices based on their specific needs~\citep{atlam2017integration}. It offers massive data storage, scalability, computational power, and infrastructure to support the extensive data processing and analytics required for IoT applications~\citep{cai2016iot}.
\subsection{Literature Review and Surveys on IoT Security and Privacy}
Research often lags behind, leaving gaps in addressing security challenges at the intersection of IoT and cloud computing.
For instance, Thakor et al.~\citep{thakor2021lightweight} highlight the challenge of securing resource-constrained devices and proposing cryptographic solutions. As IoT systems increasingly rely on cloud computing, recognizing the importance of cloud integrations and their associated security risks is urgent, as shown in Table~\ref{tab:mytable}.
However, these studies often lack systematic categorization of IoT vulnerabilities across scenarios like healthcare, children's IoT, and agriculture, while offering limited insights into open challenges and future directions in cloud-integrated IoT.

In this direction, Chen et al.~\citep{10.1145/3447625} highlighted rising security risks with the increased use of IoT devices and cloud computing in consumer applications. They identified vulnerabilities like weak passwords, unsecured communications, and poor access controls. They recommend implementing strong security measures and educating consumers about risks and protections.


\begin{table}[H] 
\caption{Comparison 
 with recent surveys.\label{tab:mytable}}
\begin{tabularx}{\textwidth}{CCCCC}
\toprule
\textbf{Publication (Year)} & \textbf{IoT Challenges} & \textbf{Cloud Challenges} & \textbf{Mitigations} & \textbf{Use Cases} \\ 
\midrule
\citep{alaba2017internet}   & \raisebox{-0.5ex}{\begin{tikzpicture} 
    \draw (0,0) circle (1.5mm);
    \fill[black] (0,0) -- (90:1.5mm) arc (90:270:1.5mm) -- cycle;
    \fill[white] (0,0) -- (270:1.5mm) arc (-90:90:1.5mm) -- cycle;
    \end{tikzpicture}} 
&   \raisebox{-0.5ex}{\begin{tikzpicture} 
    \draw (0,0) circle (1.5mm); 
    \end{tikzpicture}} 
&   \raisebox{-0.5ex}{\begin{tikzpicture} 
    \draw (0,0) circle (1.5mm);
    \fill[black] (0,0) -- (90:1.5mm) arc (90:270:1.5mm) -- cycle;
    \fill[white] (0,0) -- (270:1.5mm) arc (-90:90:1.5mm) -- cycle;
    \end{tikzpicture}} 
&   \raisebox{-0.5ex}{\begin{tikzpicture} 
    \draw (0,0) circle (1.5mm);
    \fill[black] (0,0) -- (90:1.5mm) arc (90:270:1.5mm) -- cycle;
    \fill[white] (0,0) -- (270:1.5mm) arc (-90:90:1.5mm) -- cycle;
    \end{tikzpicture}}\\
(2017) & & & & \\
\hline
\citep{tabrizi2017review}   & \raisebox{-0.5ex}{\begin{tikzpicture}
    \draw (0,0) circle (1.5mm);
    \end{tikzpicture}} 
&   \raisebox{-0.5ex}{\begin{tikzpicture}
    \draw (0,0) circle (1.5mm); 
    \end{tikzpicture}} 
&   \raisebox{-0.5ex}{\begin{tikzpicture} 
    \draw (0,0) circle (1.5mm); 
    \end{tikzpicture}} 
&   \raisebox{-0.5ex}{\begin{tikzpicture}
    \draw (0,0) circle (1.5mm);
    \fill[black] (0,0) -- (90:1.5mm) arc (90:270:1.5mm) -- cycle;
    \fill[white] (0,0) -- (270:1.5mm) arc (-90:90:1.5mm) -- cycle;
    \end{tikzpicture}}\\
(2017) & & & & \\
\hline
\citep{kumar2019internet}   & \raisebox{-0.5ex}{\begin{tikzpicture}           
    \fill[black] (0,0) circle (1.5mm); 
    \end{tikzpicture}} 
&   \raisebox{-0.5ex}{\begin{tikzpicture} 
    \draw (0,0) circle (1.5mm);
    \end{tikzpicture}} 
&   \raisebox{-0.5ex}{\begin{tikzpicture}
    \draw (0,0) circle (1.5mm);
    \fill[black] (0,0) -- (90:1.5mm) arc (90:270:1.5mm) -- cycle;
    \fill[white] (0,0) -- (270:1.5mm) arc (-90:90:1.5mm) -- cycle;
    \end{tikzpicture}} 
&   \raisebox{-0.5ex}{\begin{tikzpicture} 
    \draw (0,0) circle (1.5mm); 
    \end{tikzpicture}}\\
(2019) & & & & \\
\hline
\citep{harbi2019review}  & \raisebox{-0.5ex}{\begin{tikzpicture}             \fill[black] (0,0) circle (1.5mm); 
    \end{tikzpicture}} 
&   \raisebox{-0.5ex}{\begin{tikzpicture} 
    \draw (0,0) circle (1.5mm); 
    \end{tikzpicture}} 
&   \raisebox{-0.5ex}{\begin{tikzpicture}
    \draw (0,0) circle (1.5mm);
    \fill[black] (0,0) -- (90:1.5mm) arc (90:270:1.5mm) -- cycle;
    \fill[white] (0,0) -- (270:1.5mm) arc (-90:90:1.5mm) -- cycle;
    \end{tikzpicture}}
&   \raisebox{-0.5ex}{\begin{tikzpicture} 
    \fill[black] (0,0) circle (1.5mm); 
    \end{tikzpicture}}\\
(2019) & & & & \\
\hline
\citep{shafique2020internet}  & \raisebox{-0.5ex}{\begin{tikzpicture}
    \draw (0,0) circle (1.5mm);
    \fill[black] (0,0) -- (90:1.5mm) arc (90:270:1.5mm) -- cycle;
    \fill[white] (0,0) -- (270:1.5mm) arc (-90:90:1.5mm) -- cycle;
    \end{tikzpicture}} 
&   \raisebox{-0.5ex}{\begin{tikzpicture} 
    \draw (0,0) circle (1.5mm); \end{tikzpicture}}
&   \raisebox{-0.5ex}{\begin{tikzpicture} 
    \draw (0,0) circle (1.5mm); 
    \end{tikzpicture}} 
&   \raisebox{-0.5ex}{\begin{tikzpicture}
    \draw (0,0) circle (1.5mm);
    \fill[black] (0,0) -- (90:1.5mm) arc (90:270:1.5mm) -- cycle;
    \fill[white] (0,0) -- (270:1.5mm) arc (-90:90:1.5mm) -- cycle;
    \end{tikzpicture}}\\
(2020) & & & & \\
\hline
\citep{yousefnezhad2020security}  & \raisebox{-0.5ex}{\begin{tikzpicture}   \fill[black] (0,0) circle (1.5mm);
    \end{tikzpicture}} 
&   \raisebox{-0.5ex}{\begin{tikzpicture} 
    \draw (0,0) circle (1.5mm); 
    \end{tikzpicture}} 
&   \raisebox{-0.5ex}{\begin{tikzpicture} 
    \fill[black] (0,0) circle (1.5mm); 
    \end{tikzpicture}} 
&   \raisebox{-0.5ex}{\begin{tikzpicture}
    \draw (0,0) circle (1.5mm);
    \fill[black] (0,0) -- (90:1.5mm) arc (90:270:1.5mm) -- cycle;
    \fill[white] (0,0) -- (270:1.5mm) arc (-90:90:1.5mm) -- cycle;
    \end{tikzpicture}}\\
(2020) & & & & \\
\hline
\citep{almolhis2020security}  & \raisebox{-0.5ex}{\begin{tikzpicture}        
    \draw (0,0) circle (1.5mm); 
    \end{tikzpicture}} 
&   \raisebox{-0.5ex}{\begin{tikzpicture} 
    \fill[black] (0,0) circle (1.5mm); 
    \end{tikzpicture}} 
&   \raisebox{-0.5ex}{\begin{tikzpicture}
    \draw (0,0) circle (1.5mm);
    \fill[black] (0,0) -- (90:1.5mm) arc (90:270:1.5mm) -- cycle;
    \fill[white] (0,0) -- (270:1.5mm) arc (-90:90:1.5mm) -- cycle;
    \end{tikzpicture}} 
&   \raisebox{-0.5ex}{\begin{tikzpicture}\draw (0,0) circle (1.5mm);
    \fill[black] (0,0) -- (90:1.5mm) arc (90:270:1.5mm) -- cycle;
    \fill[white] (0,0) -- (270:1.5mm) arc (-90:90:1.5mm) -- cycle;
    \end{tikzpicture}}\\
(2020) & & & & \\
\hline
\citep{lit2021survey}  & \raisebox{-0.5ex}
{\begin{tikzpicture}    \fill[black] (0,0) circle (1.5mm); 
    \end{tikzpicture}} 
&   \raisebox{-0.5ex}{\begin{tikzpicture}\draw (0,0) circle (1.5mm);
    \fill[black] (0,0) -- (90:1.5mm) arc (90:270:1.5mm) -- cycle;
    \fill[white] (0,0) -- (270:1.5mm) arc (-90:90:1.5mm) -- cycle;
    \end{tikzpicture}} 
&   \raisebox{-0.5ex}{\begin{tikzpicture} 
    \fill[black] (0,0) circle (1.5mm); 
    \end{tikzpicture}} 
&   \raisebox{-0.5ex}{\begin{tikzpicture}\draw (0,0) circle (1.5mm);
    \fill[black] (0,0) -- (90:1.5mm) arc (90:270:1.5mm) -- cycle;
    \fill[white] (0,0) -- (270:1.5mm) arc (-90:90:1.5mm) -- cycle;
    \end{tikzpicture}}\\
 (2021) & & & & \\
 \hline
\citep{thakor2021lightweight}  & \raisebox{-0.5ex}{\begin{tikzpicture}
    \fill[black] (0,0) circle (1.5mm); 
    \end{tikzpicture}} 
&   \raisebox{-0.5ex}{\begin{tikzpicture} 
    \draw (0,0) circle (1.5mm); 
    \end{tikzpicture}} 
&   \raisebox{-0.5ex}{\begin{tikzpicture}\draw (0,0) circle (1.5mm);
    \fill[black] (0,0) -- (90:1.5mm) arc (90:270:1.5mm) -- cycle;
    \fill[white] (0,0) -- (270:1.5mm) arc (-90:90:1.5mm) -- cycle;
    \end{tikzpicture}} 
&   \raisebox{-0.5ex}{\begin{tikzpicture}\draw (0,0) circle (1.5mm);
    \fill[black] (0,0) -- (90:1.5mm) arc (90:270:1.5mm) -- cycle;
    \fill[white] (0,0) -- (270:1.5mm) arc (-90:90:1.5mm) -- cycle;
    \end{tikzpicture}}\\
 (2021) & & & & \\
 \hline
\citep{10.1145/3447625}  & \raisebox{-0.5ex}{\begin{tikzpicture}
    \draw (0,0) circle (1.5mm);
    \fill[black] (0,0) -- (90:1.5mm) arc (90:270:1.5mm) -- cycle;
    \fill[white] (0,0) -- (270:1.5mm) arc (-90:90:1.5mm) -- cycle;
    \end{tikzpicture}} 
&   \raisebox{-0.5ex}{\begin{tikzpicture} 
    \fill[black] (0,0) circle (1.5mm); 
    \end{tikzpicture}} 
&   \raisebox{-0.5ex}{\begin{tikzpicture}
    \draw (0,0) circle (1.5mm);
    \fill[black] (0,0) -- (90:1.5mm) arc (90:270:1.5mm) -- cycle;
    \fill[white] (0,0) -- (270:1.5mm) arc (-90:90:1.5mm) -- cycle;
    \end{tikzpicture}}
&   \raisebox{-0.5ex}{\begin{tikzpicture} 
    \fill[black] (0,0) circle (1.5mm);
    \end{tikzpicture}}\\
 (2021) & & & & \\
 \hline
\citep{ling2022use}  & \raisebox{-0.5ex}{\begin{tikzpicture}
    \draw (0,0) circle (1.5mm);
    \fill[black] (0,0) -- (90:1.5mm) arc (90:270:1.5mm) -- cycle;
    \fill[white] (0,0) -- (270:1.5mm) arc (-90:90:1.5mm) -- cycle;
    \end{tikzpicture}} 
&   \raisebox{-0.5ex}{\begin{tikzpicture} 
    \draw (0,0) circle (1.5mm); 
    \end{tikzpicture}} 
&   \raisebox{-0.5ex}{\begin{tikzpicture} 
    \draw (0,0) circle (1.5mm); 
    \end{tikzpicture}} 
&   \raisebox{-0.5ex}{\begin{tikzpicture} 
    \draw (0,0) circle (1.5mm); 
    \end{tikzpicture}}\\
(2022) & & & & \\
\hline
\citep{shah2022survey}  & \raisebox{-0.5ex}{\begin{tikzpicture}
    \draw (0,0) circle (1.5mm);
    \fill[black] (0,0) -- (90:1.5mm) arc (90:270:1.5mm) -- cycle;
    \fill[white] (0,0) -- (270:1.5mm) arc (-90:90:1.5mm) -- cycle;
    \end{tikzpicture}}
&   \raisebox{-0.5ex}{\begin{tikzpicture} 
    \draw (0,0) circle (1.5mm); 
    \end{tikzpicture}} 
&   \raisebox{-0.5ex}{\begin{tikzpicture} 
    \draw (0,0) circle (1.5mm); 
    \end{tikzpicture}} 
&   \raisebox{-0.5ex}{\begin{tikzpicture}
    \draw (0,0) circle (1.5mm);
    \fill[black] (0,0) -- (90:1.5mm) arc (90:270:1.5mm) -- cycle;
    \fill[white] (0,0) -- (270:1.5mm) arc (-90:90:1.5mm) -- cycle;
    \end{tikzpicture}}\\
(2022) & & & & \\
\hline
\citep{chaudhary2023ddos}  & \raisebox{-0.5ex}{\begin{tikzpicture} 
    \fill[black] (0,0) circle (1.5mm); 
    \end{tikzpicture}} 
&   \raisebox{-0.5ex}{\begin{tikzpicture} 
    \draw (0,0) circle (1.5mm); 
    \end{tikzpicture}} 
&   \raisebox{-0.5ex}{\begin{tikzpicture} 
    \fill[black] (0,0) circle (1.5mm); 
    \end{tikzpicture}} 
&   \raisebox{-0.5ex}{\begin{tikzpicture} 
    \draw (0,0) circle (1.5mm); 
    \end{tikzpicture}}\\
(2023) & & & & \\
\hline
\textbf{Our survey} & \raisebox{-0.5ex}{\begin{tikzpicture} 
    \fill[black] (0,0) circle (1.5mm); 
    \end{tikzpicture}} 
&   \raisebox{-0.5ex}{\begin{tikzpicture} 
    \fill[black] (0,0) circle (1.5mm); 
    \end{tikzpicture}} 
&   \raisebox{-0.5ex}{\begin{tikzpicture} 
    \fill[black] (0,0) circle (1.5mm); 
    \end{tikzpicture}} 
&   \raisebox{-0.5ex}{\begin{tikzpicture} 
    \fill[black] (0,0) circle (1.5mm); 
    \end{tikzpicture}}  \\
(2024) & & & & \\
\bottomrule
\end{tabularx}
\noindent{\footnotesize{Full cover: \raisebox{-0.4ex}{\tikz{\draw (0,0) circle (1.5mm); \fill[black] (0,0) circle (1.5mm);}}, partial cover: \raisebox{-0.4ex}{\tikz{\draw (0,0) circle (1.5mm); \fill[black] (0,0) -- (90:1.5mm) arc (90:270:1.5mm) -- cycle; \fill[white] (0,0) -- (270:1.5mm) arc (-90:90:1.5mm) -- cycle;}}, not covered: \raisebox{-0.4ex}{\tikz{\draw (0,0) circle (1.5mm);}}}}
\end{table}

Almolhis et al.~\citep{almolhis2020security} highlighted new security concerns when smart homeowners' data are stored and processed in third-party clouds. While these studies focus on specific domains like smart homes, they lack a unified framework for vulnerabilities or strategies for areas like children's and healthcare IoT. Our work fills this gap by systematically categorizing IoT into ten types and addressing their unique challenges. Studies on smart cities, logistics, buildings, homes, and retailing reveal inherent IoT security issues like interoperability, privacy, and scalability.~\citep{sicari2019evaluate}.
Sivaraman et al.~\citep{sivaraman2016smart} emphasized the security challenges related to IoT systems. Similarly, Harbi et al.~\citep{harbi2019review} studied the various taxonomy of security requirements related to IoT along with their mitigation approaches. Digital forensics is also impacted by IoT security concerns~\citep{stoyanova2020survey}.
While promising solutions exist~\citep{domingo2019privacy}, IoT-cloud integration is still emerging, bringing additional complexities and unaddressed security challenges as more private players enter this field.

Modern IoT cloud systems consist of large-scale systems and cloud applications with millions of users, generating vast amounts of data. Protecting these devices and the data  presents a significant challenge, especially in the absence of uniform laws and regulations for deployment and development
~\citep{10.1145/3548606.3560680}.
The marriage of IoT and cloud technology has sparked a revolution in the production of IoT devices and increased security vulnerabilities. The vast amount of data collected by these devices, often personal and sensitive, makes them prime targets for cyber attacks. This study builds on previous work by proposing a comprehensive framework that categorizes cloud-based IoT systems into ten distinct types, addressing both systemic vulnerabilities and specific domain-related security issues. Unlike the prior literature, we focus on underexplored IoT domains and highlight actionable insights for mitigating risks. Given the growing complexity of these threats, a systematic approach becomes crucial. Categorizing vulnerabilities based on their nature enables the development of targeted defense mechanisms. Through this categorization, we can design more effective security solutions to mitigate these risks and ensure the safety of \mbox{these devices.}

\subsection{Uniqueness of Our Survey}
This article introduces a comprehensive classification system for IoT security vulnerabilities, categorizing them into ten distinct groups based on the purpose and application of IoT devices.

Why Ten Categories of Cloud-Based IoT? 
 
 The rationale behind selecting ten categories lies in their ability to comprehensively encompass the diversity of use cases and their unique security challenges. We derived these categories based on an extensive review of the existing literature, current IoT applications, and the security vulnerabilities. In our study, each category showcases a significant domain where cloud-based IoT introduces unique challenges which necessitate targeted mitigation strategies.

We consider the following criteria for selecting these specific categories:

\begin{itemize}
    \item  Relevance to Major IoT Domains: each category is tied to major IoT domains, such as consumer IoT, healthcare IoT, or agriculture IoT, which are widely adopted and significant in real-world applications.
    \item Distinct Security Challenges: the categories were chosen to address scenarios posing specific vulnerabilities, such as centralized data breaches, malicious access, or \mbox{system disruptions}.
    \item Diversity of Use Cases: the ten categories collectively cover a broad spectrum of IoT use cases, ranging from critical infrastructure and healthcare to consumer devices and smart environments.
    \item Scalability and Adaptability: our category classification ensures that as IoT use cases evolve, our framework remains flexible enough to accommodate emerging domains or new categories.
\end{itemize}

For example, consumer IoT, such as voice assistants, faces unique threats like voice squatting attacks, where malicious user can exploit ambiguities in voice commands to gain unauthorized access. Similarly, healthcare IoT devices, including medical devices, are highly vulnerable to private data breaches that compromise sensitive patient information, posing significant risks to both privacy and safety. Meanwhile, industrial IoT, which powers life critical infrastructure, is particularly susceptible to attacks targeting operational integrity. This systematic approach ensures that the classification captures the most prevalent and significant IoT domains while providing a strong framework for addressing both existing vulnerabilities and emerging threats as the IoT landscape continues to evolve.

To address our research, we defined three key questions. The first question (RQ1) involves categorizing IoT devices into ten distinct groups by purpose and function, as shown in Figure~\ref{fig_2}, to identify their specific security and privacy issues. The second question (RQ2) explores various mitigation strategies for these issues and identifies remaining challenges that need further study. Finally, the third question (RQ3) assesses the feasibility and effectiveness of a standardized security solution across all categories, with a focus on finding the next best solution if a uniform approach is unworkable.

\section{Categorical Study on Cloud-Based IoT, Their Security and Privacy Challenges, Mitigation Approach, and Open Problems}\label{sec3}
This section elaborates on the security challenges associated with each of the ten categories (RQ1) and their corresponding mitigation approaches (RQ2). Additionally, we have dedicated a section to open problems that highlight security issues requiring future attention and resolution. Consequently, the following section provides responses to both RQ1 and RQ2.
\subsection{Category 1: Consumer-Oriented IoT}
\subsubsection{{Security Challenges}}
Consumer-oriented IoT devices, like voice assistants, enhance convenience but pose significant security and privacy risks. A major issue is the \textbf{voice} squatting attack (VSA), where devices dependent on third-party natural language processing can be compromised. Cloud servers like Amazon and Google do not guarantee security against vulnerabilities, allowing malicious third-party apps to access and potentially steal sensitive user data.

Precise accuracy in voice command recognition is crucial for reliable performance in voice-based systems. 
Attackers may use voice synthesis or replay attacks to mimic users, potentially accessing financial accounts~\citep{ahmed2022detecting}.
Attackers may also install malicious apps on users' phones, increasing the risk of financial loss~\citep{obaid2021assessment, zhang2019dangerous, kumar2018skill}.
Notable speech-based IoT devices such as Google Assistant, Apple's Siri, and Microsoft's Cortana are widely used, highlighting the urgent need to enhance security to protect user privacy and \mbox{prevent attacks}.

\textbf{Interoperability} issues and \textbf{unauthorized access} are key security concerns for cloud-based IoT. Interoperability problems arise when devices allow multiple apps to coexist, such as Samsung SmartThings, which is vulnerable to 20 different security threats~\citep{WinNT1}.
Samsung SmartThings is an IoT device that uses Linux OS and supports ZigBee, Z-wave, and Bluetooth. Its frequent data transmissions make it highly susceptible to cyber attacks~\citep{celik2018soteria, fernandes2016security, fernandes2017security, zhang2018homonit}. 
Attackers can exploit these channels through vulnerabilities like unsecured data transfer protocols or MITM attacks, compromising user security and privacy~\citep{sivasankari2022detection}.
These breaches allow unauthorized access to networks and remote manipulation of devices, threatening data privacy and synchronization. Additionally, source code flaws could give apps excessive privileges, enabling them to spy on other installed applications.
Reliance on a single-factor authentication mechanism, like a password, makes users prone to brute-force attacks~\citep{otoom2023deep}, where attackers can try a large number of passwords to gain access. 
Sharing the same Wi-Fi network with an attacker makes it easier for them to steal passwords and initiate unauthorized actions. Furthermore, bugs in SmartThings's code, varying programming styles, and the utilization of the ``If This, Then That'' (IFTTT) platform introduce vulnerabilities and potential command injection attacks~\citep{attoh2024towards}.
Concerns over the security of Wi-Fi networks are heightened by the vulnerabilities inherent in SmartThings systems. Notably, the trigger-action rules in the IFTTT platform can be bypassed, conflicts can arise, and actions may be repeatedly executed, leading to denial of service attacks (DDoS)~\citep{sanli2024detection}. 
Not confined to particular devices like SmartThings, DDoS attacks are recognized as particularly vulnerable targets, notably emphasized by the 2016 Mirai botnet attack, acclaimed as the most extensive of its kind in history~\citep{WinNT29}.
Ensuring robust security measures and addressing these vulnerabilities is essential to safeguard the integrity and privacy of IoT systems utilizing SmartThings and the IFTTT platform.
In addition to the vulnerabilities discussed earlier, another significant security concern arising in this category involves the potential for MITM. These attacks include a shared private key to encrypt a task linked to a specific assistant, like Amazon Alexa or Google Assistant~\citep{lit2021survey}.
 
In such attacks, hackers can intercept and decrypt sensitive information exchanged between the IoT cloud and the assistant, compromising the integrity and confidentiality of the data. 
The IoT devices sharing the same network or relying on Wi-Fi are more susceptible to MITM attacks~\citep{aziz2023securing,navas2018not}.
Many consumer-oriented devices are vulnerable to data leakage due to inadequate protection. This weakness raises another significant security issue due to inadequate protection, making the device vulnerable to data extraction via reverse engineering~\citep{sharma2022state}. 
Essential elements such as certificates, keys, algorithms, private authentication, and other sensitive personal human behaviors (such as motion and voice) are often stored within this flash memory (storage device) device, making them a prime target for malicious actors~\citep{park2019security}. 

Furthermore, many IoT consumer-oriented IoT devices are susceptible to unprotected debugging interface vulnerabilities, which can be exploited by attackers to gain unauthorized access and misuse~\citep{alladi2020consumer}.
Many IoT devices are often left unprotected by manufacturers despite being intended for device debugging purposes. Intruders can access device shells to manipulate or inject harmful commands, leading to device damage and security breaches. Issues like data leakage and debugging further compound security concerns, particularly in devices like smart cameras~\citep{obermaier2016analyzing, seralathan2018iot, tekeoglu2015investigating, xu2018internet}.

The persistent use of hardcoded passwords poses significant security risks in many IoT devices, such as smart cameras. These passwords, often stored in unprotected flash memory, are easily cracked and can only be changed by manufacturers, complicating updates or resets if compromised. Additionally, some cloud-based IoT devices have flawed authentication mechanisms, overly dependent on static identifiers like MAC addresses, which can be exploited to disrupt connections between the camera and the cloud server. 

The security of IoT devices is often compromised by the lack of authentication in data transfer from devices like cameras to the cloud, enabling attackers to easily upload videos via constructed URLs. Insecure communication protocols, such as using unencrypted HTTP instead of HTTPS and simplistic proprietary protocols, allow attackers to intercept communications, decode video streams, and access sensitive information like app SSIDs and passwords for remote camera connectivity.

Another authentication-related security issue arises due to the use of fixed hash value, which persists in many current cloud-based IoT devices including hue light bulb~\citep{mangala2021short}.
The vulnerabilities in interconnected devices, which are essential components of cloud-based IoT systems, are also contributing to the growing security concerns~\citep{geeng2019s, jang2017enabling, zeng2019understanding}.

Weak authentication practices in smart home devices can lead to unauthorized access and data breaches, especially when multiple devices are interconnected. These vulnerabilities could be exploited by third-party companies to collect user data for targeted advertising or selling purposes.
Shared use of devices like smart speakers among family members increases the risk of unintended access and privacy breaches, particularly if user data are not deleted from cloud servers when a user stops using the device. Additionally, weak authentication practices heighten vulnerability to unauthorized access and enable excessive data collection from users by cloud-based IoT devices. 

Researchers have raised concerns over the amount of data these devices send to cloud servers, often in the form of unencrypted data packets with unknown content~\citep{davis2020vulnerability,notra2014experimental}. 
This discrepancy in data size and content raises suspicions about the devices' communication. Studies have shown that even with the restriction to send only a small amount of 20 KB of encrypted data, the server can still trigger emergency notifications on the users' app, suggesting the data likely include private information collected without the user's explicit consent. 
For instance, researchers found excessive data collection loopholes in Nest smoke alarms~\citep{notra2014experimental}.
There are concerns about the device collecting potentially sensitive data beyond intended functionality, as it captures human movements and assesses user states based on lighting conditions.
Addressing these security issues is crucial to ensure the privacy of users and build trust in smart home technologies.  
We will now present some of the available solutions in the following section.
\subsubsection{{Mitigation Approaches}} 
As we acknowledge that virtual personal assistants (VPAs) are vulnerable to \textbf{voice squatting attacks (VSAs)} and \textbf{voice masquerading attacks (VMAs)}, several mitigation approaches have been proposed. One such approach developed by Zhang et al.~\citep{zhang2019dangerous} is a mitigation technique to address voice masquerading attacks in VPAs. \begin{itemize}   
\item The approach involves:
(1) Phonetic analysis to detect similarly pronounced invocation names.
(2) Natural language processing (NLP) to understand user intent and context.
(3) Machine learning to identify patterns and flag suspicious activity.\item When a risk is detected, the system captures the ongoing impersonation threat and then
alerts the user in real time to prevent data breaches.\end{itemize} This technique strengthens VPA security and protects sensitive user data. Another proposed solution includes the voice authentication technique~\citep{ren2023voice}. This work advocates for the use of a voice spoofing detection framework which integrates four modules for effectively securing speaker verification in voice-based IoT devices.
Hooda et al.~\citep{hooda2022skillfence} proposed a systems-oriented defense against voice-based confusion attacks where an attacker exploits ambiguities in voice commands to mislead VPAs into triggering malicious actions or skills instead of intended ones. Termed SkillFence by the author, it is a browser extension designed to thwart voice-based confusion attacks targeting voice assistants, such as Amazon Alexa. Concurrently, researchers are actively seeking optimal solutions to address interoperability challenges in IoT devices, including platforms like Samsung SmartThings. Similarly, Chatterjee et al.~\citep{chatterjee2017puf} developed a lightweight identity-based cryptosystem employing a physically unclonable function (PUF) suitable for IoT to enable secure authentication and message exchange among the devices.

Nazzal et al.~\citep{nazzal2022vulnerability} proposed a multi-stepped categorization of SmartThings vulnerabilities based on the platform’s component and nature of the attack and thus proposed relative mitigations.
Consumer-oriented IoTs are particularly vulnerable to \textbf{DDoS} attacks, which are currently one of the most serious virtual threats. Emerging technologies like \textbf{software-defined networking (SDN)} can help mitigate DDoS attacks for IoT devices. 
One of the mitigation approaches for DDoS attacks through SDN is \textbf{flow filtering.} The flow filtering strategy is a straightforward approach that uses centralized control to block or allow network traffic based on packet header information. 
While the use of software-defined networking (SDN) to mitigate DDoS attacks is promising, implementing SDN in consumer IoT devices faces scalability and cost barriers. Moreover, deploying flow filtering or rate-limiting approaches in resource-constrained devices could lead to bottlenecks and performance degradation.

However, it may cause delay and bottleneck issues due to statistics gathering and packet inspection~\citep{shameli2015taxonomy, salva20185g, guozi2018ddos}.
The \textbf{honeypot technique} is another mitigation approach that creates a simulated environment to collect information about malicious traffic and can be used with SDN to update detection and mitigation policies~\citep{yan2018multi}.
Security applications usually adopt a rate-limiting approach in conjunction with deep package inspection procedures. The idea implies that to prevent network overload caused by volumetric attacks (a massive volume of traffic), SDN controllers can set a traffic volume limit and reject subsequent traffic when the limit is reached, often used in combination with Deep Package Inspection procedures for enhanced security~\citep{yan2018multi}.

\textbf{Moving Target Defense (MTD)} is another DDoS mitigation approach that involves dynamically reconfiguring a network/system based on random values to prevent attackers from making the system unavailable. This includes techniques such as randomizing IP and MAC addresses to prevent DDoS attacks. However, there are concerns about the impact on performance and cost when deploying MTD on large-scale networks~\citep{ma2015defending, connell2017performance}.
MTD involves dynamically reconfiguring a network to prevent attackers from making it unavailable~\citep{connell2017performance} while another approach, Traceback, uses the packet header information to identify an attacker's origin. SDN's control plane provides a holistic view for effective mitigation solutions~\citep{chen2020ddos}.

In response to security concerns, researchers have proposed several solutions to strengthen the security of cloud-based IoT devices. One such approach is the secure IoT structural design for smart cities proposed by Bhattacharjya et al.~\citep{bhattacharjya2019secure}. This design applies to various contexts, including smart homes and the Power Internet of Things (PIoT). \begin{itemize}
\item The approach leverages a three-layered security framework:  (1) perception layer, \mbox{(2) network} layer, (3) application layer.\item The approach employs a hybrid RSA cipher, which combines public and private key encryption for secure and efficient data transmission, to ensure robust architectures in IoT devices. \item Cryptographic algorithms are foundational elements of such secure design. They provide a layer of encryption to protect data confidentiality and integrity during communication between devices and the cloud.\end{itemize}
Cryptographic algorithms can prevent security threats to a certain extent for smart homes or smart cities~\citep {abu2020towards}. However, there are certain technical issues with the application of high-end security to these IoT devices due to their low computing power, limited space, short memory, and limited battery life~\citep{qasem2024cryptography}. Thus, a lightweight encryption such as Piccolo or PICARO~\citep{thakor2021lightweight} offers a better balance between the security and resource constraints of IoT devices.
The smart home IoT devices are heavily vulnerable to private and sensitive data of consumers, and, in this direction, Gerber et al. developed \textbf{LOKI}, an interface that enables control of a smart home using local information processing and protecting smart home systems from hacking attacks ~\citep{gerber2021loki}.
\subsubsection{Open} 
 Problems
Security solutions using SDN could mitigate common IoT issues like DDoS attacks but face challenges due to IoT complexity and diverse attack methods. Additionally, manufacturers often neglect regular firmware updates for IoT devices, jeopardizing the future security of IoT and cloud ecosystems. Key concerns include the reliability of cloud providers, inconsistent firmware updates, absence of standards, privacy risks, and complex security strategies. Addressing these will necessitate cooperation among manufacturers, cloud services, policymakers, and consumers.
\subsection{Category 2: Children's IoT (Smart Toys)}
\subsubsection{{Security} Challenges}
Children's IoT devices, tailored for entertainment, development, and education, feature specialized sensors and firmware for personalized interactions with kids. These devices are subject to stringent privacy regulations like the COPPA rule~\citep{WinNT24}, which mandates confidentiality and security for online services aimed at children. Inadequate data security in these devices can lead to significant privacy risks.

Insecure communication channels, weak authentications, and limited parental control over data collection and setting further intensify these concerns. For instance, a case study by Chu et al.~\citep{chu2018security} revealed that the hydrogen tracker, a children's IoT device, violates COPPA rules.
Interestingly, cloud-based IoT toys are vulnerable not only to children but to everyone in the home who owns them. The statement ``Someone is watching through Barbie's eyes'' holds enough significance to warrant sharing it with every household member, as it highlights the vulnerability at stake~\citep{holloway2016internet}. These are open to eavesdropping attacks by malicious users~\citep {ataei2022authentication}.

Ling et el.~\citep{ling2022use} concluded that \textbf{data security} has emerged as a growing concern regarding children's IoT.
\textbf{Jamming attacks} are one of the prominent attacks found in entertainment-based drones where the malicious user intentionally captures and manipulates drone data via unauthorized transmitting channels~\citep{yang2022review}. 
These drones and vibrators also carry privacy concerns, as the attacker can access sensitive information, including videos and audio stored on cloud servers, shared by children with these toys. 
The hydrogen tracker, a type of IoT device designed for monitoring a baby's hydrogen level, exhibits security vulnerabilities both in its cloud infrastructure and user applications~\citep{apthorpe2017smart, chu2018security, holloway2016internet, hung2016glance, valente2017security}.
The access control mechanism between children's IoT devices and cloud servers has vulnerabilities, allowing attackers to intercept authentication packets. This flaw lets attackers access sensitive information, such as profile images, and manipulate cloud services to retrieve user files. The devices communicate with servers using HTTP GET and POST requests, which has led to reported security issues, including token reuse problems in systems like the hydrogen tracker.

For instance, in the hydrogen tracker, whenever the child drinks water, the user app sends this information to the cloud. Here, the attacker can capture the packets containing the authentication token and the HTTP header information and upload fake content to the cloud. These fake data can disrupt the proper functioning of the hydrogen tracker and potentially lead to misdiagnosis or improper treatment decisions.
The URL token for the user pictures contains 12 letters. On the cloud side, it returns the HTTP 301 response as long as the first three letters pair with the legitimate token. However, the cloud returns HTTP 404 for the nonexistent token. This helps to reduce the time needed to guess the correct tokens. When combined with the two security risks, the attacker can easily use brute force to exhaust the remaining 9 letters, and, hence, the user profile is leaked.

A significant security issue is the failure to delete expired files from the cloud, leading to privacy breaches when new profile images remain linked to old ones.
Referring to the security issues for the children's IoT application side, researchers primarily acknowledge three main issues. \begin{enumerate}[leftmargin=3.3em,labelsep=4mm]
\item[(1)] The 
 primary concern is the presence of third-party services integrated into the user application, posing a substantial security vulnerability. In the case of the hydrogen tracker, the mobile app utilizes four third-party analytics and performance monitoring services. Although the traffic data are encrypted, it remains uncertain whether private data are being leaked through these services.
\item[(2)] Another concern arises from the use of plain-text APIs in applications. This lack of data encryption during transmission to the cloud server heightens the risk of privacy breaches, including eavesdropping.
\item[(3)] The attacker can obtain the intermediate packets exchanged between the app and the cloud, enabling packet spoofing. Spoof packets may change the codes and cause the app to malfunction.
\end{enumerate}

The attacker can intercept packets exchanged between an app and the cloud, leading to personal data leaks. Additionally, when network errors occur, user apps send crash reports containing personal details like name, age, and weight to third-party IoT developers, inadvertently exposing sensitive information. A major security flaw in many children's IoT devices is the persistent enabling of remote access post-release, which offers attackers a straightforward way to exploit these vulnerabilities ~\citep{albrecht2015privacy}. 

Furthermore, the use of plain-text API for communication between the IoT device and the cloud significantly increases the risk of unauthorized data interception.
A prevalent security issue observed in both consumer-oriented devices and children's IoT devices involves the storage of fixed user data in the IoT memory card, which can be easily accessed by malicious individuals if the card is detached. This vulnerability raises concerns about the potential exposure of sensitive user information to unauthorized parties. 
Having an active default development account in the IoT device, which grants root privileges and has vulnerable passwords, presents a major security risk. This allows attackers to effortlessly access sensitive information and exploit the devices' vulnerabilities. In addition to these existing issues, children's IoT devices face other critical issues, including inadequate protection of the universal asynchronous receiver/transmitter (UART) interface. Commonly used for troubleshooting, the UART's conventional method makes it easy for attackers to manipulate device parameters and extract sensitive information~\citep{stanislav2015hacking}.

In addition to the previously mentioned security issues, many children's IoT devices suffer from security flaws attributed to weak authentication and the use of plain-text API. These vulnerabilities undermine the overall security of the devices, making them more susceptible to unauthorized access and potential data breaches. It is crucial to address these weaknesses by implementing stronger authentication mechanisms and ensuring secure communication protocols to protect user data and maintain the integrity of IoT systems.

\subsubsection{Mitigation Approaches}
Children's IoT has the potential to collect terabytes of sensitive personal, contextual, and usage information, which may be a subject of cybercrime. In smart toys' hardware restrictions, a specific encryption and authentication mechanism is required. In this direction, Rivera et al.~\citep{rivera2019secure} designed and built specific security mechanisms for the smart toys and also validated them on the platform.

Children lack an understanding of privacy and online safety, especially in social media and cloud environments. Rafferty et al.~\citep{rafferty2017towards} proposed a model that somehow puts the obligations for children's privacy on parents or guardians and alerts them in case of any violation.
Similarly, Yankson et al.~\citep{yankson20204p} presented an abstract forensics investigation framework focused on using non-conventional means that allow investigators to successfully ``Plan'', ``Preserve'', ``Process'', and ``Present'' (4P) as a systematic means to conduct digital \mbox{forensic analysis}. 
\subsubsection{Open Problems} Children's IoT devices pose notable security and privacy risks due to young users' lack of awareness. Many devices fail to inform parents about potential privacy breaches. Therefore, it is essential that these IoT devices include notifications or mechanisms that require parental supervision for their operation. Any forthcoming IoT devices must comply with regulations like COPPA and similar guidelines specifically designed for \mbox{children's IoT}.


\subsection{Category 3: Healthcare IoT and Medical Equipment}
\subsubsection{Security Challenges}
Healthcare IoT, including wearables, implants, and remote monitoring, improves patient outcomes, reduces costs, and personalizes care.
Post-COVID-19, interest in IoT devices, especially in healthcare, has surged. IoT healthcare solutions have proven to significantly enhance patient care and save millions of lives. Some of the examples are remote patient monitoring (RPM), decentralized clinical trials (DCT)~\citep{WinNT10}, pulse
oximeters, glucometers~\citep{ganesh2022iot, okafor2022design}, etc.

The significant shifts in IoT devices in healthcare offer staggering benefits but also elevate the risk of cyber attacks on interconnected medical devices.
Many cloud-based medical devices transmit unencrypted data, creating vulnerabilities. Attackers could manipulate these data, disrupting treatment plans and jeopardizing patient well-being\citep{WinNT11}.
Healthcare institutions utilize a range of IoT devices for accurate patient and asset tracking, commonly referred to as indoor localization facilities in healthcare~\citep{bradley2018security, farahani2018towards, mcallister2017localization}. Indoor localization facilities for healthcare face significant security vulnerabilities in both IoT and cloud server environments. Two primary concerns are prevalent. First, \textbf{weak password security} is common, with devices often using default, widely known passwords. Second, there is inadequate memory protection. Many IoT devices include built-in SSH services, compounding the risk when default passwords are not changed.

A critical security issue is the use of default passwords on devices, making them easy targets for attackers. This vulnerability can compromise entire systems and expose sensitive patient data.
The second issue is insufficient memory protection. Devices in public areas can be easily removed, and sensitive data on memory cards accessed. On the cloud side, a major vulnerability exists where communication between the cloud and server uses unencrypted plain-text APIs. Healthcare facilities transmit database credentials every few seconds without proper protection, making them susceptible to interception over unsecured Wi-Fi. Attackers can exploit this to access log files or disrupt device functionality through denial-of-service attacks~\citep{bradley2018security}.

Moreover, cloud-based pulse oximeters and glucometers, which store health data offline, are susceptible to \textbf{hijacking attacks.} In such scenarios, unauthorized access to all offline readings becomes straightforward for hackers after obtaining user credentials~\citep {flynn2020knock}. Cloud-based health image IoT systems face issues with data integrity, availability, and confidentiality due to unencrypted processing of medical image data between IoT devices and the cloud.
To address this issue, Jin et al.~\citep{jin2022p} analyzed access policies in 36 IoT devices, which included medical equipment. The authors discovered several design deficiencies within the access control mechanism of medical devices provided by established companies. 
These deficiencies, such as overly permissive settings or a lack of proper authentication, could allow several unauthorized users to access various sensitive health data, such as blood pressure, age, height, and weight, for all consumers of specific medical equipment companies, such as Biobeat and Hippokkura (used as a case study).

Many medical IoT devices are vulnerable to hardware attacks. Attackers can implant hardware trojans during chip manufacturing, compromising device functionality and integrity and posing risks to patient safety~\citep{newaz2021survey}. 
The U.S. Food and Drug Administration (FDA) has expressed serious concerns about this, issuing reports indicating instances of tampering with patients' health data that occurred after modifying the hardware of medical IoT devices~\citep{west2016wearfit, FDA}.

The use of drones in healthcare has surged, especially for collecting samples, delivering medications, and supplying resources.~\citep{wazid2020private}.
However, this deployment of drones also brings several vulnerability issues, such as replay, MITM, impersonation, and privileged-insider attacks. These vulnerabilities stem from the wireless nature of communication between deployed drones and their ground station servers. For instance, in \textbf{replay attacks}, the attacker intercepts the communication between the drone and the ground station and later replies to that communication to deceive one of the parties.
Similarly, in an MITM attack, a malicious user secretly intercepts and possibly alters the communication between two parties without their prior knowledge. Meanwhile, privileged-insider attacks involve a legitimate person with authorized access to an organization's resources and information misusing that information access for malicious purposes.
These medical devices not only enable attackers to compromise security systems but also expose patients' health data to the risk of being accessible to the public.

\subsubsection{Mitigation Approaches}
Various techniques can help mitigate security issues in medical IoT. One such approach focusing on location privacy was proposed by Alishahi's group~\citep{fathalizadeh2022privacy}.\begin{itemize}
\item This technique utilizes multiple anonymization methods, including k-anonymity, $\ell$-diversity, t-closeness, $\alpha$-anonymity, and $\delta$-presence, to offer online location-based healthcare services while ensuring robust protection of patient location data. \item However, there is a trade-off that this technique might produce some data obfuscation, impacting the accuracy of exact location tracking in some cases. \item This prioritizes user privacy but might limit the precision of location-based services. To safeguard against unauthorized access and data breaches in drone-aided medical services, it is crucial to implement reliable authentication, access control, and key management schemes in the IoT environment. \item These preventive measures ensure that only authorized users can access sensitive data transmitted by drones, protecting patient privacy and sensitive information.\end{itemize}

Additionally, incorporating Blockchain mechanisms with authentication deployment can further enhance its robustness against various types of attacks.
Blockchain technology can assist in ensuring a transparent audit by maintaining a secure ledger~\citep{arul2024iot}, making it difficult for attackers to manipulate or steal sensitive medical information.
To mitigate such issues, Bu et al.~\citep{bu2019bulwark} proposed a robust approach, Bulwark, for guarding against existing and potential communication-based attacks on implantable medical devices.
This secure protocol allows authorized third-party medical teams to securely access the internal memory device (IMD) in emergencies. Bulwark can address not only MITM but potentially other communication vulnerabilities as well.

Ma et al.~\citep{ma2024improving} proposed the use of \textbf{Time Modulated Arrays (TMAs)} to improve covert communication in IoT systems, offering a low-complexity alternative with enhanced beamforming precision and timeliness, which could be particularly valuable for drone-assisted IoT applications. Wazid et al.~\citep{wazid2020private} proposed another approach that leverages a \textbf{private-Blockchain-based technology} to establish secure communication in an IoT-enabled drone-aided healthcare environment.\begin{itemize}
\item This framework utilizes a secure voting mechanism to ensure the authenticity and integrity of data transmission. \item This mechanism involves mining, verifying, and adding blocks containing medical requirements in the P2P network, all based on the PBFT consensus algorithm. \item The security analysis performed on the proposed framework demonstrates its ability to withstand various potential attacks. Furthermore, the framework has been practically implemented through a Blockchain simulation.\end{itemize}
ML-based approaches are also one of the potential solutions for detecting attacks on healthcare systems. ML can analyze vast datasets of network and device behavior to identify patterns that deviate from normal functioning, potentially indicating malicious activity. In this direction, Saeedi et al.~\citep{saeedi2019machine} demonstrated how a decision tree algorithm can be used to detect malicious attacks on healthcare IoT devices. 

HealthGuard is a novel machine-learning-based security framework that observes the crucial signs of different connected devices of a medical IoT and correlates the crucial signs to understanding the changes in body functions of the patient to distinguish benign and malicious activities. The authors have used four different (artificial neural network, decision tree, random forest, k-nearest neighbor) ML-based detection techniques~\citep{newaz2019healthguard}.

Techniques such as Blockchain and machine learning for healthcare IoT systems offer robust solutions; however, they require high computational power and substantial storage, which are not feasible for resource-constrained medical devices.

Beyond securing data transmission, robust security solutions are also essential to safeguard medical IoT devices from hardware-level attacks. Srivastava et al.~\citep{srivastava2019light} presented an improved method for secure data transmission between a network and storage using lightweight cryptographic techniques, such as the \textbf{ARX encryption scheme.} The ring signatures have also been introduced to the communication process, which offers significant privacy benefits, including Signer's Anonymity and Signature Correctness. In addition, the same group of researchers developed GHOSTDAG, a new Blockchain protocol that uses a directed acyclic graph to monitor the health data of patients remotely. This protocol employs smart contract programs from Blockchain technology to facilitate the monitoring process~\citep{srivastava2019data}.

Security solutions for hardware attacks are specially designed to safeguard medical IoT from hardware-level attacks such as physical tampering~\citep{shrivastava2022securing}. For instance, Wu and colleagues introduced a ``golden die'' mechanism designed to identify hardware trojans with a substantial footprint, making them easily distinguishable from the standard base configuration~\citep{wu2015tpad}.
Finally, it is worth mentioning that deep learning-based 
approaches have also been implemented to enhance the security and resilience of medical IoT devices~\citep{rathore2020deep}.
Edge computing, serving as an alternative to cloud-based systems for processing data nearby, emerges as a promising solution for addressing the challenges posed by critical IoT devices. Li et al.~\citep{li2020secured} proposed solutions incorporating SDN edge computing, where edge servers authenticate medical devices before allowing them to transmit data. This approach keeps sensitive patient data localized at edge servers, minimizing the risk associated with long-distance data transfer. 
\subsubsection{{Open Problems}} Cloud-based medical IoT devices face significant security challenges, including inadequate access policies and unsecured data transmission that allow malicious interception of sensitive medical information. While AI and Blockchain offer potential solutions, they require substantial resources and computational power, complicating their implementation. Figuring out how to efficiently use these approaches remains an ongoing challenge \mbox{for researchers}.

Moreover, the transmission of data between the cloud and IoT devices occurs in plain-text, exposing patient data to unauthorized access and posing risks to data confidentiality, medical operations, and patient safety. 

Addressing these concerns requires continued research efforts in areas such as robust access control mechanisms for cloud-based medical IoT services, secure communication protocols for encrypted data transmission, and privacy-preserving data analysis techniques.
\subsection{Category 4: Industrial IoT} 
\subsubsection{{Security Challenges}}
Beyond the healthcare sector, security challenges are also prevalent in cloud-based industrial IoT (IIoT).
Industry 4.0, marked by advanced smart technologies, has transformed industrial processes but increased attack surfaces. Examples include temperature and pressure sensors. The integration of massive data from assets and sensors into cloud servers heightens future risks
~\citep{zverev2019network}. 

Smart meters, a key IIoT technology for monitoring power consumption, face significant security challenges~\citep{varga2017security, Anderson2011SmartMS, cleemput2016high, wurm2016security}. Researchers highlight two main issues: inadequate internal hardware data protection, leading to incorrect identity information and potential energy theft~\citep{wurm2016security}, and unprotected hardware interfaces, allowing attackers to access and modify memory by reactivating the debug interface. These vulnerabilities result in economic losses and compromised data integrity.

IIoT systems face security challenges beyond vulnerabilities in industrial devices. One major concern lies in the widespread susceptibility to DDoS attack~\citep{chaudhary2023ddos, horak2021vulnerability}. In this attack, devices are overwhelmed with a flood of traffic, making them unavailable to legitimate users. In the IIoT context, this can have disastrous consequences. For instance, a DDoS attack on a power plant control system could disrupt critical operations, causing blackouts and potentially damaging sophisticated instruments~\citep{soltan2018blackiot}. Compounding these issues are vulnerabilities within the analog components of some major IIoT devices. For example, a temperature-based control system relies on an analog sensor to sense temperature.  Researchers have also identified security issues related to remote tempering through electromagnetic interferance \citep{tu2019trick}. The non-linear nature of sensors in these devices allows attackers to alter the output reading, potentially disrupting critical industrial processes or causing severe safety hazards.
This vulnerability is not confined to temperature sensors, similar security risks have been identified in various other IIoTs such as medical, laboratory, and PID control applications~\citep{tu2019trick}.

Beyond vulnerabilities in industrial control systems and IIoT devices, severe security concerns are prevalent in cloud-based video surveillance systems, extensively utilized in both industrial and private settings as a means of crime prevention. In a study by 
Obermaier et al.~\citep{obermaier2016analyzing}, they examined four video surveillance systems and identified multiple security concerns, including the utilization of insecure fallback functions, weak passwords, insecure authentication processes, and the absence of robust security standards. 
There are also critical issues in video frame recognition, which has an accuracy of less than 90\% in modern cloud-based IoT devices, particularly stemming from the use of limited eigenfaces for the principal component analysis (PCA). To improve this, a recent study using a hybrid face detection system integrated with cloud-IoT and distributed computing exhibited better face recognition with accuracy above 90\%~\citep{ahamad2024hybrid}.

Finally, the Internet of Battle Things (IoBT), a novel iteration of IoT envisioned as the battlegrounds of tomorrow, comprises interconnected multi-devices capable of communication, action, and collaboration to achieve military objectives~\citep{kott2016internet}. The reason for incorporating this IoT into industrial IoT lies in the sourcing of its interconnected devices from manufacturing sectors. These devices also possess a vulnerability in terms of security, wherein the multitude of entry points provide avenues for attackers to infiltrate any connected device across the network. 

\subsubsection{{Mitigation Approaches}}
The implementation of robust hardware data protection in smart meters is presently a critical necessity for crafting resilient IIoT. To address this need, in this direction, Maritsch's group~\citep{lesjak2016hardware} proposed a hardware-secured and transparent multi-stakeholder data exchange framework for Industrial IoT systems. This framework is designed to enable secure and streamlined data sharing among various stakeholders, including manufacturers, operators, and regulators, all the while maintaining the confidentiality and integrity of the data.

To accomplish this, the authors propose the use of hardware-based security modules, such as \textbf{Trusted Platform Modules (TPMs)}, as critical components to secure the data exchange process from the local server to the remote server. TPMs can be incorporated into IIoT devices to offer a secure means of storing and processing sensitive data, including device identities, keys, and certificates. However, a holistic approach may also involve additional measures, which include encryption protocols and secure communication channels to further strengthen the data protection framework~\citep{pu2023user}. 
Beyond hardware-based security solutions, other approaches have been proposed to address security challenges in IIoT systems. Tan et al.~\citep{tan2021blockchain} proposed a Blockchain-based \textbf{Shamir's Threshold Cryptography (STC)} scheme for data protection in IIoT systems. The scheme is designed to provide secure and efficient data sharing among multiple IIoT devices while preserving data privacy \mbox{and confidentiality}.
\begin{itemize}
\item The STC scheme consists of multiple IIoT devices, a trusted authority that manages the STC keys, and a Blockchain network The IIoT devices use Shamir's secret sharing scheme to split their data into shares, which are then encrypted using the STC scheme and stored on the Blockchain network.\item The trusted authority manages the keys and controls access to the data shares.\end{itemize}

DDoS attack pose another major threat to industrial IoT. Fortunately, researchers have proposed various novel approaches to detect and mitigate DDoS attacks in IIoT~\citep{de2023intelligent, ibrahim2022ddos}. Performance indicators using security virtual network functions on network function virtualization (NFV) have been recently used to mitigate DDoS attacks in IIoT~\citep{de2023intelligent}. In this approach, various data sets were used with ML classifications to achieve very high accuracy (\textgreater99 \%) in mitigating DDoS attacks in IIoTs. The results indicate that ML models implementing tree-based structures and \textbf{XGBoost decision-making trees} are the most suitable for achieving higher accuracies. This model brings out that it is not the different sizes of the networks that influence that ML, it is the fog servers.

\textbf{`FLEAM'} is another approach to mitigate DDoS attacks by federated learning ~\citep{li2021fleam}. FLEAM consists of a set of edge devices, a coordinator device, and a cloud server. The edge devices collect data and use machine-learning algorithms to detect anomalous traffic patterns, which further indicate a DDoS attack.
The coordinator device aggregates the data from the edge devices and trains a global machine-learning model using federated learning techniques. Finally, the cloud server deploys the trained model to the edge devices, enabling them to detect and respond to DDoS attacks in real time.
Evesti's group~\citep{kuusijarvi2016mitigating} believes that different IoT, either smart home or smart industry, can be secured via trusted network edge devices.
A \textbf{rectification attack} is a type of cyber attack in which the attacker can manipulate critical temperature-based control systems. To tackle this issue, Tu et al.~\citep{tu2019trick} proposed a prototype design of a low-cost anomaly detector for critical applications to ensure the integrity of temperature sensor signals.
Stocchero et al.~\citep{stocchero2022secure} suggested a secure command and control (C2) approach for IoBT by integrating the application and network layers through the utilization of an SDN-based cybersecurity mechanism. 
\subsubsection{{Open Problems}} As the IIoT continues to evolve, emerging security and vulnerability concerns accompany the introduction of new devices. In this multifaceted IIoT environment involving multiple stakeholders, the establishment of transparent data ownership, robust access control policies, and effective data governance practices becomes imperative. Without these measures, there is a high chance that sensitive data might be compromised, leading to privacy violations.  Furthermore, the efficient handling and analysis of the vast volume of data generated by IoT devices remain persistent challenges alongside well-known issues like interoperability. These challenges restrict the wider and long-term sustainability of IIoT deployments.
\subsection{Category 5: Smart Vehicles' IoT}
 \subsubsection{{Security Challenges}}
 
The integration of IoT devices with advanced driver assistance systems (ADAS) in modern vehicles enhances driving comfort but introduces significant security challenges too~\citep{shah2022survey}. Connected vehicles use vehicle-to-vehicle (V2V) and vehicle-to-internet (V2I) communication, improving driving experiences through critical information exchange. However, this interconnection raises concerns about sensitive data leakage and \mbox{privacy risks}.

Integrating IoT with the cloud increases cyber attack vulnerabilities, especially in life-critical systems. Hackers can steal vehicle data, take control of vehicles, infiltrate secure sensors, affect critical functions like airbags and door locks, and even disable the \mbox{vehicle entirely}.

Not limited to this, by pilfering data and deploying malicious software updates, hackers can transmit false information to drivers, initiate DDoS attacks, reprogram ECUs with infected software, and download inaccurate navigation maps.

These attacks are categorized as follows: \textbf{data authenticity attacks,} which prevent unauthorized data alterations during cloud transmission, and availability attacks, including denial of service and channel interference, exploiting bandwidth and power constraints to disrupt the IoV system's functioning~\citep{tbatou2017security}.
\begin{itemize}
\item Authentication attacks on IoT vehicles (IoV) include Sybil attacks, GPS deception, masquerading, wormhole, and replay attacks. \item In Sybil attacks, attackers create fake identities to disrupt authentication and manipulate data~\citep{zhu2024sybil}. \item GPS deception leads to inaccurate location data and potential property damage. \item Masquerading involves impersonating vehicle features to infiltrate the network. \item Wormhole attacks hide the true path length between malicious nodes, causing routing errors and congestion. \item Replay attacks repetitively replay and delete messages, reducing efficiency and increasing bandwidth costs.\end{itemize}

\textbf{Data injection attack} in cloud-based vehicle IoT systems pose serious risks, including loss of control, accidents, and fatalities. These attacks occur when unauthorized access exploits outdated software, known security flaws, or insufficient cloud security. Attackers can inject or alter data, causing malfunctions or unexpected behaviors. For example, false sensor data can lead to misinterpretation and accidents, while altered software may cause sudden acceleration or braking. Addressing these vulnerabilities is crucial to fully harness the potential of IoT devices~\citep{kumar2021multimedia}.

\subsubsection{{Mitigation Approaches}}
The ever-evolving landscape of security threats in the connected car market necessitates continuous research and development of robust solutions. Researchers are actively exploring various avenues to combat cyber threats in next-generation autonomous \mbox{vehicles~\citep{taslimasa2023security}}. 

Some of these studies focus on the potential risks associated with car hacking, while others delve into encryption methods, in-vehicle network communication modules, and security enhancements for Electronic Control Units (ECUs).
ISAC with Massive MIMO transforms smart vehicles by enabling high-resolution sensing and reliable communication. Zhang et al.~\citep{zhang2024integrated} introduced a unified tensor-based framework that efficiently estimates channel and target parameters, addressing challenges such as Doppler shifts and beam squint effects. This approach enhances sensing accuracy and communication reliability, critical for applications like autonomous driving and intelligent transportation systems.

Khan et al.~\citep{khan2020cyber} proposed a mitigation approach like \textbf{secure software development practices}, which includes the adoption of writing secure codes, reviewing, threat modeling, and vulnerability testing. \textbf{Secure boot processes} and \textbf{hardware-based encryption methods} can secure the hardware of automotive cars. Smart vehicles must use network security protocols to prevent unauthorized access to the car's communication system.
Researchers are delving into the potential of machine learning and artificial intelligence to improve the security of smart vehicles.
Zewdie and Co.~\citep{zewdie2020iot} suggested that incorporating AI/ML technology into IoT security can help combat emerging cyber threats. The paper proposes several ML/AI approaches, such as predictive analysis, threat detection and response, behavior analysis, and risk assessment, which can be used to enhance IoT security.
The authors conclude that AI/ML algorithms can enhance the security of IoT devices and the cloud computing environment in which they operate by analyzing large amounts of data and detecting anomalies in real time.

In addition to AI/ML, other promising technologies are also emerging to enhance security.
Blockchain-based architecture can be another solution to enhance smart vehicle security.
Smys and Wang~\citep{smys2021security} argued that as vehicles become increasingly connected and autonomous, they become more vulnerable to cyber attacks. Therefore, they suggest that a \textbf{distributed ledger technology} like Blockchain can provide a secure and tamper-proof platform for managing data in smart vehicles.\begin{itemize}
\item The proposed architecture consists of several components, including a Blockchain network, a smart contract layer, and a data management layer. \item The Blockchain network provides a decentralized and secure platform for storing and sharing data, while the smart contract layer enables the automated execution of contracts between different entities in the system. \item The data management layer is responsible for managing data storage and access. \item By leveraging the Blockchain’s immutability and transparency, unauthorized data modification becomes easily detectable, enhancing security.\end{itemize}
Zhao et. al.~\citep{zhao2021detection} addressed the problem of \textbf{false data injection attacks} in connected and automated vehicles, which can compromise the safety and performance of the vehicle. The authors proposed a \textbf{cloud-based sandboxing approach} to detect and mitigate these attacks. The proposed approach consists of three main components: a data acquisition module, a cloud-based sandbox, and a detection module. The data acquisition module collects data from the sensors and other components of the vehicle, which are then sent to the cloud-based sandbox for analysis. The sandbox simulates the behavior of the vehicle and analyzes the incoming data for anomalies that may indicate a false data injection attack. The detection module then generates an alert if an attack is detected.

\subsubsection{{Open Problems}} While Vehicle IoT (VIoT) and its corresponding security solutions are still evolving, there remain numerous unanswered questions within this domain that demand significant attention, especially since they may pertain to critical life-related issues.
For instance, ensuring the integrity of the data transmitted between vehicles and the cloud remains a challenge. Additionally, the rise of sophisticated cyber attacks targeting VIoT systems raises serious concerns about unauthorized access and vehicle control. These issues, along with the other growing risks of personal data breaches, necessitate continued research efforts to develop robust security solutions for cloud-based VIoT.


\subsection{Category 6: Wearable IoT}
\subsubsection{{Security Challenges}}
\begin{figure}[H]
\includegraphics[width=3.5in ]{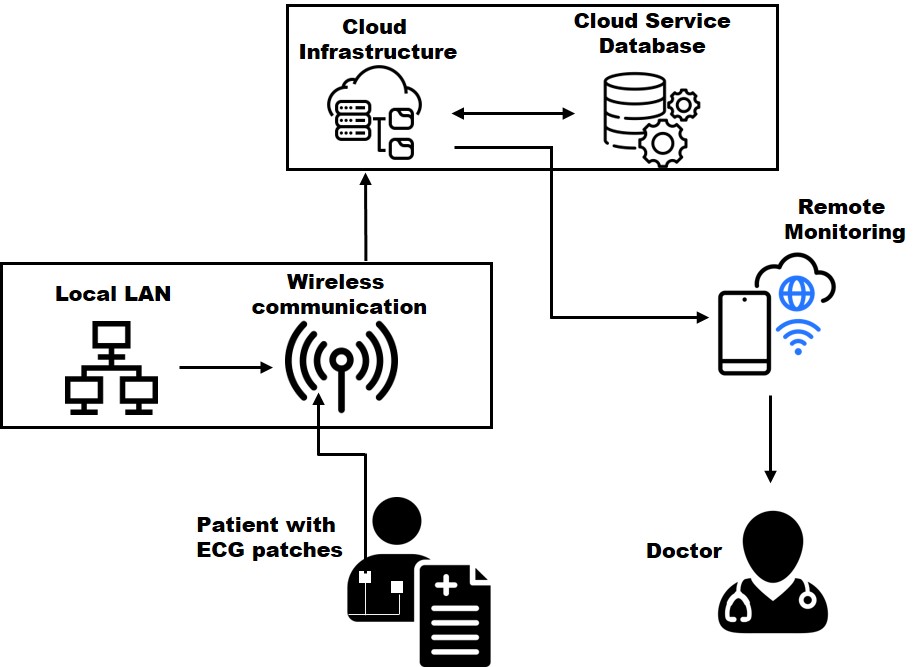}
\caption{Basic 
 architecture for cloud-based wearable IoT device.}
\label{fig_3}
\end{figure}

Wearable devices like smartwatches, medical patches, fashion items, gaming gadgets, and VR headsets are key IoT technologies, often worn close to the body and collecting various personal data.
The basic architecture of a cloud-based wearable medical IoT system involves a patient wearing a health tracker (like an ECG monitor) that transmits data wirelessly to a cloud service (see Figure~\ref{fig_3}). The cloud stores and manages the patient's health information, enabling remote monitoring.
\vspace{-5pt}

While wearable IoT devices may sometimes be classified as consumer IoT, their unique characteristics—such as continuous data collection tied to personal physiological attributes and their proximity to the human body—introduce distinct security challenges. These include the risk of identity theft from lost or stolen devices, unauthorized access to biometric data, and resource constraints that limit robust security measures. These challenges differ significantly from those faced by general consumer IoT devices, such as interoperability and multi-user access issues.

Unlike traditional computing systems, WIoT devices consist of a wide range of software, hardware, and
communication protocols. This diversity makes it challenging to establish a one-size-fits-all security solution, thereby impacting the ability to ensure confidentiality, availability, and integrity, collectively called the \textbf{CIA triad}~\citep{balas2020handbook} of the data and system functions. The diverse functionalities of WIoT devices and their intricate data lifecycle—collection, processing, storage, and transfer—expose consumers to significant security and privacy risks~\citep{lee2015risk}.
Worn on the body, these devices continuously collect environmental data, making them accessible to attackers.
The security and privacy issues of WIoT include three factors. The first one is device architecture, the second is network connectivity, and the third is data collected by wearable devices~\citep{siboni2016advanced}.
The security concerns surrounding device architecture encompass both hardware and software aspects. Due to low-resource device features (in terms of power source, memory size, and low-bandwidth communication), it may result in severe security flaws~\citep{lennon2015all}. 
From the software perspective, the basic operating system used in wearable devices is highly exposed to vulnerabilities~\citep{cusack2017assessment}.
Software security vulnerabilities also stem from the limited experience of programmers creating software for wearable IoT devices. Serious bugs in the software or applications are rarely identified, and the involvement of multiple parties in the development and deployment of WIoT means that no one is willing to take responsibility for these flaws.

Many IoT devices automate their functionalities, e.g., Google Glass devices. This often exposes wearable devices to security risks.
Since cloud-based WIoT devices are connected to the internet, they are highly exposed to traditional attacks, such as DDoS attacks, data leakage, MITM attacks, phishing attacks that lure users into revealing sensitive information, eavesdropping, side-channel attacks, and compromised attacks. In addition, many WIoT devices rely on weak authentication protocols, making them susceptible to unauthorized access and data manipulation during data transfer. 
The major concern related to wearable devices is data collection, storage, and processing. Most of the wearable IoT data include personal data that may contain sensitive information about users' habits and even financial information. Situations can become grave, specifically in the case of medical IoT, which includes a medical history, particularly heart rate, blood sugar level, and even doctor--patient conversations.

Jin et al.~\citep{jin2022p} conducted a study focusing on the security vulnerabilities of medical IoT devices, particularly those utilizing Govee technology and Amazon cloud services. Their findings revealed that simple modifications to the devices' access policies significantly compromised the confidentiality of doctor--patient communications. This vulnerability was not isolated to Govee devices alone; the authors identified similar security and privacy issues in a range of medical equipment, underscoring their susceptibility to cyber attacks. This highlights the widespread nature of security risks in WIoT devices, especially those handling sensitive medical data.
Beyond the security and privacy vulnerabilities discussed earlier, the potential for theft or loss of these WIoT devices is another serious concern.

Since these devices contain personally identifiable information (PII), losing one could expose sensitive data to attackers, who could exploit them for identity theft.
In the contemporary era, as the market for VR headset usage expands, privacy concerns are also escalating at a similar rate~\citep{giaretta2022security}. In this direction, Adams et al.~\citep{adams2018ethics} conducted an interview with 20 participants and concluded that to what extent the users and developers are concerned about the data collection practices in VR headsets.
According to the HP security report~\citep{karakaya2016secure},  smartwatches face multiple security vulnerabilities such as inadequate user authentication, absence of encryption, and insecure interfaces. Given their widespread use in the WIoT, addressing these issues should be a top priority.

In short, some of the common security vulnerability problems with wearable IoT devices include \textbf{hardcoded password issues, code injection issues, unsecured API issues, web application vulnerability issues, and lack of proper encryption} in \mbox{communication~\citep{balas2020handbook}} and most importantly, privacy violation issues.

\subsubsection{{Mitigation Approaches}}
Well, there is no doubt that it is a complex setting and nightmare as far as securities in WIoT are concerned. Building security into solutions and platforms at every component level is crucial. 
As discussed earlier, weak user authentication methods expose sensitive user data to potential attackers through passive or active attacks. Fortunately, researchers are actively working to develop solutions to address these security concerns. One such promising approach is \textbf{out-of-band biometrics authentication (OOB-BA)}, proposed by Singh et al.~\citep{mahinderjit2020novel}. \begin{itemize} \item This method utilizes enhanced out-of-band (e-OOBA) two-factor biometric authentication to create trust relationships between users and services and protect \mbox{users' data}. \item OOB-BA can provide better security and minimize information leakage, and it can be an alternative authentication model for future wearable technology adoption.\end{itemize}
Wearable IoT devices, including medical wearables like electronic skin patches,
ECG, monitors, watches, fit bits, etc., are vulnerable to cyber attacks.

Kavianpour et al.~\citep{kavianpour2021framework} propose an \textbf{``Attack-Surface-Reduction by Design''} approach to mitigate security risks in medical WIoT. This framework consists of a pronged strategy. \begin{enumerate}[leftmargin=3.3em,labelsep=4mm]
\item[(1)] Network 
 Traffic Analysis: The ML technique is used to analyze network traffic patterns of IoMT devices to identify anomalies that may indicate a potential cyber attack.
\item[(2)] Attack Surface Reduction by Design: The authors propose a set of guidelines for reducing the attack surface of IoMT devices. This includes hardening the devices' firmware and software, limiting their functionality to only what is necessary, and ensuring that only authorized users can access them.
\item[(3)] Forensics: The framework includes a forensic component to aid in the investigation of security incidents. This component uses digital forensics techniques to gather evidence of an attack and reconstruct the sequence of events leading up to \mbox{the attack}.
\item[(4)] Overall, the proposed framework aims to increase the security of networked medical devices and reduce the risk of cyber attacks that could compromise patient safety and privacy.
\end{enumerate}

To mitigate the same issue, in case of device loss/theft, Bhatia et al.~\citep{bhatia2021soft} proposed a soft computing approach for mitigating real-time abuse in IoT systems by detecting and predicting anomalies. The proposed approach uses a hybrid model that combines an online clustering algorithm with a time-series prediction model. The online clustering algorithm is used to identify anomalous data points in real time, while the time-series prediction model is used to predict future values of the IoT data.
Since these devices hold sensitive data, they need a strong authentication mechanism to make them more secure. One possible mitigation approach is continuous authentication, which is possible in the case of skin-attached medical wearable equipment. 

Zhao et al.~\citep{zhao2020trueheart} proposed a continuous authentication system based on cardiac biometrics measured from wrist-worn wearables. This approach aims to address the limitations of traditional authentication methods by using a biometric that is difficult to forge or steal and can be watched continuously without causing inconvenience to the user. The proposed system uses machine-learning algorithms to classify the users' cardiac biometrics and authenticate the user in real time.
Continuous biometric authentication methods, while effective, raise concerns about user convenience and the accuracy of measurements under varying physical conditions.
Gulhane et al.~\citep{gulhane2019security} introduced a novel risk assessment framework for VR learning environments, employing attack trees to compute a risk score for diverse threats within VRLEs.

In pursuit of striking a balance between privacy and data access, Fang et al.~\citep{fang2019fine} introduced a \textbf{fine-grained privacy-preserving access control architecture (FPAS)} tailored for smartwatches. Their approach utilizes an \textbf{identity-based authentication scheme} leveraging cryptography, coupled with the segmentation of data requesters based on \mbox{distinct attributes}.
\subsubsection{{Open Problems}} Wearable devices, often worn near the human body, store a wealth of user information, including personal sensitive biomedical data, heart rate, sleep patterns, and even user behavior. The leakage of such sensitive information represents a significant threat to user privacy and well-being, potentially leading to identity theft or manipulation of health records. 
Addressing this critical issue should be at the forefront for researchers, particularly in developing a robust data encryption method and secure communication protocols for wearable devices.
\subsection{Category 7: IoT in Smart Retail and Supply Chain.}
\subsubsection{{Security Challenges}}
The burgeoning adoption of IoT in the retail sector, projected to reach a staggering \$94 billion by 2025, presents a new wave of security challenges~\citep{WinNT26}. 
As retailers start integrating smart systems like contactless point of sale (POS), systems, mobile transactions, or smart shelves to enhance customer experience and business agility~\citep{wang2021iot, di2015towards}, these interconnected devices create a complex attack surface for malicious actors.  
The security risks in the retail IoT (RIoT) are multifaceted. \textbf{Data security} has been the prime concern as retailers face the rising threat of complex cyber attacks and the potential for data breaches. Such breaches can lead to significant brand damage and customer loyalty loss ~\citep{WinNT20}.
Major retail companies' utilization of payment platforms like Apple Pay and Google Pay is causing concerns about cyber attacks and potential financial losses. It is reported that since 2005, retailers have seen over 10,000 data breaches, mainly due to flaws and vulnerabilities in the payment system~\citep{WinNT21}.

POS systems often involve multiple components, including external hardware, software, and cloud-based services. Security vulnerabilities in such intricate networks can explode at any instance, and thus, it is very cumbersome to tackle RIoT security on a device-by-device basis. 
Cybercriminals have numerous opportunities to exploit retail systems, whether it is at the vendor's source or during on-site technology deployment.
Exploiting a security issue in the software used on POS devices or the cloud server could allow the deployment of malware or infected software. This malicious software cannot only damage the retail system but can also steal customer financial information, leading to significant financial loss and reputational damage to the brand of the company.
While RIoT can enhance the consumer--brand relationship through features like purchase history tracking, customers might feel hesitant to engage in such technologies if they fear their data are not secure. A data breach exposing customer purchase history could lead to financial losses and erode customer trust. Unreliable code and/or software of RIoT introduces an additional layer of security risk, as attackers can easily manipulate the code to access sensitive information.
The other security threat is the surveillance system of RIoT. RIoT often utilizes \textbf{customer tracking technologies}, raising privacy concerns. While tracking customer movements and purchasing habits can be beneficial for targeted marketing/advertisements, it is crucial to prioritize user privacy by ensuring data collection security, transparency, and compliance with regulations like general data protection regulation (GDPR) or country-specific data protection laws~\citep{singh2024crumbled}.
We know that RIoT tracks customers' movements and purchasing habits deliberately, as purchasing behavior correlated with consumers' movements can reveal extremely private habits. Tracking technologies like Wi-Fi logging and biometric imaging pose significant privacy risks, as they may create full behavioral profiles of consumers and could be misused for targeted marketing or discriminatory pricing~\citep{mavroudis2018eavesdropping}
\textbf{Ransomware attacks} are also prevailing in the supply chain networks due to overlapping security systems. Ransomware attacks on supply chains exploit interdependencies, causing cascading disruptions that amplify financial losses and operational delays across upstream and downstream partners. The ransom is lowest for a hub and spoke network and highest for a line network. For example, the attack on Kojima Industries disrupted Toyota’s production, showcasing the vulnerability of interconnected systems and the critical need for coordinated cybersecurity strategies~\citep{cartwright2023economics}.
\textbf{Supply chain attacks} are adversarial attacks on supply chain methods. \begin{itemize}
\item In supply chain systems, several IoTs act as network-connected sensors that may be part of cyber--physical systems. Hence, each IoT is being treated as a new door for adversaries~\citep{wang2021insecurity}. \item It is conceivable that this could lead to supply chain attacks, as the same type of attack that caused a single disruption in production could potentially be employed to impact and disrupt an entire supply chain simultaneously.\end{itemize}
Retail IoT systems are also vulnerable to DDoS attacks, which can disrupt critical systems such as POS terminals and inventory management systems. These attacks can also result in financial losses and damage to retailers' reputations.
The short lifespan of IoT devices introduces a significant security challenge~\citep{yousefnezhad2020security}. Once these devices meet their end-of-life and are no longer supported with security updates, they become vulnerable to exploitation. Intruders can potentially compromise outdated devices to gain access to sensitive data within a large supply chain, leading to confidentiality breaches, integrity, or availability of the entire system.

\subsubsection{{Mitigation Approaches}}
The new era of finding security solutions is inclining toward machine learning, AI, Blockchain, etc.
Fog computing, ML, and AI can all play important roles in enhancing the security of retail IoT devices and systems. 
Additionally, homomorphic encryption (HE) provides an innovative solution to enhance the security of Retail IoT systems by enabling computations on encrypted data without decrypting them. This ensures that sensitive customer data remain secure even during processing. For instance, \textbf{Fully Homomorphic Encryption (FHE)}, a robust subset of HE, allows complete arithmetic operations on encrypted data, thereby reducing the risk of exposure to data breaches~\citep{jaffar2022systematic}.

Fog computing involves processing data at the network's edge, closer to the devices generating the data, rather than sending all the data to the cloud for processing. This approach can improve security by reducing the amount of sensitive data that needs to be transmitted over the network, minimizing the potential attack surface for malicious actors. Additionally, fog computing allows for real-time analysis of data, enabling tasks like monitoring devices' behavior for anomalies, triggering alerts for suspicious activities, and even blocking unauthorized access attempts. The rapidly evolving landscape of RIoT security demands innovative solutions. Fog computing brings a powerful solution to enhance security, particularly within supply chains. 
In this direction, \mbox{Musa and Vidyasankar et al.~\citep{musa2017fog}} proposed a \textbf{fog computing framework} that integrates edge/cloud-computing and IoT to real-time monitoring and analysis of supply chain data. This framework also includes a security layer to enhance the efficiency and security of BlackBerry's sensitive information. To address the DDoS attack in RIoT systems, Brasilino et al.
~\citep{brasilino2019mitigating} proposed a custom hardware solution. The solution consists of two main components: \begin{itemize}
\item A hardware module that performs the actual DDoS mitigation and, \item A software framework that orchestrates the hardware module and provides an interface for configuring and managing the mitigation process.\end{itemize} The proposed custom hardware is designed to be placed between the IoT devices and the network.
The integration of IoT technologies into retail oil outlets (ROO) presents a mix of opportunities and challenges. While technologies like cloud-based IoT offer numerous advantages for streamlining operations and gathering valuable data, it also introduces new security concerns. To address this concern, Das et al.~\citep{das2022iot} propose a model for ROO by incorporating “data security” into the existent framework of technology, organization, and environment within the IoT ecosystem.
This model is effective in providing a holistic approach to secure data throughput in the lifecycle of ROO operation. 

To further strengthen Das’s data-centric model, Mohanta et al.
~\citep{mohanta2020survey} proposed the use of ML, AI, and Blockchain technology as potential solutions. Specifically, these technologies can be used for: Intrusion detection and prevention: machine learning and AI are being used to analyze network traffic and identify anomalous behavior that may indicate a security breach.
As discussed earlier, securing device identity is paramount. Blockchain technology excels in creating a secure and decentralized system for authenticating IoT devices and controlling access to sensitive data. This ensures that only authorized devices can connect to the internet, alleviating the risk of unauthorized access. 
Data encryption and privacy protection: to encrypt data and protect the privacy of sensitive information, AI and ML are being used extensively.
Threat intelligence and incident response: machine learning and AI can analyze data security and provide real-time insights into emerging threats, allowing for more effective incident response.
\subsubsection{{Open Problems}} A major challenge facing RIoT security revolves around financial transactions. Many of these transactions involving IoT devices rely on personal information stored in the cloud, thereby introducing an additional vulnerability susceptible to potential malicious attacks. These inherent vulnerabilities necessitate robust security measures to safeguard financial data and transactions within the RIoT ecosystem.

\subsection{Category 8: IoT in Agriculture}
\subsubsection{{Security Challenges}}
In the facets of changing climate and exploding global population, smart devices are emerging as an essential tool for precision agriculture ~\citep{akhter2022precision,singh2020odysseys}. However, integrating the Agriculture Internet of Things (AIoT) and cloud computing promises enhanced crop production through cost reduction, efficient monitoring of crops, and maintenance. It also introduces security concerns and major cyber attacks on critical agriculture \mbox{infrastructure~\citep{namani2020smart, patil2012internet, mekala2017survey}}.
The most common examples of cloud-based IoT applications in agriculture are (1) smart sensor-based IoT devices for monitoring crops, soil, fields, livestock, storage facilities, or generally any necessary factor that influences agriculture \mbox{production~\citep{krzywoszynskasoil, yu2021review}};
(2) smart IoT agriculture vehicles, drones, autonomous robots, and actuators (e.g., DroneSeed, SenseFly, SoilCares~\citep{refaai2022application}); (3) smart greenhouse and \mbox{hydroponics~\citep{7123563}}; (4) for data analytics, data visualization, and data management systems in agriculture; \mbox{(5) predictive} modeling and planning; (6) precision agriculture using IoT~\citep{singh2020odysseys, song2022development}.

The rapid evolution of IoT, communication technology, and digitization introduces new security risks in AIoT. In agriculture, interconnected AIoT systems collect and analyze data, creating vulnerabilities that can be exploited, leading to significant \mbox{disruption~\citep{salam2020internet, gupta2020security}}.
Mori et al.~\citep{mori2022iot} have further explored these security concerns in detail, highlighting the complexity of securing a secure environment with multiple interconnected IoT devices. AIoT relies on multiple interconnected sensors, so compromising one device can create a domino effect, making others vulnerable.
When interconnected devices such as sensors are compromised, the disruptions can propagate through the entire network, affecting water distribution, sensor readings, and automated equipment. This can significantly hinder agricultural productivity and expose unforeseen vulnerabilities in supply chains reliant on stable operations.
Data tampering is another issue where hackers can access and alter data transmitted to weather stations or cloud servers, leading to inaccurate analysis and poor decision-making by farmers.

AIoT security vulnerabilities manifest across all layers of the IoT framework.
The first application layer, where farmers connect with IoT devices, has a high risk of data theft~\citep{triantafyllou2019precision}. The layer experiences a range of security challenges, including \textbf{data theft, access control sniffing attacks, and service interruption attacks}. These threats can potentially disrupt the delivery of IoT services to farmers.~\citep{ferrag2020security, glaroudis2020survey}.
Security vulnerabilities with AIoT have a cascading effect, impacting not only IoT--farmer relationships but also retailers--stakeholders throughout the agriculture supply chain. Attacks targeting the application layer can potentially disrupt communication and data access for all connected individuals (farmers, retailers, stakeholders), jeopardizing trust and collaboration. The security vulnerabilities in the AIoT framework extend beyond the application layer. 

The middleware layer, acting as an intermediary between hardware and applications, presents significant security risks, including \textbf{MITM, SQL injection, signature wrapping, cloud malware injection, and flooding attacks.} These attacks compromise device data, disrupt access control mechanisms, and may hurt the farmer’s decision-making abilities. Thus, it is of paramount importance to explore the security of these devices, ensure process and access control, and maintain the integrity of data analysis~\citep{farooq2019survey, madushanki2019adoption}.

The cloud server connecting AIoT devices to the internet is a critical vulnerability point, susceptible to phishing, access, DDoS, data transit, and routing attacks~\citep{ahmed2018internet}.

This layer is responsible for conducting communication between device-to-device, device-to-cloud, device-to-gateway, and back-end data-sharing. A disruption at this layer could result in the device and service becoming non-functional. Moreover, if the data exchanged between IoT devices and the cloud are transmitted in plain-text without encryption, it can raise privacy concerns that impact both the IoT device and the farmers' data.

\subsubsection{{Mitigation Approaches}}
Due to the complexity of the smart agriculture environment, AIoT is vulnerable to \textbf{intrusion attacks}. These attacks involve unauthorized access to a farm's IoT network, compromising sensitive data related to crop production, livestock monitoring, or equipment management. ML can play a vital role in mitigating these attacks. \mbox{Raghuvanshi et al.~\citep{raghuvanshi2022intrusion}} proposed an intrusion detection mechanism using machine-learning techniques, designed to detect abnormal behavior in the smart irrigation system, which could be indicative of a security breach. Through the utilization of machine-learning algorithms for intrusion detection, the system can evolve to address emerging threats and enhance its accuracy progressively.
Building on this, federated learning frameworks offer a decentralized approach to training machine-learning models across multiple devices. This approach enables collaborative learning for applications like disease detection and yield forecasting while preserving sensitive agricultural data on local devices~\citep{kondaveeti2024federated}.
As machine learning continues to advance, deep learning emerges as a modern tool that offers a secure and privacy-preserving framework for smart agriculture. One such example is the \textbf{Secured Privacy-Preserving Framework (SP2F)} proposed by Kumar et al.~\citep{kumar2021sp2f}. The proposed framework is designed to protect the privacy and security of data collected by UAVs while still allowing for efficient and effective data analysis.
To address the risks posed by data tampering, integrating anomaly detection algorithms with real-time monitoring systems can help identify irregularities in weather or soil data. Additionally, deploying redundant sensors and implementing cross-verification mechanisms can maintain data accuracy and reliability, even when some systems are compromised~\citep{kethineni2023iot}.

Another approach for preserving privacy in smart agriculture involves a data aggregation scheme. Song et al.~\citep{song2020fpdp} proposed a scheme that uses the \textbf{ElGamal Cryptosystem} to enable secure aggregation of data while protecting individual data points.
\begin{itemize}\item The proposed scheme has been rigorously analyzed and designed to particularly benefit farmers to share and analyze their combined data sets without compromising the privacy of the sensor reading. \item This approach is capable of solving the privacy-preserving issue not only in smart agriculture scenarios but also in the other applications of IoT. 
\end{itemize}
Furthermore, Zhou et al.~\citep{zhou2021efficient} have proposed a \textbf{complementary privacy-preserving solution}, specifically designed for fog-based agricultural IoT systems.
\begin{itemize}
\item The scheme leverages secure multi-party computation techniques, allowing data owners to query the maximum value within a specific range without disclosing any sensitive information about their data to the fog nodes.
\item The proposed scheme is particularly useful when analyzing aggregate data at the fog, which can help to minimize computation and communication overhead while ensuring data privacy~\citep{zhou2021efficient}. 
\end{itemize}
While privacy preservation is crucial, another security challenge exists in smart agriculture, another approach involves a framework designed to detect anomalies in data collected from various sources within a smart farming system while preserving the privacy of sensitive data using homomorphic encryption techniques~\citep{chukkapalli2021privacy}. 

Looking beyond privacy, Kumar et al.~\citep{kumar2021pefl} proposed a deep privacy-encoding-based federated learning framework (PEFL) to address privacy concerns associated with federated learning in smart agriculture systems by utilizing a deep privacy-encoding mechanism.
Federated learning allows distributed training on devices, further minimizing the security risk associated with data sharing. 
Minimizing the risk of cascading failures in interconnected systems can be achieved through decentralized approaches, such as edge computing and Blockchain-based coordination. These methods help contain compromised devices and prevent disruptions from spreading across the network \citep{malik2024optimizing}.
Anidu and Dara et al.~\citep{anidu2021review} believe that standardization could be a potential solution for interoperability and consistency issues in a smart agriculture environment, whereas the Blockchain and distributed ledger technology could help in mitigating data integrity and data confidentiality issues and thus making the data storage system more secure than before.

Security monitoring is crucial to ensure the safety of smart farming. Modern technology can avail better security solutions for vulnerabilities within AIoT. One such example is the cloud-based security monitoring framework proposed by  Chaganti et al.~\citep{chaganti2022blockchain}. This framework effectively monitors device status and sensor data for anomalies, potentially identifying unauthorized access or other security breaches. Additionally, it utilizes a Blockchain-based smart contract application to securely store information about identified security anomalies. This allows for proactive mitigation for other similar attacks targeting other farms within the network.
\subsubsection{{Open Problems}}
AIoT devices are often resource-constrained, with limitations in processing power, memory, and battery life. These constraints make it challenging to implement robust security measures, leaving them widely vulnerable to unauthorized access and data breaches. These vulnerabilities could compromise sensitive agricultural information like crop yields, weather data, and farm operations. Human factors such as negligence, weak passwords, or poor security practices by farm personnel further complicate security issues. To effectively secure AIoT devices, lightweight security protocols specifically designed for resource-constrained devices are crucial. While designing these protocols, care must be taken to balance the need for strong security while still preserving the user-friendly experience for farmers, considering the inherent device limitations. 

\subsection{Category 9: IoT in Smart Grids}
\subsubsection{{Security Challenges}}
As demand for sustainable energy grows, smart grid technologies are attracting attention for their efficient energy distribution, improved grid reliability, and integration of renewable sources~\citep{al2019iot}.
Some examples of IoT in smart grids include smart switches, smart metering systems~\citep{pau2018cloud}, smart thermostats~\citep{ozgur2018iot}, smart batteries~\citep{tran2022concept}, etc.
Smart grid IoT devices, which are used to monitor, control, and optimize the distribution of energy across the power grid, are increasingly being integrated with cloud-based services to enable remote management and analysis. While cloud-based smart grid systems offer numerous benefits, they also pose several security risks that need to be addressed. Some of the security issues that arise in cloud-based smart grid IoT devices are data privacy issues, malware attacks issue, DDoS attacks, standardization, transparency issues, etc.

Smart grid systems generate huge amounts of sensitive and private data, including customer personal information, energy usage patterns, and grid performance data which are vulnerable to leaks.
These data need to be protected from unauthorized access, theft, or misuse~\citep{boroojeni2017smart}. One particular security threat in the smart grid is a \textbf{false data injection attack.} In this attack, the attacker injects false or malicious data into the system to disrupt its operation or steal information. In a smart grid, IoT devices are used to monitor and control the flow of electricity, and attackers can exploit vulnerabilities in these devices to inject false data. These false data can then cause the smart grid to malfunction or behave in unexpected ways, potentially leading to power outages or other disruptions. These are just some concerning examples of the security risks associated with cloud-based smart grids. Addressing these issues is crucial for ensuring the reliable and secure operation of the smart grid system.

Another security threat to the smart grid is a DDoS attack. Various types of  DDoS attacks can be directed at different components of the smart grid, including meters, communication networks, and control systems. Successful DDoS attacks on the smart grid infrastructure can have severe consequences, potentially resulting in power outages, data loss, and financial damages~\citep{stellios2018survey}.
Cloud-based smart grid systems are susceptible to various cyber attacks that can disrupt operations or compromise sensitive data. Malicious software malware and DDoS are major concerns for this category. Malware can be introduced through compromised IoT devices or phishing attacks on cloud service providers~\citep{gunduz2020cyber}. DDoS attacks can be used by attackers to overload cloud-based smart grid systems, leading to system crashes or unavailability. This can lead to disruptions in energy distribution and other critical services~\citep{huseinovic2020survey}.

Additional challenges exist beyond cyber attacks and DDoS attacks. Insider threats can arise if unauthorized personnel misuse this private information for personal gain, potentially engaging in illegal activities~\citep{gonen2020false}.

A lack of standardization in cloud-based smart grid systems can create vulnerabilities, as different vendors may have different security protocols or use different data formats that may not be compatible with each other.
Lack of transparency in cloud-based smart grid systems poses other vulnerabilities, as users may not have full visibility into the data being collected, processed, or shared.
Smart batteries are also prone to unauthorized access to battery data and control systems. The attackers can exploit vulnerabilities in the battery's software and hardware and could gain access to sensitive information. This could possibly result in battery failure, data theft, or other security issues.
The smart grid collects a large amount of data from consumers, which can include sensitive information such as energy consumption patterns and personal preferences. Attackers can exploit these data further to breach consumers' privacy.

\subsubsection{{Mitigation Approaches}}
The security of the smart grid necessitates a multi-layered approach that addresses vulnerabilities across cloud-based systems and individual components. Robust authentication mechanisms, data encryption, and intrusion detection systems can enhance cloud security.
The integration of post-quantum cryptography (PQC) to secure the smart grids can be a future-proof solution. With quantum computing poised to undermine traditional cryptographic methods, PQC algorithms—resilient to quantum attacks—ensure the long-term security of communications and data within smart grid infrastructures~\citep{aydeger2024towards}.

Regular security audits, employee training, and incident response plans are also important to ensure the security of these systems.
To mitigate these security concerns, manufacturers of smart batteries should implement strong encryption and authentication measures to protect data and communication channels. They should also incorporate tamper-evident features to prevent physical attacks on the battery. Moreover, routine firmware updates can help mitigate numerous security vulnerabilities and enhance the security of both the battery's software and hardware.

Security vulnerabilities in the smart grid can be life-threatening, leading to a loss of critical services, physical damage to infrastructure, and potential safety hazards for individuals. Early detection and mitigation are of the utmost importance, as vulnerabilities can disrupt critical services, damage infrastructure, and pose potential safety hazards.
Researchers are exploring various approaches to mitigate these risks. For instance, Yılmaz and Uludag ~\citep{yilmaz2021timely} proposed a framework for \textbf{anomaly detection and attack recognition} that leverages data from various sources, including IoT devices, network traffic, and system logs. The mitigation and response mechanism involves isolating the affected devices and shutting down the affected systems to prevent further damage.
To address the cybersecurity issue in smart grids, Muyeen's group~\citep{berghout2022machine} suggested the use of machine learning to tackle the issue.
They acknowledged that machine learning can be used for cybersecurity in smart grids, including intrusion detection, threat intelligence, anomaly detection, and risk assessment.

The smart grid is vulnerable to both faults and cyber attacks, and a comprehensive security scheme is necessary to ensure the reliable and secure operation of the grid. There should be multi-layer security to mitigate vulnerabilities in the smart grid against faults and cyber attacks. Chen et al.~\citep{chen2021multi} proposed a three-layer security scheme that includes the physical layer, network layer, and application layer security.
A \textbf{distributed trust management system (DTMS)} can be used to mitigate false data injection attacks (FDAs). FDAs inject malicious data that can evade traditional detection methods, thereby threatening grid reliability and causing major operational challenges~\citep{zhang2020detecting} The DTMS consists of a set of trust nodes that are responsible for verifying the data received from the sensors and measurement devices of the smart grid. The trust nodes use a distributed consensus algorithm to reach a consensus on the correctness of the data ~\citep{unsal2021enhancing}. However, DTMS has limitations, including the requirement for extensive collaboration among stakeholders, which may not always be feasible.

To address the privacy breach, the \textbf{smart generative adversarial network (GAN)} can be used to obfuscate consumer data. GAN generates synthetic data that are similar to real data but do not reveal the consumers' actual information. The synthetic data are then used for analysis and decision-making, while the real data are obfuscated to protect the privacy of the consumers ~\citep{desai2022mitigating}.
To mitigate cyber and physical attacks, big data and AI technologies can be used to analyze and predict the security risks in the smart grid.
To mitigate DDoS attacks in smart grids, various techniques like intrusion detection systems, firewalls, and load balancing mechanisms can be used. One such approach has been proposed by Chehri et al.~\citep{chehri2021security}, a framework that consists of four phases: data collection, data preprocessing, risk modeling, and risk evaluation. 
In the risk modeling phase of the four phases, diverse AI techniques like machine learning and deep learning are employed to model security risks within the smart grid.
\subsubsection{{Open Problems}} The global shift to smart grids offers significant benefits in energy integration and management, as seen in Germany's IoT-enabled grid optimizing renewable energy. However, this transition introduces security risks. As many capitals aim for IoT-based smart grids to achieve net-zero goals by 2050, the increase in devices heightens security concerns.

Key future challenges include developing robust security frameworks, ensuring data integrity and privacy, and establishing standardized protocols for smart grid technologies.

\subsection{Category 10: Cloud-Based IoT in Aviation}
\subsubsection{{Security Challenges}}
Cloud-based IoT solutions are transforming aviation by streamlining data collection, processing, and analysis in the industry.
Some of the key applications of cloud-based aviation IoT are Smart Airport~\citep{bouyakoub2017smart}, predictive maintenance~\citep{pech2021predictive}, improving customer satisfaction~\citep{andel2021iot},  safety management~\citep{wang2013aviation}, improving the overall performance by reducing emissions ~\citep{okpala2020assessing}, luggage management ~\citep{noel2021smart}, in-flight entertainment systems
~\citep{samha2020applied,wang2022intelligent}, etc.
In addressing security challenges within this category, \textbf{data security} emerges as a foremost concern. Given that cloud-based Aviation IoT systems extensively depend on data collection and transmission, safeguarding the integrity and confidentiality of these data is of prime concern. Any data breach has the potential to result in harm to aircraft, passengers, and the airline's reputation~\citep{jalali2018smart}.

A recent study by Aboti~\citep{aboti2020studies} discusses escalating security challenges and threats faced by smart airports due to the integration of IIoT and the increased use of smart devices. The integration has opened doors for vulnerabilities like malware, cyber attacks, and insider threats. The authors emphasize the importance of implementing robust cybersecurity governance in smart airports to protect against cyber threats and enhance operational practice.
Cho et al.~\citep{cho2023reviewing} provided a case study highlighting the importance of cybersecurity in commercial aviation. 
The main security issue discussed by Ukwandu et al.~\citep{ukwandu2022cyber} is the vulnerability of aviation infrastructure to cyber attacks, particularly from Advanced Persistent Threat (APT) groups that operate in collaboration with a particular state actor to steal intellectual property and intelligence to advance their domestic aerospace capabilities as well as monitor, infiltrate, and subvert other sovereign nations' capabilities.

Lazaro's group~\citep{florido2021identifying} conducted a study by combining interviews with airport security personnel and a survey of airport users to gather data on the cyber security risks and vulnerabilities of Spanish airports. The study identified several cyber security risks, including inadequate user authentication measures, vulnerabilities in airport software systems, and the use of insecure communication channels. 
In the context of aviation cyber security, the use of Electronic Flight Bags (EFBs) adds a layer of risk, as unauthorized access to EFB servers could potentially compromise the safety and security of aircraft operations.

Hackers exploit vulnerabilities in the remote desktop protocol or use \textbf{phishing attacks} to steal login credentials and gain unauthorized access to aviation IoT systems. This unauthorized access can lead to a variety of negative consequences, such as stealing sensitive data, causing equipment malfunctions, and compromising the safety of passengers and crew.
Considering the significant volume of data collected by IoT devices in the aviation industry, ensuring the privacy and confidentiality of these data becomes a major concern. Additionally, the complex environment of connected IoT devices in aviation increases vulnerability to cybersecurity threats. In the worst scenario, hackers might manipulate or disable critical equipment, leading to malfunction and potential hazards.  Overall, robust security measures must be implemented to mitigate these risks and protect the aviation sector from potential cyber attacks.

\subsubsection{{Mitigation Approaches}}
The aviation industry is facing a growing challenge from cyber attacks. To counter this challenge, researchers are exploring several approaches. One such approach has been proposed by Sam Adhikar~\citep{adhikari2021intelligent} that leverages 5G and intelligent cyber defense mechanisms to enhance the security of aviation systems, focusing on high-speed data transfer and advanced threat detection. 

Airports have robust security measures, such as the use of encryption technologies, regular security audits, and employee training programs. 
Moreover, \textbf{Blockchain-based decentralized identity management systems} can significantly enhance authentication mechanisms. It will enhance authentication in aviation IoT by decentralizing identity repositories, reducing single points of failure, and strengthening security through improved transparency and traceability~\citep{zhang2024research}.

In a study, Belkens's group's suggested approach consists of three main components \begin{enumerate}[leftmargin=3.3em,labelsep=4mm]
\item[(1)] A 
 5G-enabled communication network that provides high-speed, low-latency connectivity between aviation systems and enables real-time threat detection \mbox{and response}; \item[(2)] An intelligent cyber defense system that uses AI and ML algorithms to detect and mitigate cyber threats in real time; and \item[(3)] A set of best practices and guidelines for the implementation and operation of the cybersecurity framework~\citep{ukwandu2022cyber}.\end{enumerate}

In another measure, Lazaro et al.~\citep{florido2021identifying} recommended that a proactive and collaborative approach to cybersecurity is essential to guarantee the safety and security of Spanish airports and the well-being of passengers.
In another study, Zkik et al.~\citep{zkik2023secure} proposed a secure and reliable model for the traceability of records in the airline supply chain using Blockchain and machine-learning technologies. The model is designed to enhance the efficiency and security of the supply chain by providing real-time tracking and monitoring of records related to cargo and other goods.
ML and AI could be better alternatives to finding solutions for security attacks caused by electronic flight bags.

Bitton et al.~\citep{bitton2019machine} proposed a system that utilizes a combination of feature engineering and machine-learning algorithms to detect and classify intrusions in real time. The study focuses specifically on securing remote desktop connections, which are a common attack vector for cybercriminals seeking to gain unauthorized access to EFB servers. 
In addition to the general security measures, specific security measures need to be taken for IoT devices in the aviation sector. Such measures encompass adopting secure communication protocols, enforcing robust authentication and access control, conducting routine security audits, maintaining up-to-date firmware, and encrypting sensitive data. Through the implementation of these security measures, the aviation sector can mitigate risks linked with IoT devices, thereby guaranteeing the safe and secure integration of this technology.
\subsubsection{{Open Problems}}
Despite advancements in security, aviation IoT continues to face significant challenges, including data security risks like passenger data breaches and vulnerabilities to cyber threats such as malware and phishing attacks. Malware can compromise, steal data, and disrupt operations, while phishing can deceive passengers into revealing sensitive information, leading to financial losses. These issues threaten system integrity and pose substantial risks. To enhance the safety and security of aviation IoT systems, implementing robust security measures like encryption, access control, regular audits, and comprehensive employee training on cyber threats is essential.

\section{The Way Forward: In Search of Uniform Solutions for All Security Challenges}\label{sec4}
The rapid growth of IoT devices highlights the urgent need for global cybersecurity measures.
Consequently, 2024 emerges as a pivotal juncture prompting governments to scrutinize advancements in laws and regulations~\citep{WinNT28}. 
To accomplish this goal, numerous nations have developed distinct regulatory standards aimed at enhancing the security of IoT. For instance, the European Union introduced the Cybersecurity Act and Cyber Resilience Act to bolster cybersecurity throughout the EU, complementing existing legislation such as the GDPR and the Network and Information Security Directive~\citep{WinNT28, lata2021standards}. Likewise, in the United States, IoT regulatory standards encompass the Federal Trade Commission Act, the Cybersecurity Information Sharing Act, and COPPA~\citep{lata2021standards}. Other countries are also subject to their own nation-specific laws, with some currently in the process of formulating regulations.
Despite the nation-specific IoT regulations, there is a universal desire for a standardized security solution along with some similar ongoing work~\citep{WinNT27}. Addressing the myriad of security and privacy challenges in cloud-based IoT and creating a one-size-fits-all security solution for the diverse security and privacy challenges in cloud-based IoT is a complex task (RQ3). Various ongoing industrial collaborations are underway to achieve this goal~\citep{OCF}.
Thus, the solution to RQ3 is feasible but remains a challenging task to implement across all categories of IoT.
The rationale behind this lies in the distinct characteristics of each category of traditional and cloud-based IoT devices. For instance, consumer-oriented devices can easily scale up to accommodate more hardware resources, while medical wearable devices must remain compact and portable. Consequently, security solutions must be tailored to the specific features of each category. In our section on mitigation approaches, we delve into potential solutions, recognizing the need for category-specific security measures.
Nonetheless, we put forth an optimal approach aimed at mitigating the security issue to a certain extent. In the realm of IoT devices intended for consumer use, a complex ecosystem involving multiple third parties is typically at play, which can generally be categorized into three levels: the manufacturer (responsible for hardware components like sensors, memory cards, etc.), third-party services (involving tasks such as device code development, natural language processing, etc), and cloud services (facilitating internet connectivity, communication with end-users, etc.).
We advocate for the implementation of a multi-level protocol verification framework tailored for IoT, emphasizing rigorous security checks across manufacturing, third-party services, and cloud service levels. This approach ensures a structured and consistent security evaluation process. To cater to IoT categories with higher resource constraints, such as medical wearables or AIoT, we propose integrating adaptive verification systems driven by machine learning. These systems can dynamically adjust security measures based on real-time device usage patterns, enabling scalable and efficient implementation.
For large-scale IoT ecosystems, AI-driven modular verification systems can adapt security protocols based on device complexity. For instance, consumer-oriented IoT devices may rely on more scalable solutions like cloud-based security checks, while life-critical IoT devices, such as healthcare IoT or autonomous vehicles, would require more intensive localized verifications to ensure real-time safety and data integrity. Additionally, incentivizing manufacturers and cloud providers to adopt these frameworks through compliance rewards or certifications could accelerate adoption.
Hence, we believe this standardized verification mechanism has the potential to be a significant milestone in addressing security concerns, particularly in the context of life-critical IoT devices.
\section{Limitations} Although this paper categorizes cloud-based IoT into ten distinct types, we recognize that the field is continuously evolving, and additional categories or hybrid domains may emerge over time. To address this dynamic nature, the proposed framework is intentionally designed to be flexible and adaptable, providing a robust foundation for future research and development in enhancing the security of cloud-based IoT systems.

\section{Conclusions}
This paper introduces a novel method to analyze IoT security by categorizing devices into ten distinct areas, each with unique challenges (as shown in Table \ref{tab:SS}). The goal is to identify and address the security issues of each category and explore existing solutions, highlighting areas for future research when solutions are lacking. This study also notes that some IoT devices may not rely on cloud computing and possess their own security challenges, which are expected to increase as they integrate more with cloud services. This paper discusses the added risks, such as data breaches and DDoS attacks from cloud integration, especially for resource-limited devices like wearables, emphasizing the importance of proactive security measures in IoT design and deployment. Additionally, the possibility of implementing standardized security solutions across all categories is explored to enhance overall IoT security and protect user data. 

Penultimately, we envision an ideal world where responsible citizens take proactive measures to safeguard their digital lives. By actively updating device passwords and firmware and limiting the sharing of private data, individuals play a pivotal role in addressing security and privacy concerns. This collective effort towards adopting best practices empowers us to create a safer and more secure digital environment for all.


\begin{table}[H]
\tablesize{\footnotesize}
\caption{Short 
 summary table for each category.\label{tab:SS}}
	\begin{adjustwidth}{-\extralength}{0cm}
		\begin{tabularx}{\fulllength}{CCCCC}
			\toprule
			\textbf{Category} & \textbf{Issues} & \textbf{Mitigations} & \textbf{Use Cases} & \textbf{Ref.} \\
			\midrule
Consumer-oriented 
 IoT & VSA, third-party criticality, interoperability issue, MITM, DDoS attack, flash memory issue, data leakages, debugging interface issue, hardcoded-password issue, authentication issue, data collection issue. & SkillFence, cryptosystem, SDN, flow filtering, honeypot technique, rate limiting approach, MTD, Traceback, LOKI. & Amazon’s Alexa, Google Assistant, Samsung’s SmartThings, web camera*, Nest smoke alarm. & \citep{celik2018soteria,sivasankari2022detection,thakor2021lightweight,lit2021survey,ahmed2022detecting, obaid2021assessment, zhang2019dangerous, kumar2018skill, WinNT1, fernandes2016security, fernandes2017security, zhang2018homonit, otoom2023deep, attoh2024towards, sanli2024detection, WinNT29, aziz2023securing, navas2018not, sharma2022state, park2019security, alladi2020consumer, obermaier2016analyzing, seralathan2018iot, tekeoglu2015investigating, xu2018internet, mangala2021short, geeng2019s, jang2017enabling, zeng2019understanding, davis2020vulnerability,notra2014experimental, ren2023voice, hooda2022skillfence, chatterjee2017puf, nazzal2022vulnerability, shameli2015taxonomy, salva20185g, guozi2018ddos, yan2018multi, ma2015defending, connell2017performance, chen2020ddos, bhattacharjya2019secure, abu2020towards, qasem2024cryptography, gerber2021loki} \\ 
\midrule

Children's IoT & Eavesdropping issue, camera privacy issue, weak encryption, access control issue, POST tokens issue, jamming attack, plain-text API issue, personal information leak, authentication bypass issues, direct browsing issue, privilege escalation problem, backdoor credential leaks, remote cell access issue, unlock protected passwords issue, UART issue, authentication issue, DDoS attack. & Built-in security mechanism. & Hydrogen tracker, smart robots*, smart dolls*, hello barbie doll, BB8, baby application monitor. &  \citep{chu2018security,WinNT24, holloway2016internet, ataei2022authentication, yang2022review, hung2016glance, valente2017security, albrecht2015privacy, stanislav2015hacking, rivera2019secure, rafferty2017towards, yankson20204p,ling2022use,apthorpe2017smart} \\
\midrule

Healthcare IoT & Encryption issues, dictionary attacks for passwords, weak passwords, memory card issues, plaint-text API issues, DDoS attacks, data integrity, data confidentiality, data availability issues, replay, MITM, impersonation, and privileged-insider attacks. & Anonymization techniques, Blockchain mechanisms, bulwark, decision tree algorithm, ML-based healthGuard, cryptographic techniques, GHOSTDAG, deep learning concepts. & HRPM*, pulse oximeter*, DCT*, glucometers*, fire safety and security monitoring, indoor localization for healthcare facilities. & \citep{ganesh2022iot, WinNT10, okafor2022design,farahani2018towards, mcallister2017localization, flynn2020knock, newaz2021survey, west2016wearfit, FDA, wazid2020private, fathalizadeh2022privacy, arul2024iot, saeedi2019machine, newaz2019healthguard, srivastava2019light, srivastava2019data, shrivastava2022securing, wu2015tpad,rathore2020deep,li2020secured, bu2019bulwark,bradley2018security,jin2022p} \\

\bottomrule
\end{tabularx}
\end{adjustwidth}
\end{table}

\begin{table}[H]\ContinuedFloat
\small
\caption{{\em Cont.}}

\begin{adjustwidth}{-\extralength}{0cm}
		\begin{tabularx}{\fulllength}{CCCCC}
			\toprule
			\textbf{Category} & \textbf{Issues} & \textbf{Mitigations} & \textbf{Use Cases} & \textbf{Ref.} \\
			\midrule

Industrial IoT & Insufficient industrial hardware data protection, hardware memory dump, unprotected hardware interface, analog signal processing issue, DDoS, and device infiltration. & TPMs, STC scheme, VFNs, FLEAM. & Temperature/pressure sensors*, proximity server, smart meter, temperature-based control system*, cloud-based video surveillance systems. & \citep{chaudhary2023ddos,obermaier2016analyzing,zverev2019network, varga2017security, Anderson2011SmartMS, cleemput2016high, wurm2016security, horak2021vulnerability, soltan2018blackiot, tu2019trick, ahamad2024hybrid, kott2016internet, lesjak2016hardware, pu2023user, tan2021blockchain,  de2023intelligent, ibrahim2022ddos, li2021fleam, kuusijarvi2016mitigating,  stocchero2022secure} \\
\midrule

Smart Vehicle & Privacy threat, data authenticity attack, availability attack, Sybil attack, GPS deception attack, authentication attack, masquerading attack, wormhole attack, replay attack, data injection attack. & Secure software development practice, secure boot processes, hardware-based encryption methods, AI and ML, Blockchain-based architecture, and cloud-based sandboxing approach. & Smart vehicles. & \citep{shah2022survey,ye2004medium, tbatou2017security, zhu2024sybil, kumar2021multimedia, taslimasa2023security, khan2020cyber, zewdie2020iot, smys2021security, zhao2021detection} \\
\midrule

Wearable IoT & CIA triad, DDoS attacks, data leakage, MITM attacks, phishing attacks, eavesdropping, side-channel attacks, compromised attacks, personal information leaks in case of theft, hardcoded password issues, code injection issues, unsecured API issues, web application vulnerability issues, and lack of encryption in communication. & Out-of-band biometrics authentication (OOB BA), attack-Surface Reduction by Design, soft computing approach, continuous authentication, attack trees, FPAS. & Smart cloths*, smart wrist-wear, smart jewelry, smart foot wear*, smart eye wear, sensor patches*, medical wearable devices. & \citep{jin2022p,balas2020handbook, lee2015risk, siboni2016advanced, lennon2015all, cusack2017assessment, giaretta2022security, adams2018ethics, karakaya2016secure, mahinderjit2020novel, kavianpour2021framework, bhatia2021soft, zhao2020trueheart, gulhane2019security, fang2019fine}\\
\midrule

Smart Retail& Cyber breach, personal behavior breach, financial information leaking. & ML, AI, and Blockchain. Fog computing, cloud computing, custom hardware solutions, edge computing. & Contactless POS system, mobile transactions, or smart shelves*, IoT-based fruits and vegetable sales system*. & \citep{WinNT26, wang2021iot, di2015towards, WinNT20, WinNT21, singh2024crumbled, cartwright2023economics, wang2021insecurity, yousefnezhad2020security, musa2017fog, brasilino2019mitigating, das2022iot, mohanta2020survey} \\
\midrule
Agriculture IoT & Sensor hacked, data theft, access control sniffing attacks and service interrupt attacks, MITM, SQL injection attacks, signature wrapping attacks, cloud malware injection attacks, flooding attacks, phishing attacks, access attacks, DDoS attacks, data transit attacks, and routing attacks. & Intrusion detection mechanism using ML, data aggregation scheme, SP2F, multi-party computation techniques, homomorphic encryption technique, FPDP PEFL. & Ag vehicles, concept cars*, drones, smart sprinklers, lights, coolers, heaters*, DroneSeed, SenseFly SoilCares, moisture sensor*, allMETEO, Smart Elements, Pycno, CropX, Mothive, FarmappGrowlink, GreenIQ, Arable, Semios, SCR and Cowlar, attitude model. & \citep{akhter2022precision,singh2020odysseys, namani2020smart, patil2012internet, mekala2017survey, krzywoszynskasoil, yu2021review, refaai2022application, 7123563, song2022development, salam2020internet, gupta2020security, mori2022iot, triantafyllou2019precision, ferrag2020security, glaroudis2020survey, farooq2019survey, madushanki2019adoption, ahmed2018internet, raghuvanshi2022intrusion, kumar2021sp2f, song2020fpdp, zhou2021efficient, chukkapalli2021privacy,  kumar2021pefl, anidu2021review, chaganti2022blockchain} \\
\midrule

Smart Grid & Data privacy issues, malware attacks issue, DDos attacks, standardization issues, transparency issues, false data injection attacks, unauthorized access, insider threats. & Strong authentication mechanisms, data encryption, intrusion detection systems, regular security audits, employee training, incident response plans, anomaly detection and attack recognition, DTMS, GAN, big data and AI technologies, firewalls, and load balancing mechanisms. & Smart switches and sensors*, smart metering systems, smart inverters, grid sensors*, smart thermostats, smart batteries. & \citep{al2019iot, pau2018cloud, ozgur2018iot, tran2022concept, boroojeni2017smart, stellios2018survey, gunduz2020cyber, huseinovic2020survey, gonen2020false, yilmaz2021timely, berghout2022machine, chen2021multi, unsal2021enhancing, desai2022mitigating, chehri2021security} \\

\bottomrule
\end{tabularx}
\end{adjustwidth}
\end{table}

\begin{table}[H]\ContinuedFloat
\small
\caption{{\em Cont.}}

\begin{adjustwidth}{-\extralength}{0cm}
		\begin{tabularx}{\fulllength}{CCCCC}
			\toprule
			\textbf{Category} & \textbf{Issues} & \textbf{Mitigations} & \textbf{Use Cases} & \textbf{Ref.} \\
			\midrule

Aviation IoT & Data security, malware, cyber attacks, inadequate user authentication, insider threats, vulnerabilities in airport software systems, insecure communication channels, unauthorized access to EFB servers, equipment malfunctions, and compromising the safety of passengers and crew. & AI, Blockchain, data encryption, regular security audits, employee training programs, feature engineering, ML, secure communication protocols, strong authentication, and access control measures. & Predictive maintenance, monitoring aircraft performance and optimization operations, smart baggage, smart airport, safety management*, flyers experience, and in-flight entertainment system. & \citep{bouyakoub2017smart, pech2021predictive, andel2021iot, wang2013aviation, okpala2020assessing, noel2021smart, samha2020applied, wang2022intelligent, jalali2018smart, aboti2020studies, cho2023reviewing, ukwandu2022cyber, florido2021identifying, adhikari2021intelligent, zkik2023secure, bitton2019machine} \\
			\bottomrule
		\end{tabularx}
	\end{adjustwidth}
\end{table}

\authorcontributions{Conceptualization, N.S. and H.K.; methodology, H.K.; software, N.S.; validation, N.S., R.B., and H.K.; formal analysis, N.S.; investigation, H.K. and R.B.; resources, N.S.; data curation, N.S.; writing---original draft preparation, N.S.; writing---review and editing, R.B. and H.K.; visualization, N.S.; supervision, R.B and H.K.; project administration, H.K.; funding acquisition, H.K. All authors have read and agreed to the published version of the manuscript.}

\funding{We gratefully acknowledge the support of the National Research
Foundation of Korea (NRF), funded by the Korean government (MSIT)
(RS-2024-00451909). Additionally, this work was supported by the
Institute of Information \& Communications Technology Planning \&
Evaluation (IITP) through a grant funded by the Korean government
(MSIT) (RS-2023-00229400, Development of user authentication and
privacy-preserving technology for a secure metaverse environment).}

\institutionalreview{This study did not collect any data.}

\informedconsent{Not applicable.}

\dataavailability{Data are contained within the article.}

 \conflictsofinterest{The authors declare no conflicts of interest.}
 \begin{adjustwidth}{-\extralength}{0cm}
 \reftitle{References}


\begin{thebibliography}{999}

\bibitem[Kumar et~al.(2019)Kumar, Tiwari, and Zymbler]{kumar2019internet}
Kumar, S.; Tiwari, P.; Zymbler, M.
\newblock Internet of Things is a revolutionary approach for future technology
  enhancement: A review.
\newblock {\em J. Big Data} {\bf 2019}, {\em 6},~1--21.

\bibitem[Kamal et~al.(2020)Kamal, Aljohani, and Alanazi]{kamal2020iot}
Kamal, M.; Aljohani, A.; Alanazi, E.
\newblock IoT meets COVID-19: Status, Challenges, and Opportunities.
\newblock {\em arXiv} {\bf 2020}, arXiv:2007.12268.

\bibitem[Tabrizi and Ibrahim(2017)]{tabrizi2017review}
Tabrizi, S.S.; Ibrahim, D.
\newblock A Review on Cloud Computing and Internet of Things.
\newblock {\em Int. J. Comput. Inf. Eng.}
  {\bf 2017}, {\em 11},~493--498.

\bibitem[Alrawi et~al.(2019)Alrawi, Lever, Antonakakis, and
  Monrose]{alrawi2019sok}
Alrawi, O.; Lever, C.; Antonakakis, M.; Monrose, F.
\newblock Sok: Security evaluation of home-based iot deployments.
\newblock In Proceedings of the 2019 IEEE Symposium on Security and Privacy, San Francisco, CA, USA, 13--23 May 
 2019;  pp. 1362--1380.

\bibitem[Apthorpe et~al.(2017)Apthorpe, Reisman, and
  Feamster]{apthorpe2017smart}
Apthorpe, N.; Reisman, D.; Feamster, N.
\newblock A smart home is no castle: Privacy vulnerabilities of encrypted iot
  traffic.
\newblock {\em arXiv} {\bf 2017}, arXiv:1705.06805.

\bibitem[Bradley et~al.(2018)Bradley, El-Tawab, and
  Heydari]{bradley2018security}
Bradley, C.; El-Tawab, S.; Heydari, M.H.
\newblock Security analysis of an IoT system used for indoor localization in
  healthcare facilities.
\newblock In Proceedings of the 2018 Systems and Information Engineering Design
  Symposium (SIEDS), Charlottesville, VA, USA, 27 April 2018;  pp. 147--152.

\bibitem[Celik et~al.(2018)Celik, McDaniel, and Tan]{celik2018soteria}
Celik, Z.B.; McDaniel, P.; Tan, G.
\newblock Soteria: Automated $\{$IoT$\}$ Safety and Security Analysis.
\newblock In Proceedings of the 2018 USENIX Annual Technical Conference (USENIX
  ATC 18), Boston, MA, USA, 11--13 July 2018;  pp. 147--158.

\bibitem[Chu et~al.(2018)Chu, Apthorpe, and Feamster]{chu2018security}
Chu, G.; Apthorpe, N.; Feamster, N.
\newblock Security and privacy analyses of internet of things children’s
  toys.
\newblock {\em IEEE Internet Things J.} {\bf 2018}, {\em 6},~978--985.

\bibitem[Denning et~al.(2013)Denning, Kohno, and Levy]{denning2013computer}
Denning, T.; Kohno, T.; Levy, H.M.
\newblock Computer security and the modern home.
\newblock {\em Commun. ACM} {\bf 2013}, {\em 56},~94--103.

\bibitem[Ding and Hu(2018)]{ding2018safety}
Ding, W.; Hu, H.
\newblock On the safety of iot device physical interaction control.
\newblock In Proceedings of the 2018 ACM SIGSAC Conference
  on Computer and Communications Security, Toronto, Canada, 15--19 October 2018;  pp. 832--846.

\bibitem[Jin et~al.(2022)Jin, Xing, Fang, Jia, Yuan, and Liu]{jin2022p}
Jin, Z.; Xing, L.; Fang, Y.; Jia, Y.; Yuan, B.; Liu, Q.
\newblock P-Verifier: Understanding and Mitigating Security Risks in
  Cloud-based IoT Access Policies. 
\newblock In Proceedings of the 2022 ACM SIGSAC Conference
  on Computer and Communications Security, Los Angeles, CA, USA, 7--11 November 2022;  pp. 1647--1661.

\bibitem[Samad et~al.(2018)Samad, Alam, Mohammed, and
  Bhukhari]{samad2018internet}
Samad, A.; Alam, S.; Mohammed, S.; Bhukhari, M.
\newblock Internet of vehicles (IoV) requirements, attacks and countermeasures.
\newblock In Proceedings of the 12th INDIACom; INDIACom-2018;
  5th International Conference on “Computing for Sustainable Global
  Development” IEEE Conference, New Delhi, India, 14--16 March 2018;  pp. 1--4.

\bibitem[Michelena et~al.(2024)Michelena, Aveleira-Mata, Jove,
  Bay{\'o}n-Guti{\'e}rrez, Novais, Romero, Calvo-Rolle, and
  Al{\'a}iz-Moret{\'o}n]{michelena2024novel}
Michelena, A.; Aveleira-Mata, J.; Jove, E.; Bay{\'o}n-Guti{\'e}rrez, M.;
  Novais, P.; Romero, O.F.; Calvo-Rolle, J.L.; Al{\'a}iz-Moret{\'o}n, H.
\newblock A novel intelligent approach for man-in-the-middle attacks detection
  over internet of things environments based on message queuing telemetry
  transport.
\newblock {\em Expert Syst.} {\bf 2024}, {\em 41},~e13263.

\bibitem[Sivasankari and Kamalakkannan(2022)]{sivasankari2022detection}
Sivasankari, N.; Kamalakkannan, S.
\newblock Detection and prevention of man-in-the-middle attack in iot network
  using regression modeling.
\newblock {\em Adv. Eng. Softw.} {\bf 2022}, {\em 169},~103126.

\bibitem[Chaudhary and Mishra(2023)]{chaudhary2023ddos}
Chaudhary, S.; Mishra, P.K.
\newblock DDoS attacks in Industrial IoT: A survey.
\newblock {\em Comput. Netw.} {\bf 2023}, \emph{236}, 110015.

\bibitem[Yang et~al.(2022)Yang, Wang, Yin, Wang, and Hu]{yang2022review}
Yang, W.; Wang, S.; Yin, X.; Wang, X.; Hu, J.
\newblock A review on security issues and solutions of the Internet of Drones.
\newblock {\em IEEE Open J. Comput. Soc.} {\bf 2022}, \emph{3}, 96--110.

\bibitem[Surya(2016)]{surya2016security}
Surya, L.
\newblock Security challenges and strategies for the IoT in cloud computing.
\newblock {\em Int. J. Innov. Eng. Res. Technol. ISSN} {\bf 2016};  \emph{3}, 2394--3696.

\bibitem[Georgakopoulos et~al.(2016)Georgakopoulos, Jayaraman, Fazia, Villari,
  and Ranjan]{georgakopoulos2016internet}
Georgakopoulos, D.; Jayaraman, P.P.; Fazia, M.; Villari, M.; Ranjan, R.
\newblock Internet of Things and edge cloud computing roadmap for
  manufacturing.
\newblock {\em IEEE Cloud Comput.} {\bf 2016}, {\em 3},~66--73.

\bibitem[Zhou et~al.(2013)Zhou, Leppanen, Harjula, Ylianttila, Ojala, Yu, Jin,
  and Yang]{zhou2013cloudthings}
Zhou, J.; Leppanen, T.; Harjula, E.; Ylianttila, M.; Ojala, T.; Yu, C.; Jin,
  H.; Yang, L.T.
\newblock Cloudthings: A common architecture for integrating the internet of
  things with cloud computing.
\newblock In Proceedings of the 2013 IEEE 17th International
  Conference on Computer Supported Cooperative Work in Design (CSCWD), Whistler, BC, Canada, 27--29 June
  2013;  pp. 651--657.

\bibitem[Shafique et~al.(2020)Shafique, Khawaja, Sabir, Qazi, and
  Mustaqim]{shafique2020internet}
Shafique, K.; Khawaja, B.A.; Sabir, F.; Qazi, S.; Mustaqim, M.
\newblock Internet of things (IoT) for next-generation smart systems: A review
  of current challenges, future trends and prospects for emerging 5G-IoT
  scenarios.
\newblock {\em IEEE Access} {\bf 2020}, {\em 8},~23022--23040.

\bibitem[Atlam et~al.(2017)Atlam, Alenezi, Alharthi, Walters, and
  Wills]{atlam2017integration}
Atlam, H.F.; Alenezi, A.; Alharthi, A.; Walters, R.J.; Wills, G.B.
\newblock Integration of cloud computing with internet of things: Challenges
  and open issues.
\newblock In Proceedings of the 2017 IEEE International Conference on Internet
  of Things (iThings) and IEEE Green Computing and Communications (GreenCom)
  and IEEE Cyber, Physical and Social Computing (CPSCom) and IEEE Smart Data
  (SmartData), Exeter, UK, 21--23 June 2017;  pp. 670--675.

\bibitem[Cai et~al.(2016)Cai, Xu, Jiang, and Vasilakos]{cai2016iot}
Cai, H.; Xu, B.; Jiang, L.; Vasilakos, A.V.
\newblock IoT-based big data storage systems in cloud computing: Perspectives
  and challenges.
\newblock {\em IEEE Internet Things J.} {\bf 2016}, {\em 4},~75--87.

\bibitem[Thakor et~al.(2021)Thakor, Razzaque, and
  Khandaker]{thakor2021lightweight}
Thakor, V.A.; Razzaque, M.A.; Khandaker, M.R.
\newblock Lightweight cryptography algorithms for resource-constrained IoT
  devices: A review, comparison and research opportunities.
\newblock {\em IEEE Access} {\bf 2021}, {\em 9},~28177--28193.

\bibitem[Chen et~al.(2021)Chen, Luo, Xiang, Chen, Fan, and
  Truong]{10.1145/3447625}
Chen, F.; Luo, D.; Xiang, T.; Chen, P.; Fan, J.; Truong, H.L.
\newblock IoT Cloud Security Review: A Case Study Approach Using Emerging
  Consumer-Oriented Applications.
\newblock {\em ACM Comput. Surv.} {\bf 2021}, {\em 54}.
\newblock {\url{https://doi.org/10.1145/3447625}}.

\bibitem[Alaba et~al.(2017)Alaba, Othman, Hashem, and
  Alotaibi]{alaba2017internet}
Alaba, F.A.; Othman, M.; Hashem, I.A.T.; Alotaibi, F.
\newblock Internet of Things security: A survey.
\newblock {\em D} {\bf 2017}, {\em
  88},~10--28.

\bibitem[Harbi et~al.(2019)Harbi, Aliouat, Harous, Bentaleb, and
  Refoufi]{harbi2019review}
Harbi, Y.; Aliouat, Z.; Harous, S.; Bentaleb, A.; Refoufi, A.
\newblock A review of security in internet of things.
\newblock {\em Wirel. Pers. Commun.} {\bf 2019}, {\em
  108},~325--344.

\bibitem[Yousefnezhad et~al.(2020)Yousefnezhad, Malhi, and
  Fr{\"a}mling]{yousefnezhad2020security}
Yousefnezhad, N.; Malhi, A.; Fr{\"a}mling, K.
\newblock Security in product lifecycle of IoT devices: A survey.
\newblock {\em J. Netw. Comput. Appl.} {\bf 2020}, {\em
  171},~102779.

\bibitem[Almolhis et~al.(2020)Almolhis, Alashjaee, Duraibi, Alqahtani, and
  Moussa]{almolhis2020security}
Almolhis, N.; Alashjaee, A.M.; Duraibi, S.; Alqahtani, F.; Moussa, A.N.
\newblock The security issues in IoT-cloud: A review.
\newblock In Proceedings of the 2020 16th IEEE International Colloquium on
  Signal Processing \& Its Applications (CSPA), Langkawi, Malaysia, 28--29 February 2020;  pp. 191--196.

\bibitem[Lit et~al.(2021)Lit, Kim, and Sy]{lit2021survey}
Lit, Y.; Kim, S.; Sy, E.
\newblock A survey on amazon alexa attack surfaces.
\newblock In Proceedings of the 2021 IEEE 18th Annual Consumer Communications
  \& Networking Conference (CCNC), Las Vegas, NV, USA, 9--12 January 2021;  pp. 1--7.

\bibitem[Ling et~al.(2022)Ling, Yelland, Hatzigianni, and
  Dickson-Deane]{ling2022use}
Ling, L.; Yelland, N.; Hatzigianni, M.; Dickson-Deane, C.
\newblock The use of Internet of Things devices in early childhood education: A
  systematic review. 
\newblock {\em Educ. Inf. Technol.} {\bf 2022}, \emph{27}, 6333--6352.

\bibitem[Shah et~al.(2022)Shah, Sheth, and Doshi]{shah2022survey}
Shah, K.; Sheth, C.; Doshi, N.
\newblock A survey on iot-based smart cars, their functionalities and
  challenges.
\newblock {\em Procedia Comput. Sci.} {\bf 2022}, {\em 210},~295--300.

\bibitem[Sicari et~al.(2019)Sicari, Rizzardi, and
  Coen-Porisini]{sicari2019evaluate}
Sicari, S.; Rizzardi, A.; Coen-Porisini, A.
\newblock How to evaluate an Internet of Things system: Models, case studies,
  and real developments.
\newblock {\em Softw. Pract. Exp.} {\bf 2019}, {\em
  49},~1663--1685.

\bibitem[Sivaraman et~al.(2016)Sivaraman, Chan, Earl, and
  Boreli]{sivaraman2016smart}
Sivaraman, V.; Chan, D.; Earl, D.; Boreli, R.
\newblock Smart-phones attacking smart-homes.
\newblock In Proceedings of the 9th ACM Conference on
  Security \& Privacy in Wireless and Mobile Networks, Darmstadt, Germany, 18--2 July 2016;  pp. 195--200.

\bibitem[Stoyanova et~al.(2020)Stoyanova, Nikoloudakis, Panagiotakis, Pallis,
  and Markakis]{stoyanova2020survey}
Stoyanova, M.; Nikoloudakis, Y.; Panagiotakis, S.; Pallis, E.; Markakis, E.K.
\newblock A survey on the internet of things (IoT) forensics: Challenges,
  approaches, and open issues.
\newblock {\em IEEE Commun. Surv. Tutor.} {\bf 2020}, {\em
  22},~1191--1221.

\bibitem[Domingo-Ferrer et~al.(2019)Domingo-Ferrer, Farras, Ribes-Gonz{\'a}lez,
  and S{\'a}nchez]{domingo2019privacy}
Domingo-Ferrer, J.; Farras, O.; Ribes-Gonz{\'a}lez, J.; S{\'a}nchez, D.
\newblock Privacy-preserving cloud computing on sensitive data: A survey of
  methods, products and challenges.
\newblock {\em Comput. Commun.} {\bf 2019}, {\em 140},~38--60.

\bibitem[Ahmed et~al.(2022)Ahmed, Khan, Sheta, Tarek, Zualkernan, and
  Aloul]{ahmed2022detecting}
Ahmed, N.; Khan, J.; Sheta, N.; Tarek, R.; Zualkernan, I.; Aloul, F.
\newblock Detecting Replay Attack on Voice-Controlled Systems using Small
  Neural Networks.
\newblock In Proceedings of the 2022 IEEE 7th Forum on Research and
  Technologies for Society and Industry Innovation (RTSI), Paris, France, 24--26 August 2022;  pp.
  50--54.

\bibitem[Obaid(2021)]{obaid2021assessment}
Obaid, A.J.
\newblock Assessment of smart home assistants as an IoT.
\newblock {\em Int. J. Comput. Inf. Manuf. (IJCIM)} {\bf 2021}, {\em 1}. 

\bibitem[Zhang et~al.(2019)Zhang, Mi, Feng, Wang, Tian, and
  Qian]{zhang2019dangerous}
Zhang, N.; Mi, X.; Feng, X.; Wang, X.; Tian, Y.; Qian, F.
\newblock Dangerous skills: Understanding and mitigating security risks of
  voice-controlled third-party functions on virtual personal assistant systems.
\newblock In Proceedings of the 2019 IEEE Symposium on Security and Privacy
  (SP), San Francisco, CA, USA, 19--23 May 2019;  pp. 1381--1396.

\bibitem[Kumar et~al.(2018)Kumar, Paccagnella, Murley, Hennenfent, Mason,
  Bates, and Bailey]{kumar2018skill}
Kumar, D.; Paccagnella, R.; Murley, P.; Hennenfent, E.; Mason, J.; Bates, A.;
  Bailey, M.
\newblock Skill squatting attacks on Amazon Alexa.
\newblock In Proceedings of the 27th USENIX Security Symposium (USENIX Security
  18), Baltimore, MD, USA, 15--17 August 2018;  pp. 33--47.

\bibitem[Wollerton(2024)]{WinNT1}
\textls[-15]{Wollerton, M.
\newblock Samsung SmartThings Still Hasn’t Earned My Trust in the Smart Home.
\newblock {\em CNET} {\bf 2024}.
\newblock Available online:
  \url{https://www.cnet.com/home/smart-home/samsung-smartthings-still-hasnt-earned-my-trust-in-the-smart-home/} (accessed on 06-06-2024}). 

\bibitem[Fernandes et~al.(2016)Fernandes, Jung, and
  Prakash]{fernandes2016security}
Fernandes, E.; Jung, J.; Prakash, A.
\newblock Security analysis of emerging smart home applications.
\newblock In Proceedings of the 2016 IEEE symposium on security and privacy
  (SP), San Jose, CA, USA, 22--26 May 2016;  pp. 636--654.

\bibitem[Fernandes et~al.(2017)Fernandes, Rahmati, Jung, and
  Prakash]{fernandes2017security}
Fernandes, E.; Rahmati, A.; Jung, J.; Prakash, A.
\newblock Security implications of permission models in smart-home application
  frameworks.
\newblock {\em IEEE Secur. Priv.} {\bf 2017}, {\em 15},~24--30.

\bibitem[Zhang et~al.(2018)Zhang, Meng, Liu, Zhang, Zhang, and
  Zhu]{zhang2018homonit}
Zhang, W.; Meng, Y.; Liu, Y.; Zhang, X.; Zhang, Y.; Zhu, H.
\newblock Homonit: Monitoring smart home apps from encrypted traffic.
\newblock In Proceedings of the 2018 ACM SIGSAC Conference
  on Computer and Communications Security, Toronto, Canada, 15--19 October 2018;  pp. 1074--1088.

\bibitem[Otoom et~al.(2023)Otoom, Abdallah, et~al.]{otoom2023deep}
Otoom, A.F.; Eleisah, W.; Abdallah, E.E.
\newblock Deep learning for accurate detection of brute force attacks on IOT
  Networks.
\newblock {\em Procedia Comput. Sci.} {\bf 2023}, {\em 220},~291--298.

\bibitem[Attoh and Signer(2024)]{attoh2024towards}
Attoh, E.; Signer, B.
\newblock Towards a Write Once Run Anywhere Approach in End-User IoT
  Development.
\newblock In Proceedings of the 9th International Conference on Internet of
  Things, Bilbao, Spain, 22--25 October 2024.

\bibitem[Sanl{\i}(2024)]{sanli2024detection}
Sanl{\i}, M.
\newblock Detection and Mitigation of Denial of Service Attacks in Internet of
  Things Networks.
\newblock {\em Arab. J. Sci. Eng.} {\bf 2024}, \emph{49}, 12629--12639.

\bibitem[Cloudflare(2024)]{WinNT29}
Cloudflare.
\newblock Inside the Infamous Mirai IoT Botnet: A Retrospective Analysis.
\newblock  2024.
\newblock Available online:
  \url{https://blog.cloudflare.com/inside-mirai-the-infamous-iot-botnet-a-retrospective-analysis/} (accessed on 13-05-2024).

\bibitem[Aziz Al~Kabir et~al.(2023)Aziz Al~Kabir, Elmedany, and
  Sharif]{aziz2023securing}
Aziz Al~Kabir, M.; Elmedany, W.; Sharif, M.S.
\newblock Securing IoT devices against emerging security threats: Challenges
  and mitigation techniques.
\newblock {\em J. Cyber Secur. Technol.} {\bf 2023}, {\em
  7},~199--223.

\bibitem[Navas et~al.(2018)Navas, Le~Bouder, Cuppens, Cuppens, and
  Papadopoulos]{navas2018not}
Navas, R.E.; Le~Bouder, H.; Cuppens, N.; Cuppens, F.; Papadopoulos, G.Z.
\newblock Do not trust your neighbors! A small IoT platform illustrating a
  man-in-the-middle attack.
\newblock In Proceedings of the Ad-Hoc, Mobile, and Wireless Networks: 17th
  International Conference on Ad Hoc Networks and Wireless, ADHOC-NOW 2018,
  Saint-Malo, France, 5--7 September 2018; Springer: Cham, Switzerland, 2018; pp.
  120--125.

\bibitem[Sharma et~al.(2022)Sharma, Dyrkolbotn, {\O}verlier, Waltoft-Olsen,
  Franke, and Katsikas]{sharma2022state}
Sharma, A.; Dyrkolbotn, G.O.; {\O}verlier, L.; Waltoft-Olsen, A.J.; Franke, K.;
  Katsikas, S.
\newblock A state-of-the-art reverse engineering approach for combating
  hardware security vulnerabilities at the system and pcb level in iot devices.
\newblock In Proceedings of the 2022 IEEE Physical Assurance and Inspection of
  Electronics (PAINE), Huntsville, AL, USA, 25--27 October 2022;  pp. 1--7.

\bibitem[Park et~al.(2019)Park, Oh, and Lee]{park2019security}
Park, M.; Oh, H.; Lee, K.
\newblock Security risk measurement for information leakage in IoT-based smart
  homes from a situational awareness perspective.
\newblock {\em Sensors} {\bf 2019}, {\em 19},~2148.

\bibitem[Alladi et~al.(2020)Alladi, Chamola, Sikdar, and
  Choo]{alladi2020consumer}
Alladi, T.; Chamola, V.; Sikdar, B.; Choo, K.K.R.
\newblock Consumer IoT: Security vulnerability case studies and solutions.
\newblock {\em IEEE Consum. Electron. Mag.} {\bf 2020}, {\em 9},~17--25.

\bibitem[Obermaier and Hutle(2016)]{obermaier2016analyzing}
Obermaier, J.; Hutle, M.
\newblock Analyzing the security and privacy of cloud-based video surveillance
  systems.
\newblock In Proceedings of the 2nd ACM International
  Workshop on IoT Privacy, Trust, and Security, Xi'an, China, 30 May 2016;  pp. 22--28.

\bibitem[Seralathan et~al.(2018)Seralathan, Oh, Jadhav, Myers, Jeong, Kim, and
  Kim]{seralathan2018iot}
Seralathan, Y.; Oh, T.T.; Jadhav, S.; Myers, J.; Jeong, J.P.; Kim, Y.H.; Kim,
  J.N.
\newblock IoT security vulnerability: A case study of a Web camera.
\newblock In Proceedings of the 2018 20th International Conference on Advanced
  Communication Technology (ICACT), Chuncheon, South Korea, 11--14 February 2018;  pp. 172--177.

\bibitem[Tekeoglu and Tosun(2015)]{tekeoglu2015investigating}
Tekeoglu, A.; Tosun, A.S.
\newblock Investigating security and privacy of a cloud-based wireless IP
  camera: NetCam.
\newblock In Proceedings of the 2015 24th International Conference on Computer
  Communication and Networks (ICCCN), Las Vegas, NV, USA, 3--6 August 2015;  pp. 1--6.

\bibitem[Xu et~al.(2018)Xu, Xu, and Chen]{xu2018internet}
Xu, H.; Xu, F.; Chen, B.
\newblock Internet protocol cameras with no password protection: An empirical
  investigation.
\newblock In Proceedings of the International Conference on Passive and Active
  Network Measurement, Berlin, Germany, 26--27 March 2018; Springer: Cham, Switzerland, 2018;  pp. 47--59.

\bibitem[Mangala et~al.(2021)Mangala, Venugopal, et~al.]{mangala2021short}
Mangala, N.; Venugopal, K.;  Eswara Reddy, B.
\newblock Current Challenges in IoT Cloud Smart Applications.
\newblock In Proceedings of the 2021 IEEE International Conference on Cloud
  Computing in Emerging Markets (CCEM), New Jersey, USA, 27--30 October 
 2021;  pp. 36--40.

\bibitem[Geeng and Roesner(2019)]{geeng2019s}
Geeng, C.; Roesner, F.
\newblock Who's in control? Interactions in multi-user smart homes.
\newblock In Proceedings of the 2019 CHI Conference on Human
  Factors in Computing Systems,  2019;  pp. 1--13.

\bibitem[Jang et~al.(2017)Jang, Chhabra, and Prasad]{jang2017enabling}
Jang, W.; Chhabra, A.; Prasad, A.
\newblock Enabling multi-user controls in smart home devices.
\newblock In Proceedings of the 2017 Workshop on Internet of
  Things Security and Privacy, Glasgow, UK, 4--9 May 2017;  pp. 49--54.

\bibitem[Zeng and Roesner(2019)]{zeng2019understanding}
Zeng, E.; Roesner, F.
\newblock Understanding and Improving Security and Privacy in
  $\{$Multi-User$\}$ Smart Homes: A Design Exploration and $\{$In-Home$\}$ User
  Study.
\newblock In Proceedings of the 28th USENIX Security Symposium (USENIX Security
  19), Santa Clara, CA, USA, 14--16 August 2019;  pp. 159--176.

\bibitem[Davis et~al.(2020)Davis, Mason, and Anwar]{davis2020vulnerability}
Davis, B.D.; Mason, J.C.; Anwar, M.
\newblock Vulnerability studies and security postures of IoT devices: A smart
  home case study.
\newblock {\em IEEE Internet Things J.} {\bf 2020}, {\em
  7},~10102--10110.

\bibitem[Notra et~al.(2014)Notra, Siddiqi, Gharakheili, Sivaraman, and
  Boreli]{notra2014experimental}
Notra, S.; Siddiqi, M.; Gharakheili, H.H.; Sivaraman, V.; Boreli, R.
\newblock An experimental study of security and privacy risks with emerging
  household appliances.
\newblock In Proceedings of the 2014 IEEE Conference on Communications and
  Network Security, San Francisco, CA, USA, 29--31 October 2014;  pp. 79--84.

\bibitem[Ren et~al.(2023)Ren, Peng, Li, Xue, Lan, and Yang]{ren2023voice}
Ren, Y.; Peng, H.; Li, L.; Xue, X.; Lan, Y.; Yang, Y.
\newblock A voice spoofing detection framework for IoT systems with feature
  pyramid and online knowledge distillation.
\newblock {\em J. Syst. Archit.} {\bf 2023}, {\em 143},~102981.

\bibitem[Hooda et~al.(2022)Hooda, Wallace, Jhunjhunwalla, Fernandes, and
  Fawaz]{hooda2022skillfence}
Hooda, A.; Wallace, M.; Jhunjhunwalla, K.; Fernandes, E.; Fawaz, K.
\newblock SkillFence: A Systems Approach to Practically Mitigating Voice-Based
  Confusion Attacks.
\newblock {\em Proc. Acm Interact. Mob. Wearable Ubiquitous Technol.} {\bf 2022}, {\em 6},~1--26.

\bibitem[Chatterjee et~al.(2017)Chatterjee, Chakraborty, and
  Mukhopadhyay]{chatterjee2017puf}
Chatterjee, U.; Chakraborty, R.S.; Mukhopadhyay, D.
\newblock A PUF-based secure communication protocol for IoT.
\newblock {\em ACM Trans. Embed. Comput. Syst. (TECS)} {\bf
  2017}, {\em 16},~1--25.

\bibitem[Nazzal et~al.(2022)Nazzal, Zaid, Alalfi, and
  Valani]{nazzal2022vulnerability}
Nazzal, B.; Zaid, A.A.; Alalfi, M.H.; Valani, A.
\newblock Vulnerability classification of consumer-based IoT software.
\newblock In Proceedings of the 4th International Workshop
  on Software Engineering Research and Practice for the IoT, Pittsburgh, PA, USA, 19 May 2022;  pp. 17--24.

\bibitem[Shameli-Sendi et~al.(2015)Shameli-Sendi, Pourzandi, Fekih-Ahmed, and
  Cheriet]{shameli2015taxonomy}
Shameli-Sendi, A.; Pourzandi, M.; Fekih-Ahmed, M.; Cheriet, M.
\newblock Taxonomy of distributed denial of service mitigation approaches for
  cloud computing.
\newblock {\em J. Netw. Comput. Appl.} {\bf 2015}, {\em
  58},~165--179.

\bibitem[Salva-Garcia et~al.(2018)Salva-Garcia, Alcaraz-Calero, Wang, Bernabe,
  and Skarmeta]{salva20185g}
Salva-Garcia, P.; Alcaraz-Calero, J.M.; Wang, Q.; Bernabe, J.B.; Skarmeta, A.
\newblock 5G NB-IoT: Efficient network traffic filtering for multitenant IoT
  cellular networks.
\newblock {\em Secur. Commun. Netw.} {\bf 2018}, {\em
  2018},~1--21.

\bibitem[Guozi et~al.(2018)Guozi, Jiang, Yu, Danni, and Huakang]{guozi2018ddos}
Guozi, S.; Jiang, W.; Yu, G.; Danni, R.; Huakang, L.
\newblock DDoS attacks and flash event detection based on flow characteristics
  in SDN.
\newblock In Proceedings of the 2018 15th IEEE International Conference on
  Advanced Video and Signal Based Surveillance (AVSS), Auckland, New Zealand, 27--30 November 2018;  pp. 1--6.

\bibitem[Yan et~al.(2018)Yan, Huang, Luo, Gong, and Yu]{yan2018multi}
Yan, Q.; Huang, W.; Luo, X.; Gong, Q.; Yu, F.R.
\newblock A multi-level DDoS mitigation framework for the industrial Internet
  of Things.
\newblock {\em IEEE Commun. Mag.} {\bf 2018}, {\em 56},~30--36.

\bibitem[Ma et~al.(2015)Ma, Xu, and Lin]{ma2015defending}
Ma, D.; Xu, Z.; Lin, D.
\newblock Defending blind DDoS attack on SDN based on moving target defense.
\newblock In Proceedings of the International Conference on Security and
  Privacy in Communication Networks: 10th International ICST Conference,
  SecureComm 2014, Beijing, China, 24--26 September 2014; Springer: Cham, Swizterland, 2015;  pp. 463--480.

\bibitem[Connell et~al.(2017)Connell, Menasc{\'e}, and
  Albanese]{connell2017performance}
Connell, W.; Menasc{\'e}, D.A.; Albanese, M.
\newblock Performance modeling of moving target defenses.
\newblock In Proceedings of the 2017 Workshop on Moving
  Target Defense, Dallas, TX, USA, 30 October 2017;  pp. 53--63.

\bibitem[Chen et~al.(2020)Chen, Xiao, Liu, Jiang, and Tang]{chen2020ddos}
Chen, W.; Xiao, S.; Liu, L.; Jiang, X.; Tang, Z.
\newblock A DDoS attacks traceback scheme for SDN-based smart city.
\newblock {\em Comput. Electr. Eng.} {\bf 2020}, {\em
  81},~106503.

\bibitem[Bhattacharjya et~al.(2019)Bhattacharjya, Zhong, Wang, and
  Li]{bhattacharjya2019secure}
Bhattacharjya, A.; Zhong, X.; Wang, J.; Li, X.
\newblock Secure IoT structural design for smart homes. In {\em Smart Cities
  Cybersecurity and Privacy}; Elsevier:  Amsterdam, The Netherlands, 2019; pp. 187--201.

\bibitem[Abu-Tair et~al.(2020)Abu-Tair, Djahel, Perry, Scotney, Zia, Carracedo,
  and Sajjad]{abu2020towards}
Abu-Tair, M.; Djahel, S.; Perry, P.; Scotney, B.; Zia, U.; Carracedo, J.M.;
  Sajjad, A.
\newblock Towards secure and privacy-preserving IoT enabled smart home:
  architecture and experimental study.
\newblock {\em Sensors} {\bf 2020}, {\em 20},~6131.

\bibitem[Qasem et~al.(2024)Qasem, Thabit, Can, Naji, Alkhzaimi, Patil, and
  Thorat]{qasem2024cryptography}
Qasem, M.A.; Thabit, F.; Can, O.; Naji, E.; Alkhzaimi, H.A.; Patil, P.R.;
  Thorat, S.
\newblock Cryptography algorithms for improving the security of cloud-based
  internet of things.
\newblock {\em Secur. Priv.} {\bf 2024}, \emph{7}, e378.

\bibitem[Gerber et~al.(2021)Gerber, Heidinger, Stiegelmayer, and
  Gerber]{gerber2021loki}
Gerber, P.; Heidinger, M.; Stiegelmayer, J.; Gerber, N.
\newblock LOKI: Development of an interface for task-based, privacy-friendly
  smart home control through LOCal Information Processing: LOKI: Entwicklung
  eines Interfaces f{\"u}r die Aufgaben-basierte, Privatsph{\"a}re-freundliche
  Smart Home-Steuerung durch LOKale Informationsverarbeitung. In  Proceedings of Mensch und Computer 2021,  	Ingolstadt, Germany, 5--8 September 2021; pp. 578--581.

\bibitem[Protection(2023)]{WinNT24}
Child Protection Services. 
\newblock Children's Online Privacy Protection Rule (``COPPA'').
\newblock Federal Trade Commission. 2023.
\newblock Available online:
  \url{https://www.ftc.gov/legal-library/browse/rules/childrens-online-privacy-protection-rule-coppa} (accessed on 17-07-2023).

\bibitem[Holloway and Green(2016)]{holloway2016internet}
Holloway, D.; Green, L.
\newblock The internet of toys.
\newblock {\em Commun. Res. Pract.} {\bf 2016}, {\em
  2},~506--519.

\bibitem[Ataei~Nezhad et~al.(2022)Ataei~Nezhad, Barati, and
  Barati]{ataei2022authentication}
Ataei~Nezhad, M.; Barati, H.; Barati, A.
\newblock An authentication-based secure data aggregation method in Internet of
  Things.
\newblock {\em J. Grid Comput.} {\bf 2022}, {\em 20},~29.

\bibitem[Hung et~al.(2016)Hung, Iqbal, Huang, Melaisi, and
  Pang]{hung2016glance}
Hung, P.C.; Iqbal, F.; Huang, S.C.; Melaisi, M.; Pang, K.
\newblock A glance of child’s play privacy in smart toys.
\newblock In Proceedings of the International Conference on Cloud Computing and
  Security, Nanjing, China, 29--31 July 2016; Springer: Cham, Switzerland, 2016; pp. 217--231.

\bibitem[Valente and Cardenas(2017)]{valente2017security}
Valente, J.; Cardenas, A.A.
\newblock Security \& privacy in smart toys.
\newblock In Proceedings of the 2017 Workshop on Internet of
  Things Security and Privacy, Dallas, TX, USA, 3 November 2017;  pp. 19--24.

\bibitem[Albrecht and Mcintyre(2015)]{albrecht2015privacy}
Albrecht, K.; Mcintyre, L.
\newblock Privacy nightmare: When baby monitors go bad [opinion].
\newblock {\em IEEE Technol. Soc. Mag.} {\bf 2015}, {\em
  34},~14--19.

\bibitem[Stanislav and Beardsley(2015)]{stanislav2015hacking}
Stanislav, M.; Beardsley, T.
\newblock \emph{Hacking Iot: A Case Study on Baby Monitor Exposures and
  Vulnerabilities};
\newblock Rapid7 Report; Rapid7: Boston, MA, USA, 2015.

\bibitem[Rivera et~al.(2019)Rivera, Garc{\'\i}a, Mart{\'\i}n-Ruiz, Alarcos,
  Velasco, and Oliva]{rivera2019secure}
Rivera, D.; Garc{\'\i}a, A.; Mart{\'\i}n-Ruiz, M.L.; Alarcos, B.; Velasco,
  J.R.; Oliva, A.G.
\newblock Secure communications and protected data for a Internet of Things
  smart toy platform.
\newblock {\em IEEE Internet Things J.} {\bf 2019}, {\em
  6},~3785--3795.

\bibitem[Rafferty et~al.(2017)Rafferty, Hung, Fantinato, Peres, Iqbal, Kuo, and
  Huang]{rafferty2017towards}
Rafferty, L.; Hung, P.C.; Fantinato, M.; Peres, S.M.; Iqbal, F.; Kuo, S.Y.;
  Huang, S.C.
\newblock Towards a privacy rule conceptual model for smart toys. In
\newblock {\em Computing in Smart Toys}; Springer: Cham, Switzerland, 2017;  pp. 85--102.

\bibitem[Yankson et~al.(2020)Yankson, Iqbal, and Hung]{yankson20204p}
Yankson, B.; Iqbal, F.; Hung, P.C.
\newblock 4P-based forensics investigation framework for smart connected toys.
\newblock In Proceedings of the 15th International
  Conference on Availability, Reliability and Security, Virtual, 25--28 August 2020;  pp. 1--9.

\bibitem[Thales(2022)]{WinNT10}
Thales.
\newblock IoT in Healthcare.
\newblock Thales Group. 2022.
\newblock Available online:
  \url{https://www.thalesgroup.com/en/markets/digital-identity-and-security/iot/industries/healthcare} (accessed on 13-05-2024).

\bibitem[Ganesh et~al.(2022)Ganesh, Jeyanth, and Bevi]{ganesh2022iot}
Ganesh, K.V.S.S.; Jeyanth, S.S.; Bevi, A.R.
\newblock IOT based portable heart rate and SpO2 pulse oximeter.
\newblock {\em HardwareX} {\bf 2022}, {\em 11},~e00309.

\bibitem[Okafor et~al.(2022)Okafor, Arukalam, Ekuma, Eziefuna, Ihetu,
  Okey-Mbata, Ezeamaku, Iheaturu, and Okafor]{okafor2022design}
Okafor, S.A.; Arukalam, F.M.; Ekuma, I.C.; Eziefuna, E.O.; Ihetu, C.A.;
  Okey-Mbata, C.C.; Ezeamaku, U.L.; Iheaturu, N.C.; Okafor, A.L.
\newblock Design and Development of an Internet of things Based Glucometer with
  Wireless Transmission.
\newblock {\em J. Eng. Res. Rep.} {\bf 2022}, {\em
  22},~36--46.

\bibitem[Intellectsoft(2024)]{WinNT11}
Intellectsoft.
\newblock Biggest IoT Security Issues.
\newblock {\em Intellectsoft} {\bf 2024}.
\newblock Available online:
  \url{https://www.intellectsoft.net/blog/biggest-iot-security-issues/} (accessed on 11-05-2024).

\bibitem[Farahani et~al.(2018)Farahani, Firouzi, Chang, Badaroglu, Constant,
  and Mankodiya]{farahani2018towards}
Farahani, B.; Firouzi, F.; Chang, V.; Badaroglu, M.; Constant, N.; Mankodiya,
  K.
\newblock Towards fog-driven IoT eHealth: Promises and challenges of IoT in
  medicine and healthcare.
\newblock {\em Future Gener. Comput. Syst.} {\bf 2018}, {\em
  78},~659--676.

\bibitem[McAllister et~al.(2017)McAllister, El-Tawab, and
  Heydari]{mcallister2017localization}
McAllister, T.D.; El-Tawab, S.; Heydari, M.H.
\newblock Localization of health center assets through an iot environment
  (locate).
\newblock In Proceedings of the 2017 Systems and Information Engineering Design
  Symposium (SIEDS), Charlottesville, VA, USA, 28 April 2017;  pp. 132--137.

\bibitem[Flynn et~al.(2020)Flynn, Grispos, Glisson, and
  Mahoney]{flynn2020knock}
Flynn, T.; Grispos, G.; Glisson, W.; Mahoney, W.
\newblock Knock! knock! Who is there? Investigating data leakage from a medical
  internet of things hijacking attack. In Proceedings of the 53rd Hawaii International Conference on System Sciences, Maui, HI, USA, 7--10 January 2020.

\bibitem[Newaz et~al.(2021)Newaz, Sikder, Rahman, and Uluagac]{newaz2021survey}
Newaz, A.I.; Sikder, A.K.; Rahman, M.A.; Uluagac, A.S.
\newblock A survey on security and privacy issues in modern healthcare systems:
  Attacks and defenses.
\newblock {\em ACM Trans. Comput. Healthc.} {\bf 2021}, {\em
  2},~1--44.

\bibitem[West et~al.(2016)West, Kohno, Lindsay, and Sechman]{west2016wearfit}
West, J.; Kohno, T.; Lindsay, D.; Sechman, J.
\newblock Wearfit: Security design analysis of a wearable fitness tracker. In
\newblock {\em IEEE Center for Secure Design}; IEEE: New York, NY, USA, 2016.

\bibitem[Food and Administration(2023)]{FDA}
Food, U.; Administration, D.
\newblock Postmarket Management of Cybersecurity in Medical Devices.
\newblock {\em US Food and Drug Administration} {\bf 2023}.
\newblock Available online:
  \url{https://www.fda.gov/regulatory-information/search-fda-guidance-documents/postmarket-management-cybersecurity-medical-devices} (accessed on 09-05-2023).

\bibitem[Wazid et~al.(2020)Wazid, Bera, Mitra, Das, and Ali]{wazid2020private}
Wazid, M.; Bera, B.; Mitra, A.; Das, A.K.; Ali, R.
\newblock Private blockchain-envisioned security framework for AI-enabled
  IoT-based drone-aided healthcare services.
\newblock In Proceedings of the 2nd ACM MobiCom Workshop on
  Drone Assisted Wireless Communications for 5G and Beyond, London, UK, 25 September 2020;  pp. 37--42.

\bibitem[Fathalizadeh et~al.(2022)Fathalizadeh, Moghtadaiee, and
  Alishahi]{fathalizadeh2022privacy}
Fathalizadeh, A.; Moghtadaiee, V.; Alishahi, M.
\newblock On the privacy protection of indoor location dataset using
  anonymization.
\newblock {\em Comput. Secur.} {\bf 2022}, {\em 117},~102665.

\bibitem[Arul et~al.(2024)Arul, Alroobaea, Tariq, Almulihi, Alharithi, and
  Shoaib]{arul2024iot}
Arul, R.; Alroobaea, R.; Tariq, U.; Almulihi, A.H.; Alharithi, F.S.; Shoaib, U.
\newblock IoT-enabled healthcare systems using block chain-dependent adaptable
  services.
\newblock {\em Pers. Ubiquitous Comput.} {\bf 2024}, {\em 28},~43--57.

\bibitem[Bu et~al.(2019)Bu, Karpovsky, and Kinsy]{bu2019bulwark}
Bu, L.; Karpovsky, M.G.; Kinsy, M.A.
\newblock Bulwark: Securing implantable medical devices communication channels.
\newblock {\em Comput. Secur.} {\bf 2019}, {\em 86},~498--511.

\bibitem[Ma et~al.(2024)Ma, Ma, Lin, Zhang, Cai, Wu, and Wang]{ma2024improving}
Ma, Y.; Ma, R.; Lin, Z.; Zhang, R.; Cai, Y.; Wu, W.; Wang, J.
\newblock Improving age of information for covert communication with
  time-modulated arrays.
\newblock {\em IEEE Internet Things J.} {\bf 2024} DOI: 10.1109/JIOT.2024.3466855
. 

\bibitem[Saeedi(2019)]{saeedi2019machine}
Saeedi, K.
\newblock Machine Learning for DDOS Detection in Packet Core Network for IoT. Master's Thesis, Luleå University of Technology, Luleå, Sweden, 2019.

\bibitem[Newaz et~al.(2019)Newaz, Sikder, Rahman, and
  Uluagac]{newaz2019healthguard}
Newaz, A.I.; Sikder, A.K.; Rahman, M.A.; Uluagac, A.S.
\newblock Healthguard: A machine learning-based security framework for smart
  healthcare systems.
\newblock In Proceedings of the 2019 Sixth International Conference on Social
  Networks Analysis, Management and Security (SNAMS), Granada, Spain, 22--25 October 2019;  pp.
  389--396.

\bibitem[Srivastava et~al.(2019{\natexlab{a}})Srivastava, Crichigno, and
  Dhar]{srivastava2019light}
Srivastava, G.; Crichigno, J.; Dhar, S.
\newblock A light and secure healthcare blockchain for iot medical devices.
\newblock In Proceedings of the 2019 IEEE Canadian conference of electrical and
  computer engineering (CCECE), Edmonton, AB, Canada, 5--8 May 2019;  pp. 1--5.

\bibitem[Srivastava et~al.(2019{\natexlab{b}})Srivastava, Parizi, Dehghantanha,
  and Choo]{srivastava2019data}
Srivastava, G.; Parizi, R.M.; Dehghantanha, A.; Choo, K.K.R.
\newblock Data sharing and privacy for patient iot devices using blockchain.
\newblock In Proceedings of the Smart City and Informatization: 7th
  International Conference, iSCI 2019, Guangzhou, China, 12--15 November 2019; Springer: Cham, Switzerland, 2019;  pp. 334--348.

\bibitem[Shrivastava et~al.(2022)Shrivastava, Singh, Hasan, Islam, Abdullah,
  Aman, et~al.]{shrivastava2022securing}
Shrivastava, R.K.; Singh, S.P.; Hasan, M.K.; Islam, S.; Abdullah, S.; Aman,
  A.H.M.; Mohd Aman, A.H.
\newblock Securing Internet of Things devices against code tampering attacks
  using Return Oriented Programming.
\newblock {\em Comput. Commun.} {\bf 2022}, {\em 193},~38--46.

\bibitem[Wu et~al.(2015)Wu, Ganesan, Hu, Wong, Wong, and Mitra]{wu2015tpad}
Wu, T.F.; Ganesan, K.; Hu, Y.A.; Wong, H.S.P.; Wong, S.; Mitra, S.
\newblock TPAD: Hardware Trojan prevention and detection for trusted integrated
  circuits.
\newblock {\em IEEE Trans. Comput.-Aided Des. Integr. Circuits Syst.} {\bf 2015}, {\em 35},~521--534.

\bibitem[Rathore et~al.(2020)Rathore, Mohamed, and Guizani]{rathore2020deep}
Rathore, H.; Mohamed, A.; Guizani, M.
\newblock Deep learning-based security schemes for implantable medical devices.
  In {\em Energy Efficiency of Medical Devices and Healthcare Applications};
  Elsevier:  Amsterdam, The Netherlands, 2020; pp. 109--130.

\bibitem[Li et~al.(2020)Li, Cai, Khan, Rehman, Balasubramaniam, Sun, and
  Venu]{li2020secured}
Li, J.; Cai, J.; Khan, F.; Rehman, A.U.; Balasubramaniam, V.; Sun, J.; Venu, P.
\newblock A secured framework for sdn-based edge computing in IOT-enabled
  healthcare system.
\newblock {\em IEEE Access} {\bf 2020}, {\em 8},~135479--135490.

\bibitem[Zverev et~al.(2019)Zverev, Ag{\"u}ero, Garrido, and
  Bilbao]{zverev2019network}
Zverev, M.; Ag{\"u}ero, R.; Garrido, P.; Bilbao, J.
\newblock Network Coding for IIoT Multi-Cloud Environments.
\newblock In Proceedings of the 9th International Conference
  on the Internet of Things, Bilbao, Spain, 22--25 October 2019;  pp. 1--4.

\bibitem[Varga et~al.(2017)Varga, Plosz, Soos, and Hegedus]{varga2017security}
Varga, P.; Plosz, S.; Soos, G.; Hegedus, C.
\newblock Security threats and issues in automation IoT.
\newblock In Proceedings of the 2017 IEEE 13th International Workshop on
  Factory Communication Systems (WFCS), Trondheim, Norway, 31 May--2 June 2017;  pp. 1--6.

\bibitem[Anderson and Fuloria(2011)]{Anderson2011SmartMS}
Anderson, R.J.; Fuloria, S.
\newblock \emph{Smart Meter Security: A Survey}; University of Cambridge: Cambridge, UK,
\newblock  2011.

\bibitem[Cleemput et~al.(2016)Cleemput, Mustafa, and Preneel]{cleemput2016high}
Cleemput, S.; Mustafa, M.A.; Preneel, B.
\newblock High assurance smart metering.
\newblock In Proceedings of the 2016 IEEE 17th International Symposium on High
  Assurance Systems Engineering (HASE), Orlando, FL, USA, 7--9 January 2016;  pp. 294--297.

\bibitem[Wurm et~al.(2016)Wurm, Hoang, Arias, Sadeghi, and
  Jin]{wurm2016security}
Wurm, J.; Hoang, K.; Arias, O.; Sadeghi, A.R.; Jin, Y.
\newblock Security analysis on consumer and industrial IoT devices.
\newblock In Proceedings of the 2016 21st Asia and South Pacific design
  automation conference (ASP-DAC), Macao, China, 25--28 January 2016;  pp. 519--524.

\bibitem[Horak et~al.(2021)Horak, Strelec, Huraj, Tanuska, Vaclavova, and
  Kebisek]{horak2021vulnerability}
Horak, T.; Strelec, P.; Huraj, L.; Tanuska, P.; Vaclavova, A.; Kebisek, M.
\newblock The vulnerability of the production line using industrial IoT systems
  under ddos attack.
\newblock {\em Electronics} {\bf 2021}, {\em 10},~381.

\bibitem[Soltan et~al.(2018)Soltan, Mittal, and Poor]{soltan2018blackiot}
Soltan, S.; Mittal, P.; Poor, H.V.
\newblock $\{$BlackIoT$\}$:$\{$IoT$\}$ botnet of high wattage devices can
  disrupt the power grid.
\newblock In Proceedings of the 27th USENIX Security Symposium (USENIX Security
  18), Baltimore, MD, USA, 15--17 August 2018;  pp. 15--32.

\bibitem[Tu et~al.(2019)Tu, Rampazzi, Hao, Rodriguez, Fu, and Hei]{tu2019trick}
Tu, Y.; Rampazzi, S.; Hao, B.; Rodriguez, A.; Fu, K.; Hei, X.
\newblock Trick or heat? Manipulating critical temperature-based control
  systems using rectification attacks.
\newblock In Proceedings of the 2019 ACM SIGSAC Conference
  on Computer and Communications Security, London, UK, 11--15 November 2019;  pp. 2301--2315.

\bibitem[Ahamad and Mishra(2024)]{ahamad2024hybrid}
Ahamad, R.; Mishra, K.N.
\newblock Hybrid approach for suspicious object surveillance using video clips
  and UAV images in cloud-IoT-based computing environment.
\newblock {\em Clust. Comput.} {\bf 2024}, {\em 27},~761--785.

\bibitem[Kott et~al.(2016)Kott, Swami, and West]{kott2016internet}
Kott, A.; Swami, A.; West, B.J.
\newblock The internet of battle things.
\newblock {\em Computer} {\bf 2016}, {\em 49},~70--75.

\bibitem[Lesjak et~al.(2016)Lesjak, Bock, Hein, and
  Maritsch]{lesjak2016hardware}
Lesjak, C.; Bock, H.; Hein, D.; Maritsch, M.
\newblock Hardware-secured and transparent multi-stakeholder data exchange for
  industrial IoT.
\newblock In Proceedings of the 2016 IEEE 14th International Conference on
  Industrial Informatics (INDIN), Poitiers, France, 19--21 July 2016;  pp. 706--713.

\bibitem[Pu et~al.(2023)Pu, Lin, Chen, and He]{pu2023user}
Pu, L.; Lin, C.; Chen, B.; He, D.
\newblock User-friendly public-key authenticated encryption with keyword search
  for industrial internet of things.
\newblock {\em IEEE Internet Things J.} {\bf 2023}, \emph{10}, 13544--13555.

\bibitem[Tan et~al.(2021)Tan, Yu, Yang, and Bashir]{tan2021blockchain}
Tan, L.; Yu, K.; Yang, C.; Bashir, A.K.
\newblock A blockchain-based Shamir's threshold cryptography for data
  protection in industrial internet of things of smart city.
\newblock In Proceedings of the 1st Workshop on Artificial
  Intelligence and Blockchain Technologies for Smart Cities with 6G, New Orleans, LA, USA, 25--29 October 2021;  pp.
  13--18.

\bibitem[De~Oliveira et~al.(2023)De~Oliveira, Nogueira, dos Santos, and
  Batista]{de2023intelligent}
De~Oliveira, G.W.; Nogueira, M.; dos Santos, A.L.; Batista, D.M.
\newblock Intelligent VNF placement to mitigate DDoS attacks on industrial IoT.
\newblock {\em IEEE Trans. Netw. Serv. Manag.} {\bf 2023}, \emph{20}, 1319--1331.

\bibitem[Ibrahim et~al.(2022)Ibrahim, Abu Al-Haija, and Ahmad]{ibrahim2022ddos}
Ibrahim, R.F.; Abu Al-Haija, Q.; Ahmad, A.
\newblock DDoS attack prevention for internet of thing devices using ethereum
  blockchain technology.
\newblock {\em Sensors} {\bf 2022}, {\em 22},~6806.

\bibitem[Li et~al.(2021)Li, Lyu, Liu, Zhang, and Lyu]{li2021fleam}
Li, J.; Lyu, L.; Liu, X.; Zhang, X.; Lyu, X.
\newblock FLEAM: A federated learning empowered architecture to mitigate DDoS
  in industrial IoT.
\newblock {\em IEEE Trans. Ind. Inform.} {\bf 2021}, {\em
  18},~4059--4068.

\bibitem[Kuusij{\"a}rvi et~al.(2016)Kuusij{\"a}rvi, Savola, Savolainen, and
  Evesti]{kuusijarvi2016mitigating}
Kuusij{\"a}rvi, J.; Savola, R.; Savolainen, P.; Evesti, A.
\newblock Mitigating IoT security threats with a trusted Network element.
\newblock In Proceedings of the 2016 11th International Conference for Internet
  Technology and Secured Transactions (ICITST), Barcelona, Spain, 5--7 December 2016;  pp. 260--265.

\bibitem[Stocchero et~al.(2022)Stocchero, Silva, de~Souza~Silva, Lawisch, dos
  Anjos, and de~Freitas]{stocchero2022secure}
Stocchero, J.M.; Silva, C.A.; de~Souza~Silva, L.; Lawisch, M.A.; dos Anjos,
  J.C.S.; de~Freitas, E.P.
\newblock Secure command and control for internet of battle things using novel
  network paradigms.
\newblock {\em IEEE Commun. Mag.} {\bf 2022}, \emph{61}, 166--172.

\bibitem[Tbatou et~al.(2017)Tbatou, Ramrami, and Tabii]{tbatou2017security}
Tbatou, S.; Ramrami, A.; Tabii, Y.
\newblock Security of communications in connected cars modeling and safety
  assessment.
\newblock In Proceedings of the 2nd International Conference
  on Big Data, Cloud and Applications, Tetouan, Morocco, 29--30 March 2017;  pp. 1--7.

\bibitem[Zhu et~al.(2024)Zhu, Zeng, Weng, Han, Yang, Li, and
  Zhang]{zhu2024sybil}
Zhu, Y.; Zeng, J.; Weng, F.; Han, D.; Yang, Y.; Li, X.; Zhang, Y.
\newblock Sybil attacks detection and traceability mechanism based on beacon
  packets in connected automobile vehicles.
\newblock {\em Sensors} {\bf 2024}, {\em 24},~2153.

\bibitem[Kumar et~al.(2021)Kumar, Sharma, and Pattnaik]{kumar2021multimedia}
Kumar, R.; Sharma, R.; Pattnaik, P.K.
\newblock {\em Multimedia Technologies in the Internet of Things Environment}; Springer: Singapore, 2021; Volume 2.

\bibitem[Taslimasa et~al.(2023)Taslimasa, Dadkhah, Neto, Xiong, Ray, and
  Ghorbani]{taslimasa2023security}
Taslimasa, H.; Dadkhah, S.; Neto, E.C.P.; Xiong, P.; Ray, S.; Ghorbani, A.A.
\newblock Security issues in Internet of Vehicles (IoV): A comprehensive
  survey.
\newblock {\em Internet Things} {\bf 2023}, \emph{22}, 100809.

\bibitem[Zhang et~al.(2024)Zhang, Cheng, Wang, Lou, Gao, Wu, and
  Ng]{zhang2024integrated}
Zhang, R.; Cheng, L.; Wang, S.; Lou, Y.; Gao, Y.; Wu, W.; Ng, D.W.K.
\newblock Integrated sensing and communication with massive MIMO: A unified
  tensor approach for channel and target parameter estimation.
\newblock {\em IEEE Trans. Wirel. Commun.} {\bf 2024}, \emph{23}, 8571--8587.

\bibitem[Khan et~al.(2020)Khan, Shiwakoti, Stasinopoulos, and
  Chen]{khan2020cyber}
Khan, S.K.; Shiwakoti, N.; Stasinopoulos, P.; Chen, Y.
\newblock Cyber-attacks in the next-generation cars, mitigation techniques,
  anticipated readiness and future directions.
\newblock {\em Accid. Anal. Prev.} {\bf 2020}, {\em 148},~105837.

\bibitem[Zewdie and Girma(2020)]{zewdie2020iot}
Zewdie, T.G.; Girma, A.
\newblock IOT Security and the Role of AI/ML to Combat Emerging Cyber Threats
  in Cloud Computing Environment.
\newblock {\em Issues Inf. Syst.} {\bf 2020}, {\em 21}, 253--263.

\bibitem[Smys and Wang(2021)]{smys2021security}
Smys, S.; Wang, H.
\newblock Security enhancement in smart vehicle using blockchain-based
  architectural framework.
\newblock {\em J. Artif. Intell.} {\bf 2021}, {\em
  3},~90--100.

\bibitem[Zhao et~al.(2021)Zhao, Gill, Pisu, and Comert]{zhao2021detection}
Zhao, C.; Gill, J.S.; Pisu, P.; Comert, G.
\newblock Detection of false data injection attack in connected and automated
  vehicles via cloud-based sandboxing.
\newblock {\em IEEE Trans. Intell. Transp. Syst.} {\bf
  2021}, {\em 23},~9078--9088.

\bibitem[Balas et~al.(2020)Balas, Solanki, Kumar, and Ahad]{balas2020handbook}
Balas, V.E.; Solanki, V.K.; Kumar, R.; Ahad, M.A.R.
\newblock \emph{A Handbook of Internet of Things in Biomedical and Cyber Physical
  System};
\newblock Technical Report; Springer: Cham, Switzerland, 2020.

\bibitem[Lee et~al.(2015)Lee, Egelman, Lee, and Wagner]{lee2015risk}
Lee, L.; Egelman, S.; Lee, J.H.; Wagner, D.
\newblock Risk perceptions for wearable devices.
\newblock {\em arXiv} {\bf 2015}, arXiv:1504.05694.

\bibitem[Siboni et~al.(2016)Siboni, Shabtai, Tippenhauer, Lee, and
  Elovici]{siboni2016advanced}
Siboni, S.; Shabtai, A.; Tippenhauer, N.O.; Lee, J.; Elovici, Y.
\newblock Advanced security testbed framework for wearable IoT devices.
\newblock {\em ACM Trans. Internet Technol. (TOIT)} {\bf 2016}, {\em
  16},~1--25.

\bibitem[Lennon(2015)]{lennon2015all}
Lennon, M.
\newblock All smartwatches vulnerable to attack: HP study.
\newblock {\em Retrieved Novemb.} {\bf 2015}, {\em 28},~2015.

\bibitem[Cusack et~al.(2017)Cusack, Antony, Ward, and
  Mody]{cusack2017assessment}
Cusack, B.; Antony, B.; Ward, G.; Mody, S.
\newblock Assessment of security vulnerabilities in wearable devices. In Proceedings of the 15th Australian Information Security Management Conference, Perth, Australia, 5--6 December 2017.

\bibitem[Giaretta(2022)]{giaretta2022security}
Giaretta, A.
\newblock Security and Privacy in Virtual Reality--A Literature Survey.
\newblock {\em arXiv} {\bf 2022}, arXiv:2205.00208.

\bibitem[Adams et~al.(2018)Adams, Bah, Barwulor, Musaby, Pitkin, and
  Redmiles]{adams2018ethics}
Adams, D.; Bah, A.; Barwulor, C.; Musaby, N.; Pitkin, K.; Redmiles, E.M.
\newblock Ethics emerging: The story of privacy and security perceptions in
  virtual reality.
\newblock In Proceedings of the Fourteenth Symposium on Usable Privacy and
  Security (SOUPS 2018), Berkeley, CA, USA, 12--14 August 2018;  pp. 427--442.

\bibitem[Karakaya et~al.(2016)Karakaya, Bostan, and
  G{\"o}k{\c{c}}ay]{karakaya2016secure}
Karakaya, M.; Bostan, A.; G{\"o}k{\c{c}}ay, E.
\newblock How Secure is Your Smart Watch?
\newblock {\em Int. J. Inf. Secur. Sci.} {\bf
  2016}, {\em 5},~90--95.

\bibitem[Mahinderjit~Singh et~al.(2020)Mahinderjit~Singh, Ching, and
  Abd~Manaf]{mahinderjit2020novel}
Mahinderjit~Singh, M.; Ching, K.W.; Abd~Manaf, A.
\newblock A novel out-of-band biometrics authentication scheme for wearable
  devices.
\newblock {\em Int. J. Comput. Appl.} {\bf 2020},
  {\em 42},~589--601.

\bibitem[Kavianpour et~al.(2021)Kavianpour, Shanmugam, H~Zolait, and
  Razaq]{kavianpour2021framework}
Kavianpour, S.; Shanmugam, B.; H~Zolait, A.; Razaq, A.
\newblock A Framework to Detect Cyber-attacks against Networked Medical Devices
  (Internet of Medical Things): An Attack-Surface-Reduction by Design Approach.
\newblock {\em Int. J. Comput. Digit. Syst.} {\bf
  2021}, \emph{11}, 1289--1298.

\bibitem[Bhatia and Sangwan(2021)]{bhatia2021soft}
Bhatia, M.; Sangwan, S.R.
\newblock Soft computing for anomaly detection and prediction to mitigate
  IoT-based real-time abuse.
\newblock {\em Pers. Ubiquitous Comput.} {\bf 2021}, \emph{28}, 123--133.

\bibitem[Zhao et~al.(2020)Zhao, Wang, Liu, Chen, Cheng, and
  Yu]{zhao2020trueheart}
Zhao, T.; Wang, Y.; Liu, J.; Chen, Y.; Cheng, J.; Yu, J.
\newblock Trueheart: Continuous authentication on wrist-worn wearables using
  ppg-based biometrics.
\newblock In Proceedings of the IEEE INFOCOM 2020---IEEE Conference on Computer
  Communications, Toronto, ON, Canada, 6--9 July 2020;  pp. 30--39.

\bibitem[Gulhane et~al.(2019)Gulhane, Vyas, Mitra, Oruche, Hoefer,
  Valluripally, Calyam, and Hoque]{gulhane2019security}
Gulhane, A.; Vyas, A.; Mitra, R.; Oruche, R.; Hoefer, G.; Valluripally, S.;
  Calyam, P.; Hoque, K.A.
\newblock Security, privacy and safety risk assessment for virtual reality
  learning environment applications.
\newblock In Proceedings of the 2019 16th IEEE Annual Consumer Communications
  \& Networking Conference (CCNC), Las Vegas, NV, USA, 11--14 January 2019;  pp. 1--9.

\bibitem[Fang et~al.(2019)Fang, Li, Zhou, Zhang, and Ge]{fang2019fine}
Fang, L.; Li, M.; Zhou, L.; Zhang, H.; Ge, C.
\newblock A fine-grained user-divided privacy-preserving access control
  protocol in smart watch.
\newblock {\em Sensors} {\bf 2019}, {\em 19},~2109.

\bibitem[Now(2022)]{WinNT26}
IoT Now.
\newblock Analyst Insights on IoT in the Retail Sector.
\newblock  2022.
\newblock Available online:
  \url{https://www.iot-now.com/2022/07/18/122379-insights-on-retail-sector/} (accessed on 21-04-2024).

\bibitem[Wang et~al.(2021)Wang, Xie, and Duan]{wang2021iot}
Wang, B.; Xie, Y.; Duan, X.
\newblock An IoT Based Fruit and Vegetable Sales System: A whole system
  including IoT based integrated intelligent scale and online shop.
\newblock In Proceedings of the 2021 5th International
  Conference on Cloud and Big Data Computing, Liverpool, UK, 13--15 August 2021;  pp. 109--115.

\bibitem[Di~Rienzo et~al.(2015)Di~Rienzo, Garzotto, Cremonesi, Fr{\`a}, and
  Valla]{di2015towards}
Di~Rienzo, A.; Garzotto, F.; Cremonesi, P.; Fr{\`a}, C.; Valla, M.
\newblock Towards a smart retail environment.
\newblock In Proceedings of the Adjunct Proceedings of the 2015 ACM
  International Joint Conference on Pervasive and Ubiquitous Computing and
  Proceedings of the 2015 ACM International Symposium on Wearable Computers, Osaka, Japan, 7--1 September
  2015;  pp. 779--782.

\bibitem[Networks(2024)]{WinNT20}
Nozomi Networks.
\newblock IoT Cybersecurity for Retail Organizations.
\newblock 2024.
\newblock Available online:
  \url{https://www.nozominetworks.com/industries/retail-cybersecurity} (accessed on 21-05-2024).

\bibitem[Mitzner(2022)]{WinNT21}
Mitzner, D.
\newblock IoT And The Rise Of Cyber Security Threats In Retail.
\newblock {\em Forbes} {\bf 2022}.
\newblock Available online:
  \url{https://www.forbes.com/sites/dennismitzner/2022/09/14/self-checkouts-iot-and-the-rise-of-retail-cyber-security-threats/?sh=683e47b157f5} (accessed on 05-11-2024).

\bibitem[Singh et~al.(2024)Singh, Do, Fouad, Kim, and Kim]{singh2024crumbled}
Singh, N.; Do, Y.; Fouad, Y.Y.I.; Kim, J.; Kim, H.
\newblock Crumbled Cookie Exploring E-commerce Websites Cookie Policies with
  Data Protection Regulations.
\newblock {\em ACM TWEB} {\bf 2024}, https://doi.org/10.1145/3708515.

\bibitem[Mavroudis and Veale(2018)]{mavroudis2018eavesdropping}
Mavroudis, V.; Veale, M.
\newblock Eavesdropping whilst you're shopping: Balancing personalisation and
  privacy in connected retail spaces.
\newblock In Proceedings of the Living in the Internet of Things: Cybersecurity
  of the IoT---2018, London, UK, 28--29 March 2018;  pp. 1--10.

\bibitem[Cartwright and Cartwright(2023)]{cartwright2023economics}
Cartwright, A.; Cartwright, E.
\newblock The economics of ransomware attacks on integrated supply chain
  networks.
\newblock {\em Digit. Threat. Res. Pract.} {\bf 2023}, \emph{4}, 56.

\bibitem[Wang et~al.(2021)Wang, Tu, Lei, Xie, Li, Chou, Hsieh, Hu, Xiao, and
  Peng]{wang2021insecurity}
Wang, S.; Tu, G.H.; Lei, X.; Xie, T.; Li, C.Y.; Chou, P.Y.; Hsieh, F.; Hu, Y.;
  Xiao, L.; Peng, C.
\newblock Insecurity of operational cellular IoT service: New vulnerabilities,
  attacks, and countermeasures.
\newblock In Proceedings of the 27th Annual International
  Conference on Mobile Computing and Networking, New Orleans, LA, USA, 25--29 October 2021;  pp. 437--450.

\bibitem[Jaffar and Elmedany(2022)]{jaffar2022systematic}
Jaffar, J.A.; Elmedany, W.
\newblock A systematic review of homomorphic encryption applications in
  Internet of Things.
\newblock In Proceedings of the 6th Smart Cities Symposium (SCS 2022), Hybrid Conference, Bahrain, 6--8 December
  2022; Volume 2022, pp. 489--497.

\bibitem[Musa and Vidyasankar(2017)]{musa2017fog}
Musa, Z.; Vidyasankar, K.
\newblock A fog computing framework for blackberry supply chain management.
\newblock {\em Procedia Comput. Sci.} {\bf 2017}, {\em 113},~178--185.

\bibitem[Brasilino and Swany(2019)]{brasilino2019mitigating}
Brasilino, L.R.; Swany, M.
\newblock Mitigating ddos flooding attacks against iot using custom hardware
  modules.
\newblock In Proceedings of the 2019 Sixth International Conference on Internet
  of Things: Systems, Management and Security (IOTSMS), Granada, Spain, 22--25 October 2019;  pp.
  58--64.

\bibitem[Das(2022)]{das2022iot}
Das, S.
\newblock An IoT business model for public sector retail oil outlets.
\newblock {\em Inf. Technol. People} {\bf 2022}, {\em
  35},~2344--2367.

\bibitem[Mohanta et~al.(2020)Mohanta, Jena, Satapathy, and
  Patnaik]{mohanta2020survey}
Mohanta, B.K.; Jena, D.; Satapathy, U.; Patnaik, S.
\newblock Survey on IoT security: Challenges and solution using machine
  learning, artificial intelligence and blockchain technology.
\newblock {\em Internet Things} {\bf 2020}, {\em 11},~100227.

\bibitem[Akhter and Sofi(2022)]{akhter2022precision}
Akhter, R.; Sofi, S.A.
\newblock Precision agriculture using IoT data analytics and machine learning.
\newblock {\em J. King Saud Univ.-Comput. Inf. Sci.} {\bf 2022}, {\em 34},~5602--5618.

\bibitem[Singh and Singh(2020)]{singh2020odysseys}
Singh, N.; Singh, A.N.
\newblock Odysseys of agriculture sensors: Current challenges and forthcoming
  prospects.
\newblock {\em Comput. Electron. Agric.} {\bf 2020}, {\em
  171},~105328.

\bibitem[Namani and Gonen(2020)]{namani2020smart}
Namani, S.; Gonen, B.
\newblock Smart agriculture based on IoT and cloud computing.
\newblock In Proceedings of the 2020 3rd International Conference on
  Information and Computer Technologies (ICICT), San Jose, CA, USA, 9--12 March 2020;  pp. 553--556.

\bibitem[Patil et~al.(2012)Patil, Al-Gaadi, Biradar, and
  Rangaswamy]{patil2012internet}
Patil, V.; Al-Gaadi, K.; Biradar, D.; Rangaswamy, M.
\newblock Internet of things (Iot) and cloud computing for agriculture: An
  overview.
\newblock In Proceedings of Agro-Informatics and Precision Agriculture (AIPA
  2012), Hyderabad, India, 1--3 August 2012;  pp. 292--296.

\bibitem[Mekala and Viswanathan(2017)]{mekala2017survey}
Mekala, M.S.; Viswanathan, P.
\newblock A Survey: Smart agriculture IoT with cloud computing.
\newblock In Proceedings of the 2017 International Conference on
  Microelectronic Devices, Circuits and Systems (ICMDCS), Vellore, India, 10--12 August 2017;  pp.
  1--7.

\bibitem[Krzywoszynska()]{krzywoszynskasoil}
Krzywoszynska, A.
\newblock Soil 
 care: Understanding soil needs, responsibilities, and
  attentiveness through the concept of care, \emph{Preprint, DOI: 10.13140/RG.2.2.13789.74728}.

\bibitem[Yu et~al.(2021)Yu, Gao, R~Shamshiri, Tao, Ren, Zhang, and
  Su]{yu2021review}
Yu, L.; Gao, W.; R~Shamshiri, R.; Tao, S.; Ren, Y.; Zhang, Y.; Su, G.
\newblock Review of research progress on soil moisture sensor technology. \emph{Int.} J. Agric. Biol. Eng. {\bf
  2021}, \emph{14}, 32--42.

\bibitem[Refaai et~al.(2022)Refaai, Dattu, Gireesh, Dixit, Sandeep,
  Christopher, et~al.]{refaai2022application}
Refaai, M.R.; Dattu, V.S.; Gireesh, N.; Dixit, E.; Sandeep, C.; Christopher,
  D.
\newblock Application of IoT-Based Drones in Precision Agriculture for Pest
  Control.
\newblock {\em Adv. Mater. Sci. Eng.} {\bf 2022}, {\em
  2022}, 1160258.

\bibitem[Al-Fuqaha et~al.(2015)Al-Fuqaha, Guizani, Mohammadi, Aledhari, and
  Ayyash]{7123563}
Al-Fuqaha, A.; Guizani, M.; Mohammadi, M.; Aledhari, M.; Ayyash, M.
\newblock Internet of Things: A Survey on Enabling Technologies, Protocols, and
  Applications.
\newblock {\em IEEE Commun. Surv. Tutor.} {\bf 2015}, {\em
  17},~2347--2376.
\newblock {\url{https://doi.org/10.1109/COMST.2015.2444095}}.

\bibitem[Song et~al.(2022)Song, Ma, Li, Qi, and Liu]{song2022development}
Song, C.; Ma, W.; Li, J.; Qi, B.; Liu, B.
\newblock Development Trends in Precision Agriculture and Its Management in
  China Based on Data Visualization.
\newblock {\em Agronomy} {\bf 2022}, {\em 12},~2905.

\bibitem[Salam and Salam(2020)]{salam2020internet}
Salam, A.; Salam, A.
\newblock Internet of things for sustainability: Perspectives in privacy,
  cybersecurity, and future trends. In
\newblock {\em Internet of Things for Sustainable Community Development:
  Wireless Communications, Sensing, and Systems}; Springer: Cham, Switzerland, 2020;  pp. 299--327.

\bibitem[Gupta et~al.(2020)Gupta, Abdelsalam, Khorsandroo, and
  Mittal]{gupta2020security}
Gupta, M.; Abdelsalam, M.; Khorsandroo, S.; Mittal, S.
\newblock Security and privacy in smart farming: Challenges and opportunities.
\newblock {\em IEEE Access} {\bf 2020}, {\em 8},~34564--34584.

\bibitem[Mori et~al.(2022)Mori, Kundaliya, Naik, and Shah]{mori2022iot}
Mori, H.; Kundaliya, J.; Naik, K.; Shah, M.
\newblock IoT technologies in smart environment: Security issues and future
  enhancements.
\newblock {\em Environ. Sci. Pollut. Res.} {\bf 2022}, {\em
  29},~47969--47987.

\bibitem[Triantafyllou et~al.(2019)Triantafyllou, Sarigiannidis, and
  Bibi]{triantafyllou2019precision}
Triantafyllou, A.; Sarigiannidis, P.; Bibi, S.
\newblock Precision agriculture: A remote sensing monitoring system
  architecture.
\newblock {\em Information} {\bf 2019}, {\em 10},~348.

\bibitem[Ferrag et~al.(2020)Ferrag, Shu, Yang, Derhab, and
  Maglaras]{ferrag2020security}
Ferrag, M.A.; Shu, L.; Yang, X.; Derhab, A.; Maglaras, L.
\newblock Security and privacy for green IoT-based agriculture: Review,
  blockchain solutions, and challenges.
\newblock {\em IEEE Access} {\bf 2020}, {\em 8},~32031--32053.

\bibitem[Glaroudis et~al.(2020)Glaroudis, Iossifides, and
  Chatzimisios]{glaroudis2020survey}
Glaroudis, D.; Iossifides, A.; Chatzimisios, P.
\newblock Survey, comparison and research challenges of IoT application
  protocols for smart farming.
\newblock {\em Comput. Netw.} {\bf 2020}, {\em 168},~107037.

\bibitem[Farooq et~al.(2019)Farooq, Riaz, Abid, Abid, and
  Naeem]{farooq2019survey}
Farooq, M.S.; Riaz, S.; Abid, A.; Abid, K.; Naeem, M.A.
\newblock A Survey on the Role of IoT in Agriculture for the Implementation of
  Smart Farming.
\newblock {\em IEEE Access} {\bf 2019}, {\em 7},~156237--156271.

\bibitem[Madushanki et~al.(2019)Madushanki, Halgamuge, Wirasagoda, and
  Ali]{madushanki2019adoption}
Madushanki, A.R.; Halgamuge, M.N.; Wirasagoda, W.S.; Ali, S.
\newblock Adoption of the Internet of Things (IoT) in agriculture and smart
  farming towards urban greening: A review.
\newblock {\em Int. J. Adv. Comput. Sci. Appl.} {\bf 2019}, {\em 10}, 11--28.

\bibitem[Ahmed et~al.(2018)Ahmed, De, and Hussain]{ahmed2018internet}
Ahmed, N.; De, D.; Hussain, I.
\newblock Internet of Things (IoT) for smart precision agriculture and farming
  in rural areas.
\newblock {\em IEEE Internet Things J.} {\bf 2018}, {\em
  5},~4890--4899.

\bibitem[Raghuvanshi et~al.(2022)Raghuvanshi, Singh, Sajja, Pallathadka,
  Asenso, Kamal, Singh, and Phasinam]{raghuvanshi2022intrusion}
Raghuvanshi, A.; Singh, U.K.; Sajja, G.S.; Pallathadka, H.; Asenso, E.; Kamal,
  M.; Singh, A.; Phasinam, K.
\newblock Intrusion detection using machine learning for risk mitigation in
  IoT-enabled smart irrigation in smart farming.
\newblock {\em J. Food Qual.} {\bf 2022}, {\em 2022},~1--8.

\bibitem[Kondaveeti et~al.(2024)Kondaveeti, Sai, Athar, Vatsavayi, Mitra, and
  Ananthachari]{kondaveeti2024federated}
Kondaveeti, H.K.; Sai, G.B.; Athar, S.A.; Vatsavayi, V.K.; Mitra, A.;
  Ananthachari, P.
\newblock Federated Learning for Smart Agriculture: Challenges and
  Opportunities.
\newblock In Proceedings of the 2024 Third International Conference on
  Distributed Computing and Electrical Circuits and Electronics (ICDCECE), Ballari, India, 26--27 April 2024;  pp. 1--7.

\bibitem[Kumar et~al.(2021)Kumar, Kumar, Tripathi, Gupta, Gadekallu, and
  Srivastava]{kumar2021sp2f}
Kumar, R.; Kumar, P.; Tripathi, R.; Gupta, G.P.; Gadekallu, T.R.; Srivastava,
  G.
\newblock SP2F: A secured privacy-preserving framework for smart agricultural
  Unmanned Aerial Vehicles.
\newblock {\em Comput. Netw.} {\bf 2021}, {\em 187},~107819.

\bibitem[Kethineni and Gera(2023)]{kethineni2023iot}
Kethineni, K.; Gera, P.
\newblock IoT-based privacy-preserving anomaly detection model for smart
  agriculture.
\newblock {\em Systems} {\bf 2023}, {\em 11},~304.

\bibitem[Song et~al.(2020)Song, Zhong, Wang, Su, Tan, and Liu]{song2020fpdp}
Song, J.; Zhong, Q.; Wang, W.; Su, C.; Tan, Z.; Liu, Y.
\newblock FPDP: Flexible privacy-preserving data publishing scheme for smart
  agriculture.
\newblock {\em IEEE Sens. J.} {\bf 2020}, {\em 21},~17430--17438.

\bibitem[Zhou et~al.(2021)Zhou, Zheng, Guan, Peng, and Lu]{zhou2021efficient}
Zhou, M.; Zheng, Y.; Guan, Y.; Peng, L.; Lu, R.
\newblock Efficient and privacy-preserving range-max query in fog-based
  agricultural IoT. 
\newblock {\em Peer-Peer Netw. Appl.} {\bf 2021}, {\em
  14},~2156--2170.

\bibitem[Chukkapalli et~al.(2021)Chukkapalli, Ranade, Mittal, and
  Joshi]{chukkapalli2021privacy}
Chukkapalli, S.S.L.; Ranade, P.; Mittal, S.; Joshi, A.
\newblock A Privacy-Preserving Anomaly Detection Framework for Cooperative
  Smart Farming Ecosystem.
\newblock In Proceedings of the 2021 Third IEEE International Conference on
  Trust, Privacy and Security in Intelligent Systems and Applications
  (TPS-ISA), Atlanta, GA, USA, 13--15 December 2021;  pp. 340--347.

\bibitem[Kumar et~al.(2021)Kumar, Gupta, and Tripathi]{kumar2021pefl}
Kumar, P.; Gupta, G.P.; Tripathi, R.
\newblock PEFL: Deep privacy-encoding-based federated learning framework for
  smart agriculture.
\newblock {\em IEEE Micro} {\bf 2021}, {\em 42},~33--40.

\bibitem[Malik et~al.(2024)Malik, Hussain, Shah, Saleem, alsanoosy, and
  Chaudhary]{malik2024optimizing}
Malik, J.A.; Hussain, A.; Shah, H.; Saleem, M.; alsanoosy, T.; Chaudhary,
  U.M.D.
\newblock Optimizing Agricultural Risk Management with Hybrid Block-Chain and
  Fog Computing Architectures for Secure and Efficient Data Handling. In {\em
  Computational Intelligence in Internet of Agricultural Things}; Springer: Cham, Switzerland,
  2024; pp. 309--337.

\bibitem[Anidu and Dara(2021)]{anidu2021review}
Anidu, A.; Dara, R.
\newblock A review of data governance challenges in smart farming and potential
  solutions.
\newblock In Proceedings of the 2021 IEEE International Symposium on Technology
  and Society (ISTAS), Waterloo, ON, Canada, 28--31 October 2021;  pp. 1--8.

\bibitem[Chaganti et~al.(2022)Chaganti, Varadarajan, Gorantla, Gadekallu, and
  Ravi]{chaganti2022blockchain}
Chaganti, R.; Varadarajan, V.; Gorantla, V.S.; Gadekallu, T.R.; Ravi, V.
\newblock Blockchain-based cloud-enabled security monitoring using Internet of
  Things in smart agriculture.
\newblock {\em Future Internet} {\bf 2022}, {\em 14},~250.

\bibitem[Al-Turjman and Abujubbeh(2019)]{al2019iot}
Al-Turjman, F.; Abujubbeh, M.
\newblock IoT-enabled smart grid via SM: An overview.
\newblock {\em Future Gener. Comput. Syst.} {\bf 2019}, {\em
  96},~579--590.

\bibitem[Pau et~al.(2018)Pau, Patti, Barbierato, Estebsari, Pons, Ponci, and
  Monti]{pau2018cloud}
Pau, M.; Patti, E.; Barbierato, L.; Estebsari, A.; Pons, E.; Ponci, F.; Monti,
  A.
\newblock A cloud-based smart metering infrastructure for distribution grid
  services and automation.
\newblock {\em Sustain. Energy, Grids Netw.} {\bf 2018}, {\em
  15},~14--25.

\bibitem[{\"O}zg{\"u}r et~al.(2018){\"O}zg{\"u}r, Akram, Challenger, and
  Da{\u{g}}deviren]{ozgur2018iot}
{\"O}zg{\"u}r, L.; Akram, V.K.; Challenger, M.; Da{\u{g}}deviren, O.
\newblock An IoT based smart thermostat.
\newblock In Proceedings of the 2018 5th International Conference on Electrical
  and Electronic Engineering (ICEEE), Istanbul, Turkey, 3--5 May 2018;  pp. 252--256.

\bibitem[Tran et~al.(2022)Tran, Panchal, Khang, Panchal, Fraser, and
  Fowler]{tran2022concept}
Tran, M.K.; Panchal, S.; Khang, T.D.; Panchal, K.; Fraser, R.; Fowler, M.
\newblock Concept review of a cloud-based smart battery management system for
  lithium-ion batteries: Feasibility, logistics, and functionality.
\newblock {\em Batteries} {\bf 2022}, {\em 8},~19.

\bibitem[Boroojeni et~al.(2017)Boroojeni, Amini, and
  Iyengar]{boroojeni2017smart}
Boroojeni, K.G.; Amini, M.H.; Iyengar, S.S.
\newblock {\em Smart Grids: Security and Privacy Issues}; Springer: Cham, Switzerland,
  2017; Volume 221.

\bibitem[Stellios et~al.(2018)Stellios, Kotzanikolaou, Psarakis, Alcaraz, and
  Lopez]{stellios2018survey}
Stellios, I.; Kotzanikolaou, P.; Psarakis, M.; Alcaraz, C.; Lopez, J.
\newblock A survey of iot-enabled cyberattacks: Assessing attack paths to
  critical infrastructures and services.
\newblock {\em IEEE Commun. Surv. Tutor.} {\bf 2018}, {\em
  20},~3453--3495.

\bibitem[Gunduz and Das(2020)]{gunduz2020cyber}
Gunduz, M.Z.; Das, R.
\newblock Cyber-security on smart grid: Threats and potential solutions.
\newblock {\em Comput. Netw.} {\bf 2020}, {\em 169},~107094.

\bibitem[Huseinovi{\'c} et~al.(2020)Huseinovi{\'c}, Mrdovi{\'c}, Bicakci, and
  Uludag]{huseinovic2020survey}
Huseinovi{\'c}, A.; Mrdovi{\'c}, S.; Bicakci, K.; Uludag, S.
\newblock A survey of denial-of-service attacks and solutions in the smart
  grid.
\newblock {\em IEEE Access} {\bf 2020}, {\em 8},~177447--177470.

\bibitem[G{\"o}nen et~al.(2020)G{\"o}nen, Sayan, Y{\i}lmaz, {\"U}st{\"u}nsoy,
  and Karacay{\i}lmaz]{gonen2020false}
G{\"o}nen, S.; Sayan, H.H.; Y{\i}lmaz, E.N.; {\"U}st{\"u}nsoy, F.;
  Karacay{\i}lmaz, G.
\newblock False data injection attacks and the insider threat in smart systems.
\newblock {\em Comput. Secur.} {\bf 2020}, {\em 97},~101955.

\bibitem[Aydeger et~al.(2024)Aydeger, Zeydan, Yadav, Hemachandra, and
  Liyanage]{aydeger2024towards}
Aydeger, A.; Zeydan, E.; Yadav, A.K.; Hemachandra, K.T.; Liyanage, M.
\newblock Towards a quantum-resilient future: Strategies for transitioning to
  post-quantum cryptography.
\newblock In Proceedings of the 2024 15th International Conference on Network
  of the Future (NoF), Castelldefels, Spain, 2--4 October 2024;  pp. 195--203.

\bibitem[Y{\i}lmaz and Uludag(2021)]{yilmaz2021timely}
Y{\i}lmaz, Y.; Uludag, S.
\newblock Timely detection and mitigation of IoT-based cyberattacks in the
  smart grid.
\newblock {\em J. Frankl. Inst.} {\bf 2021}, {\em
  358},~172--192.

\bibitem[Berghout et~al.(2022)Berghout, Benbouzid, and
  Muyeen]{berghout2022machine}
Berghout, T.; Benbouzid, M.; Muyeen, S.
\newblock Machine learning for cybersecurity in smart grids: A comprehensive
  review-based study on methods, solutions, and prospects.
\newblock {\em Int. J. Crit. Infrastruct. Prot.}
  {\bf 2022}, \emph{38}, 100547.

\bibitem[Chen et~al.(2021)Chen, Mohamed, Dampage, Rezaei, Salmen, Obaid, and
  Annuk]{chen2021multi}
Chen, J.; Mohamed, M.A.; Dampage, U.; Rezaei, M.; Salmen, S.H.; Obaid, S.A.;
  Annuk, A.
\newblock A multi-layer security scheme for mitigating smart grid vulnerability
  against faults and cyber-attacks.
\newblock {\em Appl. Sci.} {\bf 2021}, {\em 11},~9972.

\bibitem[Zhang et~al.(2020)Zhang, Wang, and Chen]{zhang2020detecting}
Zhang, Y.; Wang, J.; Chen, B.
\newblock Detecting false data injection attacks in smart grids: A
  semi-supervised deep learning approach.
\newblock {\em IEEE Trans. Smart Grid} {\bf 2020}, {\em 12},~623--634.

\bibitem[Unsal et~al.(2021)Unsal, Ustun, Hussain, and Onen]{unsal2021enhancing}
Unsal, D.B.; Ustun, T.S.; Hussain, S.S.; Onen, A.
\newblock Enhancing cybersecurity in smart grids: False data injection and its
  mitigation.
\newblock {\em Energies} {\bf 2021}, {\em 14},~2657.

\bibitem[Desai et~al.(2022)Desai, Sabar, Alhadad, Mahmood, and
  Chilamkurti]{desai2022mitigating}
Desai, S.; Sabar, N.R.; Alhadad, R.; Mahmood, A.; Chilamkurti, N.
\newblock Mitigating consumer privacy breach in smart grid using
  obfuscation-based generative adversarial network.
\newblock {\em Math. Biosci. Eng} {\bf 2022}, {\em 19},~3350--3368.

\bibitem[Chehri et~al.(2021)Chehri, Fofana, and Yang]{chehri2021security}
Chehri, A.; Fofana, I.; Yang, X.
\newblock Security risk modeling in smart grid critical infrastructures in the
  era of big data and artificial intelligence.
\newblock {\em Sustainability} {\bf 2021}, {\em 13},~3196.

\bibitem[Bouyakoub et~al.(2017)Bouyakoub, Belkhir, Bouyakoub, and
  Guebli]{bouyakoub2017smart}
Bouyakoub, S.; Belkhir, A.; Bouyakoub, F.M.; Guebli, W.
\newblock Smart airport: An IoT-based airport management system.
\newblock In Proceedings of the International Conference on
  Future Networks and Distributed Systems, Cambridge, UK, 19--20 July 2017;  pp. 1--7.

\bibitem[Pech et~al.(2021)Pech, Vrchota, and
  Bedn{\'a}{\v{r}}]{pech2021predictive}
Pech, M.; Vrchota, J.; Bedn{\'a}{\v{r}}, J.
\newblock Predictive maintenance and intelligent sensors in smart factory.
\newblock {\em Sensors} {\bf 2021}, {\em 21},~1470.

\bibitem[Andel et~al.(2021)Andel, {\v{S}}im{\'a}k, {\v{S}}kult{\'e}ty, and
  Nemec]{andel2021iot}
Andel, J.; {\v{S}}im{\'a}k, V.; {\v{S}}kult{\'e}ty, F.; Nemec, D.
\newblock IoT-based Data Acquisition Unit for aircraft and road vehicle.
\newblock {\em Transp. Res. Procedia} {\bf 2021}, {\em
  55},~969--976.

\bibitem[Wang et~al.(2013)Wang, Liu, Li, Li, and Zhang]{wang2013aviation}
Wang, X.k.; Liu, Y.y.; Li, Z.y.; Li, M.f.; Zhang, S.l.
\newblock Aviation equipment maintenance safety management based on the
  technology of IOT.
\newblock In Proceedings of the 19th International Conference on Industrial
  Engineering and Engineering Management: Engineering Management, Changsha, China, 27--29 October 2013; Springer: Berlin/Heidelberg, Germany,
  2013;  pp. 1185--1194.

\bibitem[Okpala et~al.(2020)Okpala, Parajuli, Nnaji, and
  Awolusi]{okpala2020assessing}
Okpala, I.; Parajuli, A.; Nnaji, C.; Awolusi, I.
\newblock Assessing the feasibility of integrating the internet of things into
  safety management systems: A focus on wearable sensing devices.
\newblock In Proceedings of the Construction Research Congress 2020: Computer
  Applications, Tempe, AZ, USA, 8--10 March 2020; American Society of Civil Engineers: Reston, VA, USA, 2020;  pp.
  236--245.

\bibitem[Noel et~al.(2021)Noel, Navya, Likitha, Manjula, and
  Priya]{noel2021smart}
Noel, S.; Navya, M.; Likitha, D.; Manjula, K.; Priya, S.K.
\newblock A Smart IoT based real-time system to Minimize Mishandled Luggage at
  Airports.
\newblock In Proceedings of the 2021 5th International Conference on Computing
  Methodologies and Communication (ICCMC), Erode, India, 8--10 April 2021;  pp. 377--384.

\bibitem[Samha et~al.(2020)Samha, Alghamdi, Albader, and
  Alshammri]{samha2020applied}
Samha, A.K.; Alghamdi, N.; Albader, H.; Alshammri, G.H.
\newblock Applied Internet of Things in Saudi Arabia Airports.
\newblock In Proceedings of the 2020 16th International Computer Engineering
  Conference (ICENCO), Cairo, Egypt, 29--30 December 2020;  pp. 103--111.

\bibitem[Wang et~al.(2022)Wang, Feng, Gao, Jiang, and
  Tang]{wang2022intelligent}
Wang, J.; Feng, J.; Gao, Y.; Jiang, X.; Tang, J.
\newblock A Intelligent Cabin Design Based on Aviation Internet of Things.
\newblock In Proceedings of the 5th China Aeronautical
  Science and Technology Conference, Jiaxing, China, 3 November 2022; 
 Springer: Singapore, 2022;  pp. 963--969.

\bibitem[Jalali and Zeinali(2018)]{jalali2018smart}
Jalali, R.; Zeinali, S.
\newblock Smart Flight Security in Airport Using IOT (Case Study: Airport of
  Birjand).
\newblock {\em Int. J. Comput. Sci. Softw. Eng.} {\bf 2018}, {\em 7},~142--147.

\bibitem[Aboti(2020)]{aboti2020studies}
Aboti, C.D.
\newblock Studies of Challenges to Mitigating Cyber Risks in IoT-Based
  Commercial Aviation.
\newblock {\em Int. J. Sci. Res. Dev.}
  {\bf 2020}, {\em 7},~133--139.

\bibitem[Cho et~al.(2023)Cho, Park, et~al.]{cho2023reviewing}
Cho, S.H.; Park, S.Y.
\newblock Reviewing the Utilization of Smart Airport Security.
\newblock {\em J. Korean Soc. Aviat. Aeronaut.} {\bf
  2023}, {\em 31},~172--177.

\bibitem[Ukwandu et~al.(2022)Ukwandu, Ben-Farah, Hindy, Bures, Atkinson,
  Tachtatzis, Andonovic, and Bellekens]{ukwandu2022cyber}
Ukwandu, E.; Ben-Farah, M.A.; Hindy, H.; Bures, M.; Atkinson, R.; Tachtatzis,
  C.; Andonovic, I.; Bellekens, X.
\newblock Cyber-security challenges in aviation industry: A review of current
  and future trends.
\newblock {\em Information} {\bf 2022}, {\em 13},~146.

\bibitem[Florido-Ben{\'\i}tez(2021)]{florido2021identifying}
Florido-Ben{\'\i}tez, L.
\newblock Identifying cyber security risks in Spanish airports.
\newblock {\em Cyber Secur. Peer-Rev. J.} {\bf 2021}, {\em
  4},~267--291.

\bibitem[Adhikari(2021)]{adhikari2021intelligent}
Adhikari, S.
\newblock Intelligent Cyber Defense in 5G Augmented Aviation Cybersecurity
  Framework.
\newblock In Proceedings of the AIAA Scitech 2021 Forum, Virtual, 11--15 January 2021; p. 0661.

\bibitem[Zhang(2024)]{zhang2024research}
Zhang, C.
\newblock Research on Transparency and Optimization of Aviation Materials
  Supply Chain based on Blockchain.
\newblock In Proceedings of the 2024 IEEE 13th International Conference on
  Communication Systems and Network Technologies (CSNT), Jabalpur, India, 6--7 April 2024;  pp.
  334--339.

\bibitem[Zkik et~al.(2023)Zkik, Sebbar, Nejjari, Lahlou, Fadi, and
  Oudani]{zkik2023secure}
Zkik, K.; Sebbar, A.; Nejjari, N.; Lahlou, S.; Fadi, O.; Oudani, M.
\newblock Secure Model for Records Traceability in Airline Supply Chain Based
  on Blockchain and Machine Learning. In {\em Digital Transformation and
  Industry 4.0 for Sustainable Supply Chain Performance}; Springer: Cham, Switzerland, 2023; pp.
  141--159.

\bibitem[Bitton and Shabtai(2019)]{bitton2019machine}
Bitton, R.; Shabtai, A.
\newblock A machine learning-based intrusion detection system for securing
  remote desktop connections to electronic flight bag servers.
\newblock {\em IEEE Trans. Dependable Secur. Comput.} {\bf
  2019}, {\em 18},~1164--1181.

\bibitem[GROUP(2024)]{WinNT28}
Thales Group.
\newblock IoT Cybersecurity: Regulating the Internet of Things.
\newblock {\em THALES GROUP} {\bf 2024}.
\newblock Available online:
  \url{https://www.thalesgroup.com/en/markets/digital-identity-and-security/iot/inspired/iot-regulations} (accessed on 13-05-2024).

\bibitem[Lata and Kumar(2021)]{lata2021standards}
Lata, M.; Kumar, V.
\newblock Standards and regulatory compliances for IoT security.
\newblock {\em Int. J. Serv. Sci. Manag. Eng. Technol. (IJSSMET)} {\bf 2021}, {\em 12},~133--147.

\bibitem[Foundation(2022)]{WinNT27}
Open Connectivity Foundation.
\newblock OCF Solving the IoT Standards Gap.
\newblock {\em Open Connectivity Foundation} {\bf 2022}.
\newblock Available online:
  \url{https://openconnectivity.org/what-to-expect-from-an-interoperable-iot/} (accessed on 13-03-2023).

\bibitem[OCF(2023)]{OCF}
Open Connectivity Foundation.
\newblock OCF Solving the IoT Standards Gap.
\newblock 2023.
\newblock Available online: \url{https://openconnectivity.org/} (accessed on 25-04-2024).

\bibitem[Ye et~al.(2004)Ye, Heidemann, and Estrin]{ye2004medium}
Ye, W.; Heidemann, J.; Estrin, D.
\newblock Medium access control with coordinated adaptive sleeping for wireless
  sensor networks.
\newblock {\em IEEE/ACM Trans. Netw.} {\bf 2004}, {\em
  12},~493--506.

\end{thebibliography}
\PublishersNote{}
 \end{adjustwidth}
\end{document}